\documentclass[aps,pre,superscriptaddress,groupedaddress]{revtex4}
\usepackage{enumitem}
\usepackage{graphicx,amssymb,natbib,bm}
\usepackage[caption=false]{subfig}
\usepackage{float}
\newcommand{\n}{{\mathbf{n}}}
\newcommand{\N}{{\mathbf{N}}}
\newcommand{\A}{{\mathbf{A}}}
\begin{document}
\title[Stochastic pair approximation]{Stochastic pair approximation treatment of the noisy voter model}
\author{A. F. Peralta}
\address{IFISC, Instituto de F{\'\i}sica Interdisciplinar y Sistemas Complejos (UIB-CSIC), Campus Universitat de les Illes Balears, 07122-Palma de Mallorca, Spain}
\author{A. Carro}
\address{Institute for New Economic Thinking at the Oxford Martin School, University of Oxford, OX2 6ED, UK}
\address{Mathematical Institute, University of Oxford, OX2 6GG, UK}
\author{M. San Miguel}
\address{IFISC, Instituto de F{\'\i}sica Interdisciplinar y Sistemas Complejos (UIB-CSIC), Campus Universitat de les Illes Balears, 07122-Palma de Mallorca, Spain}
\author{R. Toral}
\address{IFISC, Instituto de F{\'\i}sica Interdisciplinar y Sistemas Complejos (UIB-CSIC), Campus Universitat de les Illes Balears, 07122-Palma de Mallorca, Spain}
\begin{abstract}
We present a full stochastic description of the pair approximation scheme to study binary-state dynamics on heterogeneous networks. Within this general approach, we obtain a set of equations for the dynamical correlations, fluctuations and finite-size effects, as well as for the temporal evolution of all relevant variables. We test this scheme for a prototypical model of opinion dynamics known as the noisy voter model that has a finite-size critical point. Using a closure approach based on a system size expansion around a stochastic dynamical attractor we obtain very accurate results, as compared with numerical simulations, for stationary and time dependent quantities whether below, within or above the critical region. We also show that finite-size effects in complex networks cannot be captured, as often suggested, by merely replacing the actual system size $N$ by an effective network dependent size $N_{{\rm eff}}$.
\end{abstract}
\maketitle

\section{Introduction}

From classical problems in statistical physics \cite{Gunton1983,Marro1999} to questions in biology and ecology \cite{Hopfield1982,Clifford1973,Crawley1987}, and over to the spreading of opinions and diseases in social systems \cite{Anderson1991,PastorSatorras2001,Watts2002,Castellano2009,Castellano2012}, stochastic binary-state models have been widely used to study the emergence of collective phenomena in systems of stochastically interacting components. In general, these components are modeled as binary-state variables ---spin up or down--- sitting at the nodes of a network whose links represent the possible interactions among them. While initial research focused on the limiting cases of a well-mixed population, where each of the components is allowed to interact with any other, and regular lattice structures, later works turned to more complex and heterogeneous topologies \cite{Albert2002,Newman2003,Barrat2008,Newman2010}. A most important insight derived from these more recent works is that the macroscopic dynamics of the system can be greatly affected by the particular topology of the underlying network. In the case of systems with critical behavior, different network characteristics have been shown to have a significant impact on the critical values of the model parameters \cite{Lambiotte2007,Gleeson2011,Gleeson2013}, such as the critical temperature of the Ising model \cite{Dorogovtsev2002,Leone2002,VianaLopes2004} and the epidemic threshold in models of infectious disease transmission \cite{Boguna2003,Durrett2010,Castellano2010,Parshani2010,Satorras2015}. Thus, the identification of the particular network characteristics that have an impact on the dynamics of these models, as well as the quantification of their effect, are of paramount importance.

The first theoretical treatments introduced for the study of stochastic, binary-state dynamics on networks relied on a global-state approach \cite{Marro1999, Alfarano2009}, i.e. they focused on a single variable ---for example, the number of nodes in one of the two possible states--- assumed to represent the whole state of the system. In order to write a master equation for this global-state variable, some approximation is required to move from the individual particle transition rates defining the model to some effective transition rates depending only on the chosen global variable. Within this global-state approach, the \emph{effective field} approximation assumes that all nodes have the same rate of switching to the other state, such that local densities can be replaced by their global equivalents and all details of the underlying network are completely lost. As a consequence, results are only accurate for highly connected homogeneous networks, closer to fully connected topologies.

More recent theoretical treatments have proposed a node-state approach, departing from a master equation for the $N$-node probability distribution and, thus, taking into account the state of each individual node. From this master equation, general evolution equations for the moments and correlations of the relevant dynamical variables can be derived. Depending on the particular functional form of the individual particle transition rates, this system of dynamical equations can be open ---an infinite hierarchy of equations, each of them depending on higher-order correlation functions--- or closed. For open systems of equations, different schemes have been proposed for their closure: a \emph{mean-field} approximation \cite{Satorras2015,Chakrabarti2008,Gomez2010}, in which all moments are replaced by products of individual averages, and a \emph{van~Kampen} type system-size expansion~\cite{vanKampen1981,Lafuerza2013}. Other variants of the moments expansion have been previously considered in the literature with different types of moment closure assumptions \cite{Keeling,Rand,Demirel,Bauch}, see also the review \cite{Kuehn} of these techniques. Even when the system of equations is closed, further approximations are required to deal with the appearance of terms depending on the specific network structure. In particular, an \emph{annealed network} approximation has been successfully used in a number of models \cite{Sood2005B,Sood2008,Vilone2004,Dorogovtsev2008,Guerra2010,Sonnenschein2012,Carro2016} in order to deal with these terms. This approximation involves replacing the original network by a complementary, weighted, fully connected network, whose weights are proportional to the probability of each pair of nodes being connected. In this way, highly precise analytical solutions can be obtained for all relevant dynamical variables.

A third group of analytical treatments has sought to establish an intermediate level of detail in its description of the dynamics, in-between the global-state and the node-state approaches. Similar to the global-state method, though significantly improving on it, a reduced set of variables is chosen to describe the state of the system. In particular, information is aggregated for nodes of a given degree. In this manner, some essential information about the network structure is kept ---different equations for nodes with different degrees--- while the tractability of the problem is greatly improved. As with the global-state approach, in order to write a master equation for this reduced set of variables, some further approximation is required to translate the individual particle transition rates that define the model into some effective transition rates depending only on the chosen set of variables. The number of these variables, as well as their nature, define different levels of approximation, which can be ordered from more to less detailed as: Approximate Master Equation (AME) \cite{Gleeson2011,Gleeson2013}, Pair Approximation (PA) \cite{Vazquez2008,Vazquez2008b,Vazquez2010,Oliveira1993,Dickman1986,Pugliese2009,Ferreira2014} and Heterogeneous Mean Field (HMF) \cite{Satorras2015,Sood2008,Vespignani2011,Ferreira2011}. Many works based on these approaches have further neglected dynamical correlations, fluctuations and finite-size effects, which corresponds to a mean field type of analysis. While there have been attempts to relax these restrictions and deal with finite-size effects, these efforts have been usually based on an adiabatic elimination of variables \cite{Sood2008,Vazquez2008,Pugliese2009,Ferreira2014,Ferreira2011,Ferreira2011B,Peralta2018}, whose range of validity is limited.

In this paper we consider this intermediate level of description focusing on the Pair Approximation that chooses as a reduced set of relevant variables the number of nodes with a given degree in a given state and in the number of links connecting nodes in different states. We develop a full stochastic treatment of the pair approximation scheme in which we avoid mean-field analysis or adiabatic 
elimination of variables. In particular, after writing a master equation for the reduced set of variables, we obtain effective 
transition rates by introducing the pair approximation as described in Ref.~\cite{Vazquez2008} and elaborating on the assumptions 
used on this approach as well as on their quantitative implications.
Furthermore, after writing general equations for the moments and correlations, we propose two different strategies for their closure and solution. The first one, that we term \textbf{S1PA}, is based on a van~Kampen type system-size expansion~\cite{vanKampen1981,moments} around the deterministic solution of the dynamics, that is, around the lowest order term of the expansion which corresponds to the thermodynamic limit of infinite system size. This approach allows us to derive linear equations for the correlation matrix and first-order corrections to the average values. The second strategy, that we term \textbf{S2PA}, is based on a system-size expansion around a dynamical attractor, and it allows us to obtain expressions not only for the stationary value of all relevant quantities but also for their dynamics.

Although our methodology is very general for any set of transition rates, only for the first closure strategy \mbox{---\textbf{S1PA}---} one can apply straightforwardly our general method to any model under study, to end up with the desired equations. The second one \mbox{---\textbf{S2PA}---} requires specific rates for the analysis to proceed. Thus, for the sake of concreteness, we focus on the noisy voter model as an explicit example for which we carry out a full analytical treatment. The noisy voter model is a variant of the original voter model \cite{Clifford1973,Holley1975} which, apart from pairwise interactions in which a node copies the sate of a randomly selected neighbor, it also includes random changes of state. It is a paradigmatic example of a stochastic binary-state model, with applications in the study of non-equilibrium systems in a wide range of fields, and it has been studied by, at least, five mutually independent strands of research, largely unaware of each other. Namely, random processes in genetics \cite{Moran1958}, percolation processes in strongly correlated systems \cite{Lebowitz1986}, heterogeneous catalytic chemical reactions \cite{Fichthorn1989,Considine1989}, herding behavior in financial markets \cite{Kirman1993}, and probability theory \cite{Granovsky1995}. The behavior of the noisy voter model is characterized by the competition between two mechanisms: On the one hand, the pairwise copy interactions tend to order the system, driving it towards a homogeneous configuration ---all spins in the same state, whether up or down---. On the other hand, the random changes of state tend to disorder the system, pulling it away from the homogeneous configurations. These homogeneous configurations are the absorbing states of the ordinary voter model, which lose their absorbing character due to the random changes of state. Depending on the values of the model parameters, which set the relative strength of one mechanism over the other, the system can be found in a mostly ordered regime ---dominated by pairwise interactions--- or a mostly disordered regime ---dominated by noise---. Furthermore, a noise-induced, finite-size transition can be observed between these two behavioral regimes. The finite-size character of the transition is due to the fact that the critical point tends to zero in the thermodynamic limit of large system sizes. Although most of the initial literature about the noisy voter model focused only on regular lattices \cite{Lebowitz1986,Granovsky1995} and a fully-connected network \cite{Kirman1993,Alfarano2008}, recent studies have addressed more complex topologies, both from an effective field perspective \cite{Alfarano2009,Alfarano2013,Diakonova2015} and using an annealed network approximation \cite{Carro2016}. In particular, while the effective field approach was only able to broadly capture the effect of the network size and mean degree on the results of the model for highly homogeneous and connected networks, the annealed approximation \cite{Carro2016} was, in addition, able to reproduce the impact of the degree heterogeneity ---variance of the underlying degree distribution--- on the critical point of the transition and the temporal correlations with a high level of accuracy, as well as the main effects on the local order parameter, though with significantly less accuracy. Further studies of the noisy voter model include the effects of zealots \cite{Nagi}, nonlinear group interactions \cite{Peralta2018} and ageing \cite{Oriol}.

When applied to the noisy voter model, the first of the approaches proposed in this paper ---\textbf{S1PA}--- is able to improve on the accuracy of previous methods only for the region above the critical point, while it fails in its proximity and in the region below it. This failure is due to the existence of a mathematical divergence as the noise intensity approaches zero in the analytical expressions obtained for the fluctuations and the local order parameter. The second approach proposed in this paper ---\textbf{S2PA}---, on the contrary, leads to highly accurate results for all relevant variables and for all values of the parameters, whether below, within, or above the critical region, thus significantly improving on previous methods. Furthermore, this approach is not limited to stationary values, but rather it is able to provide highly accurate analytical expressions for their time evolution. Finally, this second approach allows us to show that finite size effects in a complex network can not be simply reduced to replacing the system size by an effective one, as most previous studies suggest \cite{Sood2008,Vazquez2008,Pugliese2009,Ferreira2014,Ferreira2011}.

Summing up, this paper contributes to the analysis of stochastic binary-state models in three main ways. First, it presents a general methodology with clearly identified and well-justified arguments and approximations, including an analytical assessment of their validity ranges. Second, our approach allows to find analytical expressions for the correlations, fluctuations and finite-size effects, as well as for the temporal evolution of all relevant variables, without needing to resort to crude adiabatic elimination of variables. Third, the results obtained for the noisy voter model represent a significant improvement in accuracy, being valid both above and below the critical point as well as in the critical region.

The paper is organized as follows: in Section~\ref{sec:model} we introduce the stochastic model and define the main quantities of interest namely, the fluctuations of the density of nodes in a particular state and the density of active links. In Section~\ref{sec:previous} we revisit some of the previous theoretical treatments introduced for the study of the stochastic binary-state dynamics in the two limiting cases of a global-state approach and a node-state approach. In the same section we explain the main approximations required to close and eventually solve the dynamical equations for the moments and correlations of the relevant dynamical variables. In Section~\ref{sec:spa} we introduce our stochastic description based on a master equation for a reduced set of dynamical variables. In this section we use the pair approximation to derive the hierarchy of equations for the moments and correlations of the dynamical variables. Two different closure schemes are presented and discussed in Sections~\ref{sec:expansion1} and~\ref{sec:expansion2} based, respectively, on an expansion around the deterministic solution and around a stochastic dynamical attractor. The main results of both schemes as well as other approaches in the literature are compared against the results of numerical simulations in Section~\ref{sec:numerical} in the steady state. In Section~\ref{sec:dynamics} we extend our method to determine the dynamical evolution of the correlation function and its comparison with numerical simulations. Finally, in Section~\ref{sec:conclusions} we present the main conclusions of our study. The technical details of the solutions of some equations are presented in \ref{app:ckk} and \ref{app:riccati}. In \ref{app:linear} we present a simple example using linear equations in which one of our closure schemes can be carried out in full detail. In \ref{app:link} we explain the method of adiabatic elimination to obtain a partial solution of some of the variables under study.

\section{Model}\label{sec:model}

We consider an undirected network consisting of $N$ nodes. Links between nodes are mapped into the usual (symmetric) adjacency matrix $\A=(A_{ij})$ as $A_{ij}=1$ if nodes $i$ and $j$ are connected, and $A_{ij}=0$ otherwise. Connected nodes are ``neighbors'' of each other. Each node $i=1,\dots,N$ holds a time-dependent binary variable $n_i=0,1$. We do not adopt in this paper any specific interpretation, but the state $n_i=1$ (resp. 0) will be labeled as ``{\slshape up}'' (resp. ``{\slshape down}''), and a link between nodes in different states will be defined as {\slshape active}. A stochastic dynamics is introduced by which the node variable $n_i(t)$ can switch its state at time $t$, either $0\to 1$ with rate $r_i^+$, or $1\to 0$ with rate $r_i^-$. The rates $r_i^\pm(\n)$ depend, in general, on the whole set of node states $\n=(n_1,\dots,n_N)$ but are assumed to not depend explicitly on time. Although our method is rather general, we will use throughout the paper the noisy voter (Kirman) model rates as an archetypal example
\begin{equation}
\label{ind_rates}
r_i^+=a+\frac{h}{k_i}\sum_jA_{ij}n_j,\quad r_i^-=a+\frac{h}{k_i}\sum_jA_{ij}(1-n_j),
\end{equation}
being $k_i=\sum_j{A_{ij}}$ the degree (number of connected nodes) of node $i$. The term proportional to $h$ represents a ``herding'' mechanism by which the transition rates are proportional to the fraction of neighbor nodes holding the opposite state, while the term proportional to $a$ represents random jumps between states at a constant rate and is also known as the ``noise term''.

We introduce the probability $P(\n;t)$ of finding node configuration $\n$ at time $t$. This probability can be understood at different levels, as the detailed configuration $\n(t)$ depends on (i) the initial condition, (ii) the realization of the stochastic process, and (iii) the network realization $\A$. For a given network configuration, let $N_k=\#\{i|\sum_j A_{ij}=k\}$ be the number of nodes with degree $k$. By definition, $N_k=0$ for $k<k_{\rm min}$ and for $k>k_{\rm max}$, the minimum and maximum values of the degree for that given network. We introduce
\begin{equation}
P_k=\frac{N_k}{N},\quad \mu_{m}=\sum_{k=k_{\rm min}}^{k_{\rm max}} P_k k^{m},
\end{equation}
as the probability $P_k$ that a randomly chosen node has degree $k$, and the $m-$moment $\mu_{m}$ of the degree distribution, and for brevity $\mu \equiv \mu_{1}$. The variance of the degree distribution is $\sigma_k^2=\mu_2-\mu^2$ and we introduce the degree heterogeneity $\kappa=\sigma_k^2/\mu^2$.

We will focus on the time evolution of the first moment $\langle m (t)\rangle$ and variance $\sigma^2[m(t)]=\langle m^2 (t)\rangle-\langle m (t)\rangle^2$ of the rescaled density of nodes in the up-state, as well as on the average $\langle \rho(t)\rangle$ of the density of active links. The precise definitions of these quantities are
\begin{eqnarray}
m&=&\frac{2}{N}\sum_{i} n_i - 1,\label{mdef}\\
\rho&=&\frac{\frac12\sum_{ij}A_{ij}(n_i(1-n_j)+(1-n_i)n_j)}{\frac12\sum_{ij}A_{ij}}\equiv\frac{L}{\frac12 \mu N},\label{rhodef}
\end{eqnarray}
where $L$ is the number of active links, and the denominator $\frac12N\mu=\frac12\sum_{ij}A_{ij}$ is the total number of links. While the ``magnetization'' $m(t)\in [-1,1]$ describes the global state of the system, $\rho(t)\in[0,1]$ can be considered as a measure of disorder, corresponding the value $\rho=1/2$ to a random distribution of states, and $\rho=0$ to full order or consensus where all nodes are in the same state, either $n_i=0$ or $n_i=1$. All averages $\langle F[\n(t)] \rangle$ are understood with respect to the probability $P(\n;t)$
\begin{equation}
\label{average_def}
\langle F[\n(t)] \rangle=\sum_{n_1=0,1; \cdots; n_N=0,1}F(\n)P(\n;t).
\end{equation}
Results can be later averaged over the ensemble of networks with a given degree distribution, always after performing the average in Eq.~\ref{average_def}.

We also introduce the correlation functions
\begin{eqnarray}
C_{ij}(t)&=&\langle \Delta_i(t)\Delta_j(t)\rangle=\langle n_i(t)n_j(t)\rangle-\langle n_i(t)\rangle\langle n_j(t)\rangle,\\
C_{ijk}(t)&=&\langle \Delta_i(t)\Delta_j(t)\Delta_k(t)\rangle,\\
\cdots\cdots&\cdots&\cdots\cdots\nonumber
\end{eqnarray}
where $\Delta_i=n_{i}-\langle n_{i}\rangle$.
As $n_i^2=n_i$, note that $C_{ii}=\langle n_i\rangle-\langle n_i\rangle^2$ and that
\begin{eqnarray}
\sigma^2[m(t)]&=&\frac{4}{N^2}\sum_{i,j}C_{ij},\label{s2fromsigma}\\
\langle\rho(t)\rangle&=&\frac{\sum_{ij}A_{ij}\left(\langle n_i\rangle(1-\langle n_j\rangle)-C_{ij}\right)}{\frac12\mu N}.\label{rhofromsigma}
\end{eqnarray}

\section{Node-state and Global-state approaches}\label{sec:previous}

\subsection{The node-state approach}

For general rates $r_i^\pm$, the probability function satisfies the master equation~\cite{vanKampen1981,Toral2014}
\begin{equation}
\label{me1}
\frac{\partial P(\n;t)}{\partial t}= \sum_{i=1}^N\left\{(E_i^+-1)\left[n_ir_i^-P(\n;t)\right]+(E_i^--1)\left[(1-n_i)r_i^+P(\n;t)\right]\right\},
\end{equation}
being $E_i$ the step operator $E_{i}^{\pm} f(n_1, ..., n_i, ..., n_{N}) = f(n_1, ..., n_i \pm 1, ..., n_{N})$. From the master equation it is possible to get general evolution equations for the moments as~\footnote{For a step-by-step derivation, see the Supplementary Information of Ref.~\cite{Carro2016}}
\begin{eqnarray}
\frac{d\langle n_i(t)\rangle}{dt}&=& \langle r_i^+\rangle -\langle (r_i^++r_i^-)n_i\rangle,\label{nit}\\
\frac{dC_{ij}(t)}{dt}&=&\langle(r_i^-+r_i^+)n_i\Delta_j\rangle+\langle(r_j^-+r_j^+)n_j\Delta_i\rangle+\langle r_i^+\Delta_j\rangle +\langle r_j^+\Delta_i\rangle,\quad i\ne j.
\label{sijt}
\end{eqnarray}

To proceed forward, one would need to solve these equations, introducing explicit expressions for the transition rates $r_i^\pm$ and finding solutions for $\langle n_i(t)\rangle$ and $C_{ij}(t)$, which can then be introduced in equations (\ref{mdef}), (\ref{s2fromsigma}), and (\ref{rhofromsigma}) in order to obtain expressions for the desired quantities $\langle m (t)\rangle$, $\sigma^2[m(t)]$, and $\langle \rho(t)\rangle$. When introducing explicit expressions for $r_i^\pm$ for any given model, however, one might find two different cases depending on whether the resulting system of equations is closed or not, which will determine the need for different types of approximations:
\begin{itemize}
 \item \textbf{Non-closed system of equations:} With the remarkable exception of the noisy voter model, the set of equations~(\ref{nit}-\ref{sijt}) is, in general, not closed. In particular, when the explicit expressions for $r_i^\pm$ are introduced, the right-hand-side of the equations will contain higher-order correlation functions, $C_{i_1i_2\dots i_m}$ with $m\ge 3$. While it is also possible to obtain evolution equations for these higher-order correlation functions, the resulting hierarchy of equations is, in principle, infinite. Thus, some approximations are needed in order to close this hierarchy. In general, there are two basic schemes for the closure of equations (\ref{nit}) and (\ref{sijt}) (see \cite{Kuehn} for alternative techniques):
 \begin{itemize}
 \item \textbf{Mean-field approximation:} In the mean-field approximation, all moments are replaced by products of individual averages, e.g.
 \begin{eqnarray}
 \langle n_i(t)n_j(t)\rangle &\approx& \langle n_i(t)\rangle\langle n_j(t)\rangle,\label{mf1}\\
 \langle n_i(t)n_j(t)n_k(t)\rangle &\approx& \langle n_i(t)\rangle \langle n_j(t)\rangle\langle n_k(t)\rangle,\nonumber\\
 &\cdots&\nonumber
 \end{eqnarray}
 Within this mean-field approximation, Eq.~(\ref{nit}) is closed and, as $C_{ij}=0$, Eq.~(\ref{sijt}) is not needed.
 \item \textbf{System-size scaling hypothesis of the correlation functions:} The usual van~Kampen system-size expansion~\cite{vanKampen1981} splits a relevant dynamical variable as $x(t)=\langle x(t)\rangle +\Omega^{1/2}\xi(t)$, being $\Omega$ a large parameter (typically the system size or volume). This assumes that the dynamical variable scales as $x(t)\sim \Omega$. When trying to apply this approach to the master equation~(\ref{me1}) one faces the fact the dynamical variables $n_i(t)=0,1$ do not scale with system size $N$. A more sophisticated expansion was developed in Ref.~\cite{Lafuerza2013} with the main ansatz that the correlation functions scale as $C_{i_1i_2\dots i_m}=\langle \Delta_{i_1}\Delta_{i_2}\cdots\Delta_{i_m}\rangle=O(\Omega^{-m/2})$ for all $i_1,i_2,\dots, i_m$ different. Once this ansatz is assumed, it turns out that one can still use the mean-field results (\ref{mf1}) in the evolution equations for the mean values (\ref{nit}), thus finding for $\langle n_i(t)\rangle$ the same result as the mean-field approximation. Regarding the correlation $C_{ij}$, however, this approach leads to non-trivial evolution equations, allowing one to improve on the mean-field approach.
 \end{itemize}
 \item \textbf{Closed system of equations:} In the particular case of the noisy voter model, and due to its linear transition rates, Eqs. (\ref{nit}) and (\ref{sijt}) are already closed, and thus there is no need for any closure approximation. Of course, closing the equations is just the first step towards solving them. Typically, after closing the equations one finds factors of the form $\sum_{j}A_{ij}n_j$, whose complexity for a general lattice structure makes further analytical progress very difficult. These terms can be dealt with by assuming the:
 \begin{itemize}
 \item \textbf{Annealed approximation:} This approach replaces the original adjacency matrix $A_{ij}$ by a complementary, weighted, fully connected adjacency matrix $\widetilde{A}_{ij}$, whose weights are proportional to the probability of each pair of nodes being connected~\cite{Vilone2004,Dorogovtsev2008,Guerra2010,Sonnenschein2012}. For uncorrelated networks of the configuration ensemble this can be written as
 \begin{equation}
 \widetilde{A}_{ij} \approx \frac{k_ik_j}{N \mu},
 \end{equation}
 where the normalization of the weights is chosen so that both the degree sequence and the total number of links remain unchanged~\cite{Newman2003B,Boguna2004,Sood2008,Bianconi2009}. This approximation reduces the aforementioned sums to
 \begin{equation}
 \sum_{j=1}^NA_{ij}f_j \approx \sum_{j=1}^N \widetilde{A}_{ij} f_j = \frac{k_i}{N \mu}\sum_{j=1}^Nk_jf_j,
 \end{equation}
 being $f_j[n_j,k_j]$ any function that depends on the degree $k_j$ of the node $j$ and/or on the node variable $n_j$. In the particular case that $f_j[n_j,k_j]=f[k_j]$, a further simplification is possible in terms of the degree distribution $P_k=N_k/N$ as
 \begin{equation}
 \frac{1}{N}\sum_{j=1}^Nk_jf_j=\sum_k P_kk f(k) \, .
 \end{equation}
 There is one case in which the annealed approximation is exact. This is when the adjacency matrix is $A_{ij}=1,\,\forall i,j$, meaning that every node is connected to every other node, a situation known as {\bfseries all-to-all} coupling or fully connected network. The results of the treatment in terms of the full set of node-states and the annealed approximation have been described in Ref.~\cite{Carro2016}.
 \end{itemize}
\end{itemize}
The main results for the noisy voter model using the mean-field and the annealed approximations will be compared in Section~\ref{sec:numerical} with our new treatment using a stochastic pair approximation scheme. Before that, let us explain in the next subsection yet another, very simple, approximation which is capable of capturing the main phenomenological features of the noisy voter model.

\subsection{The global-state approach}

Instead of the full state $\n=(n_1,\dots,n_N)$, this approach proposes a much coarser description, in which the only relevant variable is the number of nodes in the up state $N_1=\#\{i | n_i=1\} $. This is clearly an important reduction with respect to the $N$ variables needed in the $\n$ representation, leading to a much simpler analytical treatment but also to less accurate results in general network topologies and exact results only for fully connected networks. As far as the global variable $N_1$ is concerned, there are only two possible outcomes of the stochastic process: $N_1\to N_1+1$ and $N_1\to N_1-1$. Let $P(N_1;t)$ be the probability that the number of nodes in the up state takes the value $N_1$ at time t. It obeys the master equation
\begin{eqnarray}\label{me2}
\frac{\partial P(N_1;t)}{\partial t}&=& \left(E_{N_1}^+-1\right)\left[W^-P(N_1;t)\right]+ \left(E_{N_1}^--1\right)\left[W^+P(N_1;t)\right],
\end{eqnarray}
where $W^+$ and $W^-$ are effective rates, depending only on the global variable $N_1$, and the step operator now acts as $E_{N_1}^{\pm}F(N_1)=F(N_1\pm 1)$. From here we derive equations for the average value and the second moment of the global variable as
\begin{eqnarray}
\frac{d\langle N_1\rangle}{dt}&=&\langle W^+-W^-\rangle,\label{eq:N1}\\
\frac{d\langle N_1^2\rangle}{dt}&=&\langle (1-2N_1)W^-+(1+2N_1)W^+\rangle\label{eq:N2}.
\end{eqnarray}
Some approximation is needed to find the effective rates $W^+$ and $W^-$ in such a way that they depend only on the $N_1$ variable and in a closed form. The simplest approximation is to assume that all nodes have the same rate of switching to the other state, such that we can replace the local density of nodes in the up state by the global density of nodes in the up state,
\begin{equation}
\frac{\sum_j A_{ij}n_j}{k_i} \approx \frac{N_1}{N} \, ,
\end{equation}
leading to
\begin{eqnarray}
W^+&=&(N-N_1)\left(a+h\frac{N_1}{N}\right),\label{WpN1}\\
W^-&=&N_1\left(a+h\frac{N-N_1}{N}\right)\label{WmN1}.
\end{eqnarray}
Note that in this {\bfseries effective-field} approximation, the details of the network (any dependence on the connectivity matrix) are completely lost. In the case of a fully connected network, where each node is connected to every other node, the effective-field approximation is exact.

Once the effective-field approximation has been assumed, no further approximations are needed, for after replacing Eqs. (\ref{WpN1}--\ref{WmN1}) in Eqs. (\ref{eq:N1}--\ref{eq:N2}), the evolution equations for the mean value $\langle m(t)\rangle$ and the variance $\sigma^2[m(t)]$ of \mbox{$m=2 N_1/N-1$} satisfy
\begin{eqnarray}
\label{first_FC_temp}
\frac{d\langle m\rangle}{dt}&=&-2a\langle m\rangle,\\
\label{second_FC_temp}
\frac{d\sigma^2[m]}{dt}&=&-2\left(2a+\frac{h}{N}\right)\sigma^2[m]-\frac{2h}{N}\langle m\rangle^2+\frac{4a+2h}{N}.
\end{eqnarray}
The density of active links $\rho$ can be calculated using the all-to-all ansatz as the probability of finding a node in the up state times the probability of finding one of its neighbors in the down state or vice versa \mbox{$\rho= 2 \frac{N_1}{N} \frac{N-N_1}{N-1} = \frac{N}{N-1}\frac{1- m^2}{2}$}. In the steady state we have
\begin{eqnarray}
\label{first_FC}
\langle m\rangle_{\rm st} &=& 0, \phantom{\frac{1}{1}}\\
\label{second_FC}
\sigma^2_{\rm st}[m]&=&\frac{2a+h}{2aN+h},\\
\langle \rho\rangle_{\rm st}&=&\frac{aN}{2aN+h}.\label{rho_FC}
\end{eqnarray}
Further analysis shows that the noisy voter model has a transition separating regions in which the steady state probability $P_{\rm st}(N_1)$ has a single peak at $N_1=\frac{N}{2}$ and regions in which it has two peaks at $N_1=0,N$. The critical condition of the transition is thus that the probability distribution becomes flat $P_{{\rm st}}(N_1)=\frac{1}{N+1}$ for $N_1=0,\dots,N$. This condition leads to a variance $\sigma_{\rm st}^2[m]=\frac{N+2}{3N}$, from which we can obtain the critical value $a_c = h/N$, such that $P_{\rm st}(N_1)$ is double peaked for $a<a_c$ and single peaked for $a>a_c$ (see \cite{Kirman1993,Alfarano2008} for a different method of finding this critical value). These predictions for the transition point and the detailed dependence of $\sigma_{\rm st}^2[m]$ and $\langle\rho\rangle_{\rm st}$ on the system parameters will be analyzed in the next sections.

As the critical value $a_c=h/N$ depends on $N$, the single to double peaked transition observed for $P_{\rm st}(N_1)$ is said to be a finite-size-transition~\cite{Toral2007}. Certainly, the transition disappears in the thermodynamic limit $N\to\infty$ where, for all values of $a$, the distribution $P_{\rm st}(N_1)$ is always single peaked. Therefore, from the perspective of a strict statistical mechanics formulation, the transition does not exist. Nevertheless, the behavior near $a=0$ bears many similarities with a true phase transition, as there are valid scaling laws, divergence of fluctuations with characteristic exponents, critical slowing down, etc., as we will show in the next sections. For this reason, and being clearly an abuse of language, it is customary to refer to the small $a\sim h/N$ values as the ``critical region''. It is also possible to identify the critical dimension $d_c=2$ above which the ``critical'' exponents are independent of dimension~\cite{Granovsky1995}.

As shown in Eq. (\ref{second_FC}), which, we recall, is exact in the all-to-all coupling scenario, when $a/h=O(N^{-1})$, the fluctuations scale as $\sigma[m]\sim O(N^{0})$, while far away from the critical point, $a/h=O(1)$, they scale as $\sigma[m]\sim O(N^{-1/2})$. As the critical dimension of this model is $d_c=2$, this scaling with system size is the correct one in the all-to-all coupling.
This simple global approach captures one of the main difficulties we will encounter when closing the equations for the moments in more complicated setups, namely, the fact that the fluctuations scale differently near the critical point and far away from it. A standard van~Kampen approach~\cite{vanKampen1981} consisting in splitting $m(t)$ in deterministic $\phi(t)$ and stochastic $\xi(t)$ contributions $m(t)=\phi(t)+N^{-1/2}\xi(t)$, will lead invariably to fluctuations of the order $\sigma[m]=O(N^{-1/2})$, and it is hence not appropriate near the critical point.

\section{Stochastic pair approximation}\label{sec:spa}

In this section, we introduce the stochastic pair approximation, starting from a master equation description of the dynamics and using an intermediate level of detail in our characterization, in-between the node-state and the global-state approaches. Our method is a full stochastic treatment of the pair approximation considered by Vazquez and Egu\'{i}luz~\cite{Vazquez2008} in their study of the (noiseless) voter model. We elaborate further on this approach and we apply it to the case of the noisy voter model. We derive in this section the evolution equations for the moments and correlations of the dynamical variables and leave for the next two sections the closure of the hierarchy and the solution of the equations.

As explained, the starting point is not the full node-state configuration $\n=(n_1,\dots,n_N)$, but a reduced set of variables $\{\N_1,L\}$, where $L$ is the number of active links and $\N_1=(N_{1,k_{\rm min}},N_{1,k_{\rm min}+1},\dots, N_{1,k_{\rm max}})$ with $N_{1,k}$ the number of nodes in the up state with degree $k$. The total number of variables in this description is thus $k_{\rm max}-k_{\rm min}+2$, still much smaller than the $N$ variables needed in the $\n$ representation.

In the elementary process $n_i=0 \to n_i=1$, we have $N_{1,k_i} \to N_{1,k_i}+1$, being $k_i$ the degree of node $i$. At the same time, the number of active links $L$ can vary in different amounts $L\to L+k_i-2q_i$ depending on the number $q_i\in(0,k_i)$ of neighbors of node $i$ in the up state. Similarly, in the elementary process $n_i=1\to n_i=0$ it is $N_{1,k_i} \to N_{1,k_i}-1$ and $L$ can vary in different amounts $L\to L-k_i+2q_i$. Introducing $P(\N_1,L ; t)$ as the probability of the state $\{\N_1,L\}$ at time $t$, we derive the master equation
\begin{eqnarray}
\label{me3}
 \frac{\partial P(\N_1,L ; t)}{\partial t} &=& \sum_{k} \sum_{q=0}^{k} \bigg\{ \left( E_{N_{1,k}}^+ E_{L}^{k-2q} - 1 \right) \left[ W^{(-,k,q)} P(\N_1,L ; t) \right] \nonumber\\
 &+& \left( E^{-}_{N_{1,k}} E_{L}^{-k+2q} - 1 \right) \left[ W^{(+,k,q)} P(\N_1,L ; t) \right] \bigg\},
\end{eqnarray}
with $E_{N_{1,k}}$ and $E_{L}$ the step operators acting on $N_{1,k}$ and $L$, respectively. From this master equation and following standard techniques, one can derive the equations for the moments and correlations of the rescaled variables \mbox{$m_k=2\frac{ N_{1,k}}{N_k}-1$} and $\rho=\frac{L}{\frac12\mu N}$ as
\begin{eqnarray}
\frac{d\langle m_k\rangle}{dt}&=&\left\langle F_k\right\rangle,\label{dtmk}\\
\frac{d\langle \rho\rangle}{dt}&=&\left\langle F_\rho\right\rangle,\label{dtrhoL}\\
\frac{d\langle m_k m_{k'}\rangle}{dt}&=&\left\langle m_{k'}F_k\right\rangle+\left\langle m_{k}F_{k'}\right\rangle+\delta_{k,k'} \frac{\left\langle G_k\right\rangle}{N},\label{dtmkmk}\\
\frac{d\langle m_k \rho \rangle}{dt}&=&\left\langle \rho F_k\right\rangle+\left\langle m_{k}F_{\rho}\right\rangle+ \frac{\left\langle G_{k,\rho}\right\rangle}{N},\label{dtmkrho}\\
\frac{d\langle \rho^2 \rangle}{dt}&=& 2\left\langle \rho F_{\rho} \right\rangle + \frac{\left\langle G_{\rho}\right\rangle}{N},\label{dtrho2}
\end{eqnarray}
where
\begin{eqnarray}
\label{def_Fk}
F_{k} &=& \frac{2}{N_{k}} \sum_{q=0}^{k} \left[ W^{(+,k,q)} - W^{(-,k,q)} \right],\\
\label{def_Frho}
F_{\rho} &=& \frac{2}{\mu N} \sum_{k} \sum_{q=0}^{k} (k-2 q) \left[ W^{(+,k,q)} - W^{(-,k,q)} \right],\\
\label{def_Gk}
G_k &=& \frac{4}{P_{k} N_{k}} \sum_{q=0}^{k} \left[ W^{(+,k,q)} + W^{(-,k,q)} \right],\\
\label{def_Gkrho}
G_{k,\rho} &=& \frac{4}{\mu N_{k}} \sum_{q=0}^{k} (k-2 q) \left[ W^{(+,k,q)} + W^{(-,k,q)} \right],\\
\label{def_Gho}
G_{\rho} &=& \frac{4}{\mu^2 N} \sum_{k} \sum_{q=0}^{k} (k-2 q)^2 \left[ W^{(+,k,q)} + W^{(-,k,q)} \right].
\end{eqnarray}

As in the global state approach, some approximation is needed in order to find the effective rates $W^{(\pm,k,q)}$ that appear in the master equation. Writing $q_i$ as $q$ and $k_i$ as $k$, the rates of the elementary processes $n_i=0\to n_i=1$ and $n_i=1\to n_i=0$ are, respectively,
\begin{eqnarray}
\label{elementary_rates1}
R_{k,q}^{+} &=& a + h \frac{q}{k} \, ,\\
\label{elementary_rates2}
R_{k,q}^{-} &=& a + h \frac{k-q}{k} \, .
\end{eqnarray}
Therefore, the total rate of the process $N_{1,k} \to N_{1,k} + 1$, $L\to L + k-2q$ is calculated as the product of the elementary rate $R_{k,q}^{+}$, the number of nodes in the down state among the population with degree $k$, and the fraction (probability) of the population of nodes in the down state with degree $k$ that have $q$ neighbors in the up state $P_{0}(k,q)$, i.e.
\begin{eqnarray}
W^{(+,k,q)}&=& (N_{k}-N_{1,k}) \cdot P_{0}(k,q) \cdot R^+_{k,q} \, ,
\end{eqnarray}
A similar reasoning leads to the total rate of the process $N_{1,k} \to N_{1,k} - 1$, $L\to L - k + 2q$,
\begin{eqnarray}
W^{(-,k,q)}&=& N_{1,k} \cdot P_{1}(k,q) \cdot R^-_{k,q} \, .
\end{eqnarray}

In general, those probabilities depend on the detailed configuration $\n$ and adjacency matrix $\A$, but we need to write them in terms only of the description variables $\{\N_1,L\}$. We then make the {\bfseries pair approximation}~\cite{Vazquez2008}
\begin{eqnarray}
\label{binomp}
P_{0}(k,q)&\approx&{k \choose q}c_0^q (1-c_0)^{k-q} \, ,\\
\label{binomm}
P_{1}(k,q)&\approx&{k \choose q} c_1^{k-q} (1-c_1)^q \, ,
\end{eqnarray}
where
\begin{equation}
\label{binom_prob}
c_n=\frac{L}{\sum_{k=k_{\rm min}}^{k_{\rm max}}k N_{n,k}}=\frac{\rho}{\frac{2}{\mu N}\sum_{k=k_{\rm min}}^{k_{\rm max}}k N_{n,k}} \, , \quad n=0,1 \, ,
\end{equation}
is the ratio between the total number of active links and the total number of links connected to nodes in state $n=0,1$. This $c_{0/1}$ can be interpreted as the single event probability that a node in the state $0/1$ has a neighbor in the opposite state $1/0$. The pair approximation assumes that the probability of having $q$ nodes in the up state among the $k$ nodes connected to a node in state $0/1$ is a sequence of $q$ independent processes and thus binomial. The pair approximation is known to be accurate for uncorrelated networks \cite{Gleeson2013,Pugliese2009,Vazquez2008}.

For the noisy voter model, encoded in the transition rates (\ref{elementary_rates1}) and (\ref{elementary_rates2}) using the known moments of the binomial distribution $\sum_{q=0}^k q P_{0/1}(k,q )= k c_{0/1}$ and $\sum_{q=0}^k q^2 P_{0/1}(k,q) = k c_{0/1} (1+(k-1)c_{0/1})$, after lengthy but straightforward algebra we arrive at
\begin{eqnarray}
\label{Fk}
 F_{k} &=& - 2 a m_k + \frac{2 h \rho}{1-m_L^2} (m_L-m_k),\\
\label{Frho}
 F_{\rho} &=& 2 a + 2 (h-2 a) \rho -\frac{4 h}{1-m_L^2} \rho^2
+ \frac{1}{\mu} \frac{4 h \rho}{1-m_L^2} \left[ \frac{1+m_L^2 - 2 m m_L}{1-m_L^2} \rho - 1 + m m_L \right],\\
\label{Gk}
 G_{k} &=& \frac{4}{P_k} \left( a + \frac{h \rho}{1-m_L^2} (1-m_k m_L) \right),\\
\label{Gkrho}
 G_{k,\rho} &=& \frac{4}{\mu} \left( (h(k-2)-2ak) \frac{m_L-m_k}{1-m_L^2} \rho - 2 h (k-1) \frac{m_k - 2 m_L + m_k m_L^2}{(1-m_L^2)^2}\rho^2 \right),
\end{eqnarray}
with the link (or ``degree-weighted'') magnetization $m_L$ defined as
\begin{equation}
\label{mL}
m_L=\frac{1}{\mu}\sum_{k=k_{\rm min}}^{k_{\rm max}} P_k k m_k.
\end{equation}
The complete expression for $G_\rho$ being too long, only a simplified version valid in the stationary state will be displayed later [see Eq. (\ref{G_{rho}})].

Eqs. (\ref{dtmk}--\ref{dtrho2}) together with Eqs. (\ref{Fk}--\ref{mL}) are the basis of our subsequent analysis. They are not yet closed. Note, however, that $m_L(t)$ satisfies the exact equation
\begin{equation}
\label{dmL}
\frac{d\langle m_L\rangle}{dt}=-2a\langle m_L\rangle,
\end{equation}
whose solution $\langle m_L(t)\rangle=\langle m_L(0)\rangle e^{-2at}$ indicates that $\langle m_L\rangle_{{\rm st}} =0$. Remarkably, $\langle m_L\rangle$ is self-governed and fulfills an equation equivalent to $\langle m \rangle$ in the effective-field approximation [see Eq. (\ref{first_FC_temp})]. This will not be the only similarity between these two approximations and variables, as we will show in section \ref{sec:expansion2:b}.

\section{Expansion around the deterministic solution}\label{sec:expansion1}

\subsection{The deterministic solution}\label{sub_deter}

In order to close and eventually solve Eqs. (\ref{dtmk}--\ref{dtrho2}), we first study the deterministic solution using a mean-field assumption, i.e., neglecting all correlations. Introducing the notation \mbox{$\phi_k(t)=\langle m_k(t)\rangle$}, \mbox{$\phi_{\rho}(t)=\langle \rho(t)\rangle$}, \mbox{$\phi_L(t) = \langle m_L(t)\rangle =\phi_L(0) e^{-2a t}$}, and \mbox{$\phi(t)=\langle m(t)\rangle = \sum_{k} P_k \phi_k(t)$}, and substituting Eqs. (\ref{Fk}) and (\ref{Frho}) into Eqs. (\ref{dtmk}) and (\ref{dtrhoL}), we find
\begin{eqnarray}
\label{determ_m}
 \dot \phi_{k} &=& - 2 a \phi_{k} + 2 h \frac{\phi_\rho}{1-\phi_L^2} (\phi_L-\phi_{k}) \, ,\\
\label{determ_rho}  \dot \phi_{\rho} &=& 2 a + 2 (h-2 a) \phi_{\rho} -\frac{4 h}{1-\phi_L^2} \phi_{\rho}^2 + \frac{1}{\mu} \frac{4 h \phi_{\rho}}{1-\phi_L^2} \left( \frac{1+\phi_L^2 - 2 \phi \phi_L}{1-\phi_L^2} \phi_{\rho} - 1 + \phi \phi_L \right) \, ,
\end{eqnarray}
where the dot indicates a time derivative. The equation for $\phi$ would be identical to Eq. (\ref{determ_m}) with the exception of changing $\phi_{k} \rightarrow \phi$. This system of equations has only one (stable) fixed point $\phi_{k}=0$, $\phi_{\rho} = \xi$, where $\xi$ is the positive solution of the quadratic equation
\begin{equation}
\label{xi_eq}
2 h \frac{\mu-1}{\mu} \xi^2 - \left( h \frac{\mu-2}{\mu} - 2 a \right) \xi - a= 0 \, ,
\end{equation}
namely
\begin{equation}
\label{xi}
\xi = \frac{h (\mu-2)-2 \mu a}{4 h (\mu-1)} + \sqrt{\left(\frac{h (\mu-2)-2 \mu a}{4 h (\mu-1)} \right)^2+\frac{\mu a}{2h (\mu-1)}} \, .
\end{equation}
For $a=0$ this is $\xi=\frac{\mu-2}{2(\mu-1)}$. The other solution (unstable) of Eq. (\ref{xi_eq}) is negative and thus non-physical, except in the case $a=0$, which is $\phi_\rho=0$. The stability of the fixed point $\phi_{k}=0$ and $\phi_{\rho}=\xi$ can be assessed using the elements of the Jacobian matrix $\mathbf{J}$, defined as $J_{a,b} = \left. - \frac{\partial{\dot \phi_a}}{\partial \phi_{b}}\right|_{{\phi_k=0, \atop \phi_\rho=\xi}}$, namely
\begin{eqnarray}
\label{Jacobian_vkk}
J_{k,k'} &=& 2 (a + h \xi) \delta_{k,k'} -2 h \xi \frac{P_{k'} k'}{\mu} \, ,\\
\label{Jacobian_krho}
J_{k,\rho} &=& J_{\rho,k}= 0,\\
\label{Jacobian_rho2}
J_{\rho,\rho} &=& 2 (2 a - h) + 8 h \xi + \frac{4 h}{\mu} (1-2 \xi) \, .
\end{eqnarray}
For future reference, let us also mention here the coefficients of the Hessian matrices $\mathbf {H}$, evaluated at the fixed point, $H^a_{b,c}= \left.\frac{\partial^2\dot \phi_a}{\partial \phi_{b}\partial \phi_c}\right|_{{\phi_k=0, \atop \phi_\rho=\xi}}$, namely
\begin{eqnarray}
\label{hessian}
H^k_{k',\rho}&=& 2 h \left( \frac{P_{k'} k'}{\mu}-\delta_{k,k'} \right) \, , \\
H^\rho_{k,k'}&=&\frac{4 h \xi}{\mu^2} P_{k} P_{k'}\left[(1-2\xi) (k+k')+ \left( -\xi + \frac{3 \xi -1}{\mu}\right) 2kk'\right] \, ,\\
H^\rho_{\rho,\rho}&=& -8 h \frac{\mu-1}{\mu} \, ,\\
H^k_{k',k''}&=&H^k_{\rho,\rho}=H^\rho_{k,\rho}=0 \, .
\end{eqnarray}

\begin{figure}[ht]
\centering
\includegraphics[width=0.70\textwidth]{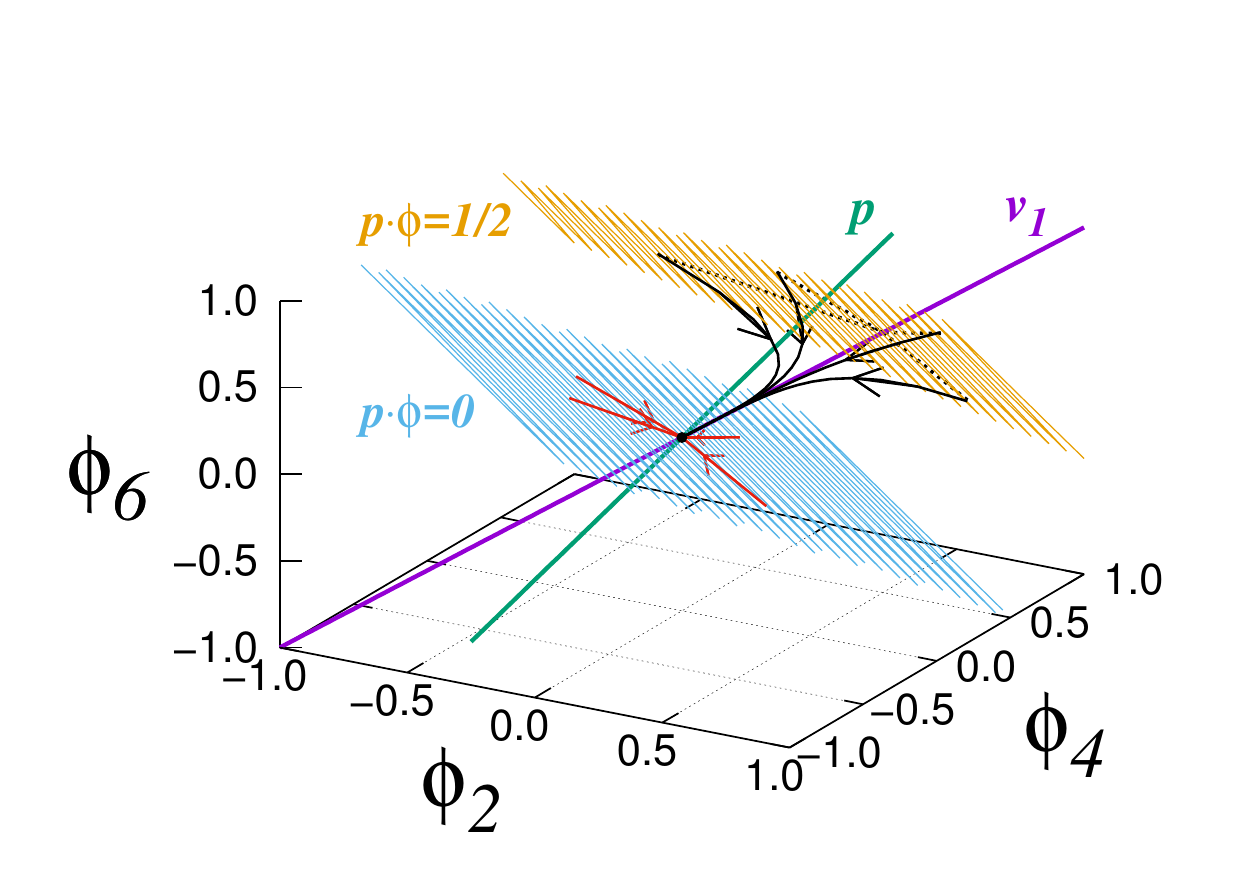}
\caption{Simple example of different deterministic trajectories in the linear approximation with $h=1$, $a=0.01$, $\mu=4$, degree sequence $k=2,4,6$ and normal vector $\mathbf{p} = (1/8,1/2,3/8)$.}
\label{fig_plane}
\end{figure}

Note that the cross-terms $J_{k,\rho}$ and $J_{\rho,k}$ are zero, which means that the deterministic dynamics of $\phi_{k}$ and $\phi_{\rho}$ are uncoupled ---at least in the linear regime---. The eigenvalues of the $\mathbf{J}$ matrix (obviating the trivial $\phi_{\rho}$ part) are (i) $2 a$ (slow) and (ii) $2 (a + h \xi)$ (fast). Both of them are positive, which means that the fixed point is a stable node (note that the Jacobian matrix has been defined with a minus sign). The first eigenvalue has multiplicity one while the second one has multiplicity $d-1$, where $d=k_{{\rm max}}-k_{{\rm min}}+1$ is the dimensionality of the space $\bm{\phi} = \lbrace \phi_{k} \rbrace$. The corresponding eigenvectors are (i) $\mathbf{v_{1}}=(1, ...., 1)$ and (ii) the vectors $\mathbf{v_{2}}$ contained in the plane $\mathbf{p} \cdot \mathbf{v_{2}}=0$, where $\mathbf{p} = (k_{{\rm min}}P_{k_{{\rm min}}} , ..., k_{{\rm max}}P_{k_{{\rm max}}} )/\mu$ is the normal vector to the plane. The linear deterministic dynamics is then given by ${\bm{\phi}}(t) = \phi_{L}(0) \mathbf{v_1} e^{-2a t} + ({\bm{\phi}}(0) - \phi_{L}(0) \mathbf{v_1}) e^{-2(a+h\xi)t}$, where $\phi_{L}(0)=\mathbf{p} \cdot {\bm{\phi}}(0)$, which has a clear geometric interpretation. As shown in Fig.~\ref{fig_plane}, in the early stages of the dynamics, the fast part dominates and trajectories evolve close to the parallel plane $\mathbf{p} \cdot {\bm{\phi}}(t)=\phi_{L}(0)$. Later, the slow part plays the important role and trajectories bend following the direction $\mathbf{v_1}$, an effect which is more pronounced the larger the separation of time scales is.

\subsection{Expansion around the deterministic solution}

The first approach that we will consider to close and solve Eqs. (\ref{dtmk}--\ref{dtrho2}) is a system size expansion around the deterministic trajectory. We term this approach as {\bfseries S1PA}. In the spirit of van~Kampen's expansion~\cite{vanKampen1981}, we propose here the expansion
\begin{eqnarray}
\label{expansion_detk}
m_{k}(t) &=& \phi_{k}(t) + N^{-1/2} \gamma_{k}(t) + N^{-1} \eta_{k}(t) + \dots \, , \\
\label{expansion_dettho}
\rho(t) &=& \phi_{\rho}(t) + N^{-1/2} \gamma_{\rho}(t) + N^{-1} \eta_{\rho}(t) + \dots \, ,
\end{eqnarray}
where $\gamma_{k}, \,\gamma_{\rho},\, \eta_{k},\, \eta_{\rho}$ are stochastic processes, while $\phi_{k}, \,\phi_{\rho}$ are the deterministic terms as given by the solution of Eqs. (\ref{determ_m}) and (\ref{determ_rho}).

If we introduce the ansatz (\ref{expansion_detk}) and (\ref{expansion_dettho}) into Eqs. (\ref{dtmk}--\ref{dtrho2}) and equate powers of $N$, we are able to close the system of equations for the moments\footnote{The traditional van~Kampen's expansion does not include the second order terms $\eta_{k,\rho}$ and it is applied directly to the master equation (\ref{me3}). We will, however, do it at the level of the equations for the moments (\ref{dtmk}--\ref{dtrho2}), and it can be proven \cite{moments} that only when we include these second order terms, the expansion leads to the correct results.}.
From now on, we restrict ourselves to the stationary state and set all time derivatives to zero. For the stochastic variables $\gamma_k$ and $\gamma_\rho$ we find $\langle \gamma_k \rangle_{{\rm st}}=0$ and $\langle \gamma_\rho \rangle_{{\rm st}}=0$ for the average values, and the following set of equations for the correlation matrix $C_{a,b} \equiv \langle \gamma_{a} \gamma_{b} \rangle_{{\rm st}}$
\begin{eqnarray}
\label{correlationskk}
\sum_{k''} \left[ C_{k',k''} J_{k,k''} + C_{k,k''} J_{k',k''} \right] + C_{k',\rho} J_{k,\rho} + C_{k,\rho} J_{k',\rho} &=& \delta_{k,k'} G_{k}^{\rm st} ,\\
\label{correlationskrho}
\sum_{k''} \left[ C_{\rho,k''} J_{k,k''} + C_{k,k''} J_{\rho,k''} \right] + C_{\rho,\rho} J_{k,\rho} + C_{k,\rho} J_{\rho,\rho} &=& G_{k,\rho}^{\rm st} ,\\
\label{correlationsrhorho}
2 \sum_{k''} C_{\rho,k''} J_{\rho,k''} + 2 C_{\rho,\rho} J_{\rho,\rho} &=& G_{\rho}^{\rm st},
\end{eqnarray}
with $G_{k}$, $G_{k,\rho}$ and $G_{\rho}$ evaluated at the steady state
\begin{eqnarray}
\label{G_{kk}}
G_{k}^{\rm st} &=& \frac{4(a + h \xi)}{P_{k}} ,\\
\label{G_{krho}}
G_{k,\rho}^{\rm st} &=& 0 ,\\
\label{G_{rho}}
G_{\rho}^{\rm st} &=& - \frac{8 \xi}{\mu^2} \bigg( a \Big[\mu_2-2\mu+2\xi(\mu-\mu_2)\Big] \nonumber\\
&&+ h \Big[\mu-2+(\mu_2-7\mu+6)\xi-2(\mu_2-3 \mu+2)\xi^2 \Big] \bigg) \, .
\end{eqnarray}
Although the system of linear equations (\ref{correlationskk}--\ref{correlationsrhorho}) might be difficult to solve in general, the decoupling $J_{k,\rho}=J_{\rho,k}=G_{k,\rho}^{\rm st}=0$, which can be shown to be a general consequence of the up-down symmetry in the rates $R_{k,q}^{-}=R_{k,k-q}^{+}$, in addition to the simplicity of the terms $J_{k,k'}$ for the noisy voter model, simplifies significantly the problem. As shown in \ref{app:ckk}, the solution is
\begin{eqnarray}
C_{\rho,\rho}&=&G_{\rho}^{\rm st}/2J_{\rho,\rho} \, ,\\
C_{k,\rho}&=&0 \, ,\\
C_{k,k'} &=& \frac{h\xi}{2a+h\xi} \left( \frac{h\xi}{a}\frac{\mu_2}{\mu^2}+\frac{k+k'}{\mu}\right) + \frac{1}{P_{k}}\delta_{k,k'} \, .
\label{Ckk}
\end{eqnarray}

\begin{figure}[ht]
\centering
\subfloat[]{\label{fig_var:a}\includegraphics[width=0.45\textwidth]{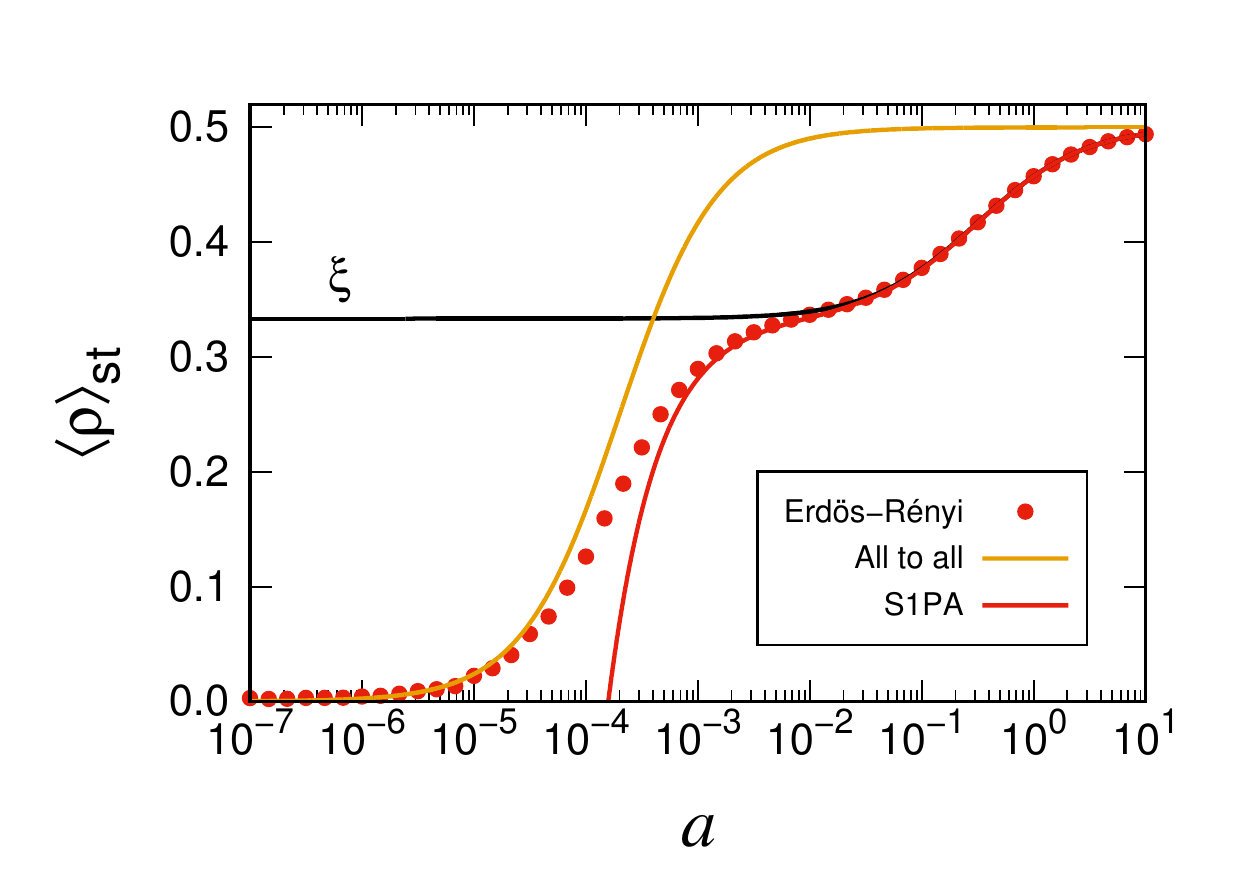}}
\subfloat[]{\label{fig_var:b}\includegraphics[width=0.45\textwidth]{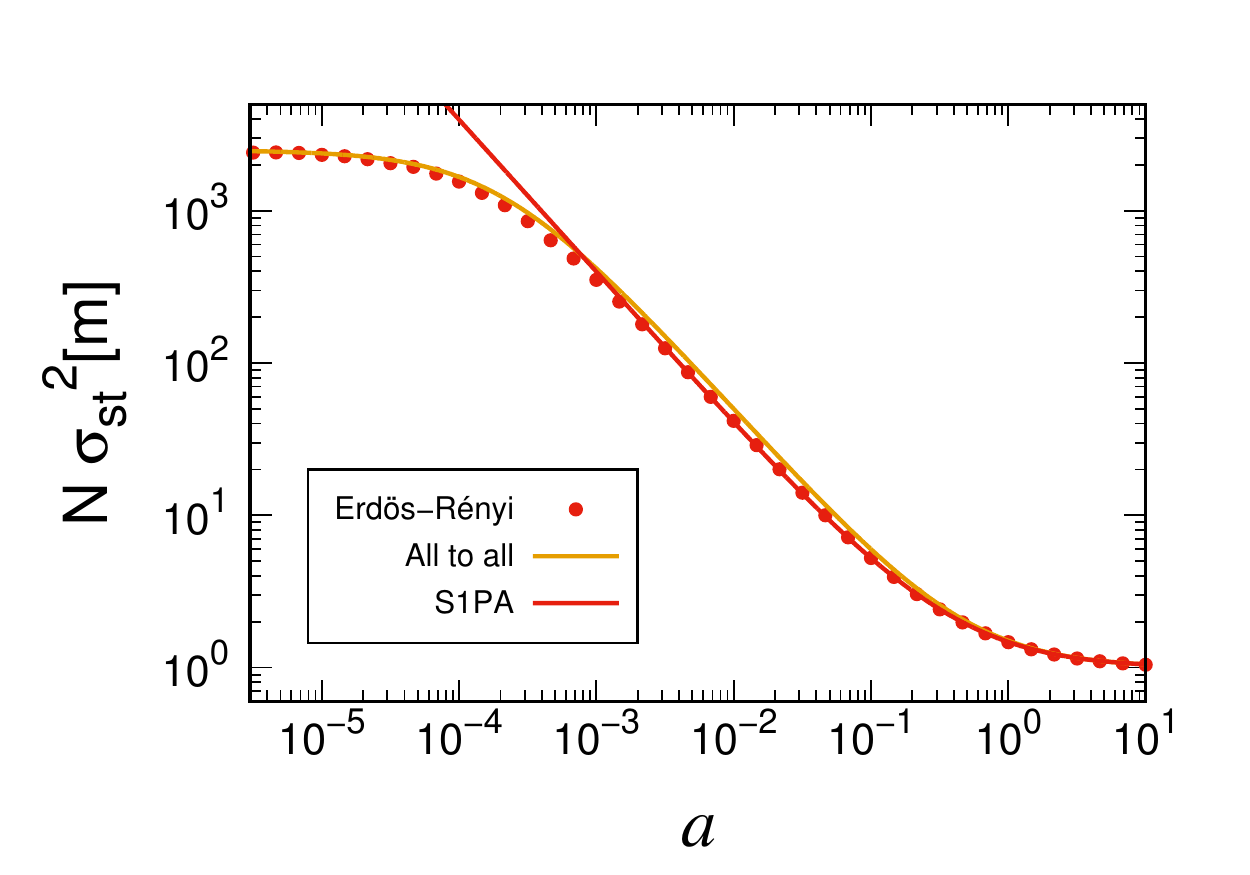}}
\caption{Stationary average density of active links (Fig. \ref{fig_var:a}) and variance of the magnetization (Fig. \ref{fig_var:b}) as a function of the parameter $a$, for an Erd\"{o}s-R\'enyi graph with fixed $h=1$, $\mu=4$ and $N=2500$, $a_c \approx h/N=4\times 10^{-4}$. Points correspond to numerical simulations, and different lines, as indicated in the legend, are the analytical results corresponding to the global all-to-all approach, Eqs. (\ref{second_FC}) and (\ref{rho_FC}), the stochastic pair approximation combined with a van~Kampen expansion (S1PA), Eqs. (\ref{variance_mag}) and (\ref{rhoN1}), and the prediction for the density of active links which is obtained neglecting fluctuations, $\langle \rho \rangle_{\rm st} =\xi$ from Eq. (\ref{xi}).}
\label{fig_var}
\end{figure}

The variance of the global magnetization $m$ can now be obtained from Eq. (\ref{s2fromsigma}), expressed in the form \mbox{$\sigma^2[m]=\sum_{k,k'} P_{k} P_{k'} C_{k,k'}$},
\begin{equation}
\label{variance_mag}
N\sigma_{{\rm st}}^2[m]=\frac{h^2 \xi^2}{2a+h\xi}\frac{N}{a N_{\rm eff}}+\frac{2h\xi}{2a+h\xi}+1 \, ,
\end{equation}
where we have defined an effective system size $N_{{\rm eff}}\equiv N \mu^2/\mu_2$. For a given finite network, $N_{\rm eff}$ adopts a finite value. In the large $N$ limit, however, $N_{{\rm eff}}$ can be characterized by a non-trivial dependence on $N$, as the moments of the distribution can diverge. For instance, in a Barab\'asi-Albert network it is $N_{\rm eff}\sim N/\log N$. From Eq. (\ref{variance_mag}) we obtain that, in the limit $a\to 0$, the fluctuations diverge as $N\sigma_{{\rm st}}^2[m] \sim \frac{h\xi}{a} \frac{\mu_2}{\mu^2}$. In the same limit, the correlations behave as $C_{k,k'} \sim N \sigma_{{\rm st}}^2[m]$, independent of $k,k'$, which is strongly related to the geometric picture of the linear deterministic dynamics that we described earlier, where variables fluctuate along the slow direction $\mathbf{v_1}$ with an amplitude $N^{1/2} \sigma_{{\rm st}}[m]$. In a homogeneous network, where $\mu_2=\mu^2$, we find $N\sigma_{{\rm st}}^2[m]=\frac{a+h\xi}{a}$, which in the limit $N\to\infty$ coincides with the effective-field result in Eq. (\ref{second_FC}) whenever $\xi=1/2$, which happens, for instance, in the limit of large connectivity $\mu\to\infty$. On the other hand, below or around the critical point, when $a=O(N^{-1})$, the van~Kampen expansion fails to reproduce the all-to-all result and, hence, it provides the wrong scaling dependence with system size near the critical region (as can be seen in Fig.~\ref{fig_var:b}). The critical point can be obtained from the condition of a flat probability distribution of the magnetization, $\sigma_{\rm st}^2[m]=\frac{N+2}{3N}$, which leads to
\begin{equation}\label{ac_S1PA}
a_c=\frac{3h}{2 N_{\rm eff}}\frac{\mu-2}{\mu-1}+O(N^{-2}) \, .
\end{equation}

It is also possible to calculate the $O(N^{-1})$ stochastic corrections to the average value of the description variables defined in Eqs. (\ref{expansion_detk}) and (\ref{expansion_dettho}), which are led by $\langle \eta_k \rangle_{{\rm st}}$ and $\langle \eta_\rho \rangle_{{\rm st}}$. In the stationary state, the equations for $\langle \eta_k \rangle_{{\rm st}}$ and $\langle \eta_\rho \rangle_{{\rm st}}$ read
\begin{eqnarray}
\label{nu_k}
 \sum_{k'} J_{k,k'} \langle \eta_{k'} \rangle_{{\rm st}} + J_{k,\rho} \langle \eta_{\rho}\rangle_{{\rm st}} &=& \frac{1}{2} \sum_{k',k''} C_{k',k''} H^k_{k',k''} + \sum_{k'} C_{k',\rho}H^k_{k', \rho} + \frac{1}{2} C_{\rho,\rho} H^k_{\rho,\rho} \, ,\\
\label{nu_rho}
 \sum_{k'} J_{\rho,k'} \langle \eta_{k'} \rangle_{{\rm st}} + J_{\rho,\rho} \langle \eta_{\rho}\rangle_{{\rm st}} &=& \frac{1}{2} \sum_{k',k''} C_{k',k''}H^\rho_{k',k''} + \sum_{k'} C_{k',\rho} H^\rho_{k',\rho} + \frac{1}{2} C_{\rho,\rho} H^\rho_{\rho,\rho} \, .
\end{eqnarray}
Again, due to the particularities of the model, these equations greatly simplify, leading to
\begin{eqnarray}
\label{nu_k_s}
\sum_{k'} J_{k,k'} \langle \eta_{k'} \rangle_{{\rm st}} &=& 0 \, ,\\
\label{nu_rho_s}
 J_{\rho,\rho} \langle \eta_{\rho}\rangle_{{\rm st}} &=& \frac{1}{2} \sum_{k',k''} C_{k',k''}H^\rho_{k',k''} + \frac{1}{2} C_{\rho,\rho}H^\rho_{\rho,\rho} \, ,
\end{eqnarray}
from where we obtain $\langle \eta_{k} \rangle_{{\rm st}}=0$, a result arising again from the up-down symmetry of the model, and
\begin{eqnarray}
\label{nu_rhO_results}
\langle \eta_{\rho}\rangle_{{\rm st}} =& (J_{\rho,\rho})^{-1} \left[ \frac{4 h \xi(1-2\xi)}{\mu(2a+h\xi)}\left[h \xi D+2(a+h\xi)\right] \right. \nonumber\\
&+ 4 \left. \left( - h \xi^2 + \frac{h \xi}{\mu} (3 \xi -1) \right) D + \frac{C_{\rho,\rho}}{2} H^\rho_{\rho,\rho} \right] \, ,
\end{eqnarray}
where $D$ is given by Eq. (\ref{D}). Finally, the density of active links in the stationary state is
\begin{equation}
\label{rhoN1}
\langle \rho \rangle_{\rm st} =\xi+N^{-1}\langle \eta_{\rho}\rangle_{{\rm st}} \, .
\end{equation}
In the all-to-all limit $\mu_2\to\mu^2\to\infty$, we have $\langle \eta_{\rho}\rangle_{{\rm st}}=-\frac{h}{4a}$ and $\langle \rho \rangle_{\rm st}$ coincides with the effective-field result in Eq. (\ref{rho_FC}) to the order $O(N^{-1})$. On the other hand, in the limit of $a\to 0$ we have
\begin{equation}
\label{limit1}
\lim_{a \rightarrow 0}\langle \eta_{\rho}\rangle_{{\rm st}}=-\frac{h\xi^2}{a}\frac{\mu_2}{\mu^2} \, ,
\end{equation}
which leads to unphysical results for the density of active links, as, by definition, it should be $\rho\ge 0$.

In Fig.~\ref{fig_var} we test the validity of these analytical results for a wide range of values of $a$ for an Erd\"os-R\'enyi network with $N=2500$ nodes and average connectivity $\mu=4$ (a more detailed comparison for other network types is postponed to the next section). The results in this figure suggest that the expansion S1PA around the deterministic solution works perfectly for $a \gg a_{c} =h N^{-1}$, while it fails below the critical point $a \ll a_{c}$, where it is not able to reproduce even qualitatively the observed dependence. We also plot the prediction for the density of active links which is obtained neglecting fluctuations, $\langle \rho \rangle_{\rm st} =\xi$, which performs very poorly, specially for small $a$. In the same panel we show the results of the simple global approach, Eqs. (\ref{first_FC}) and (\ref{second_FC}), which is surprisingly accurate for the variance of the magnetization but not so good for the density of active links. Nevertheless, we note that the global approach, equivalent to the all-to-all approximation, completely disregards the details of the network connectivity and, as shown in the the next sections, is not capable of explaining the differences observed between different networks. The reason for the failure near the critical region of the stochastic pair approximation combined with a van~Kampen expansion lies in the ansatz stated in Eqs. (\ref{expansion_detk}) and (\ref{expansion_dettho}), which requires $\sigma^2[m]$ to scale as $\sim N^{-1}$ and also to be small enough for the expansion to be accurate, which is only true in the limit $a \gg a_{c}$ [see Eq. (\ref{second_FC})]. This is strongly related to the associated deterministic dynamics and to the slow eigendirection $\mathbf{v_1}$ having an infinite time scale as $a \rightarrow 0$, which eventually leads to large fluctuations following this direction. If we want an expansion to be also valid in the critical region, we have to explore the full set of nonlinear deterministic dynamical equations (\ref{determ_m}--\ref{determ_rho}) and find a solution (in this case a privileged slow trajectory) around which the dynamics has a finite time scale in the entire parameter region. Then, we can propose an expansion similar to Eqs. (\ref{expansion_detk}) and (\ref{expansion_dettho}) around this solution.
This is the method that we are going to apply in the next section and whose procedure is made clear in \ref{app:linear} with a simple example of a set of linear equations.

\section{Expansion around a stochastic dynamical attractor}\label{sec:expansion2}

\subsection{The dynamical attractor}\label{sec:expansion2:a}

As explained before, the second method we propose to tackle Eqs. (\ref{dtmk}--\ref{dtrho2}) starts by finding a particular deterministic solution around which we can include stochastic corrections. The solution of Eqs. (\ref{determ_m}) and (\ref{determ_rho}) depends on the initial conditions $\phi_k(0)$, $\phi_{\rho}(0)$ besides the time $t$. We are interested, however, in a particular solution that we will name $\phi_k^*(t)$, $\phi_\rho^*(t)$ and that depends on time only through $\phi_{L}(t)$, namely $\phi_k^*(\phi_L(t))$, $\phi_\rho^*(\phi_L(t))$, i.e., in a particular solution such that the system reacts instantly to the self-governed variable $\phi_L(t)=\phi_L(0)e^{-2at}$. This solution has to fulfill the constraints $\phi^* = \pm 1$ and $\phi_\rho^*=0$ when $\phi_{L}=\pm 1$, which arise from the definition of the variables when all the nodes are up or down. By inspection, we note that $\phi_k^* = \phi_L$ is an exact solution of Eq. (\ref{determ_m}) that fulfills these constraints, while the dependence $\phi_{\rho}^*(\phi_{L})$ is not so trivial. Introducing $\phi=\phi_{L}$ in Eq. (\ref{determ_rho}) and changing the time derivative $\frac{d}{dt} = -2 a \phi_{L} \frac{d\,}{d \phi_{L}}$, we get
\begin{equation}
\label{determ_rho2}
 -2 a \phi_L \frac{d \phi_\rho^*}{d \phi_L} = 2 a + 2 (h - 2 a) \phi_{\rho}^* - \frac{4 h}{1-\phi_{L}^2} \phi_\rho^{*2}-\frac{4 h}{\mu} \phi_{\rho}^* \left( 1-\frac{\phi_{\rho}^*}{1-\phi_L^2}\right) ,
\end{equation}
a Riccati type of equation whose exact solution, carried out in \ref{app:riccati}, depends on a combination of Gauss hypergeometric functions. A simpler form, more amenable to analytical treatment, is also derived in \ref{app:riccati} as
\begin{equation}
\label{rho_proposed}
\phi_{\rho}^{*} = \xi (1-\phi_L^2) + O (N^{-1}) \, .
\end{equation}
In order to show that this simpler $\phi_{\rho}^{*}$ is a stable attractor of the deterministic dynamics, we linearize Eqs. (\ref{determ_m}) and (\ref{determ_rho}) around the particular trajectory $\phi_k = \phi_k^{*} + \delta_k$, $\phi_{\rho} = \phi_{\rho}^{*} + \delta_{\rho}$, finding
\begin{eqnarray}
\label{linearizing_a}
\dot \delta_k &= - 2(a + h \xi) \delta_k \, , \\ \label{linearizing_b}
\dot \delta_{\rho} &= - J_{\rho, \rho} \delta_{\rho} + \frac{4 h}{\mu} \xi (1-2\xi) \phi_L \delta + 2 a (1-2\xi) \phi_{L}^2 \, ,
\end{eqnarray}
with $\delta(t)=\sum_k P_k\delta_k(t)$ and $J_{\rho,\rho}$ given by Eq. (\ref{Jacobian_rho2}). The solution of these linear equations is
\begin{eqnarray}
\label{delta_k}
 \delta_k(t) &=& \delta_k(0) e^{-2(a+h\xi)t}, \\
\label{delta_t}
 \quad \delta(t) &=& \delta(0) e^{-2(a+h\xi)t}, \\
\label{delta_rho}
 \delta_{\rho}(t) &=& c e^{-J_{\rho,\rho} t} + \frac{4 h}{\mu} \frac{\xi(1-2\xi)}{J_{\rho,\rho}-4a-2h\xi} \phi_L(t) \delta(t) + \frac{2 a (1-2\xi)}{J_{\rho,\rho}-4a} \phi_L(t)^2,
\end{eqnarray}
where $c$ is a constant determined by the initial conditions. The time scale of $\delta_k$ is fast (compared to the time scale of $\phi_L(t)$) and corresponds to the second eigenvalue calculated in Section~\ref{sub_deter}, while $\delta_{\rho}$ has three terms: the first two are fast and the last one is slow (for $a\ll 1$). It is important to note that this last slow term appears because we used the approximate expression of the trajectory in Eq. (\ref{rho_proposed}), instead of the exact one. According to the discussion in \ref{app:riccati}, this term can be neglected, since it is of order $O(N^{-1})$ as long as the approximation is accurate. We thereby conclude that this trajectory is an attractor of the dynamics and that the attractor is reached in a time scale which is longer than the time scale within the attractor. This approach captures a critical feature that the expansion around the deterministic solution overlooked. While $\phi_{k}(t)$ is exactly the same for both approaches, $\phi_{\rho}(t)$ depends on $\phi_{L}(t)$ following Eq. (\ref{rho_proposed}), and so the dynamics are not uncoupled as in the previous case.

\subsection{Expansion around the dynamical attractor}\label{sec:expansion2:b}

\begin{figure}[ht]
\centering
\subfloat[]{\label{fig:attractor:a}\includegraphics[width=0.45\textwidth]{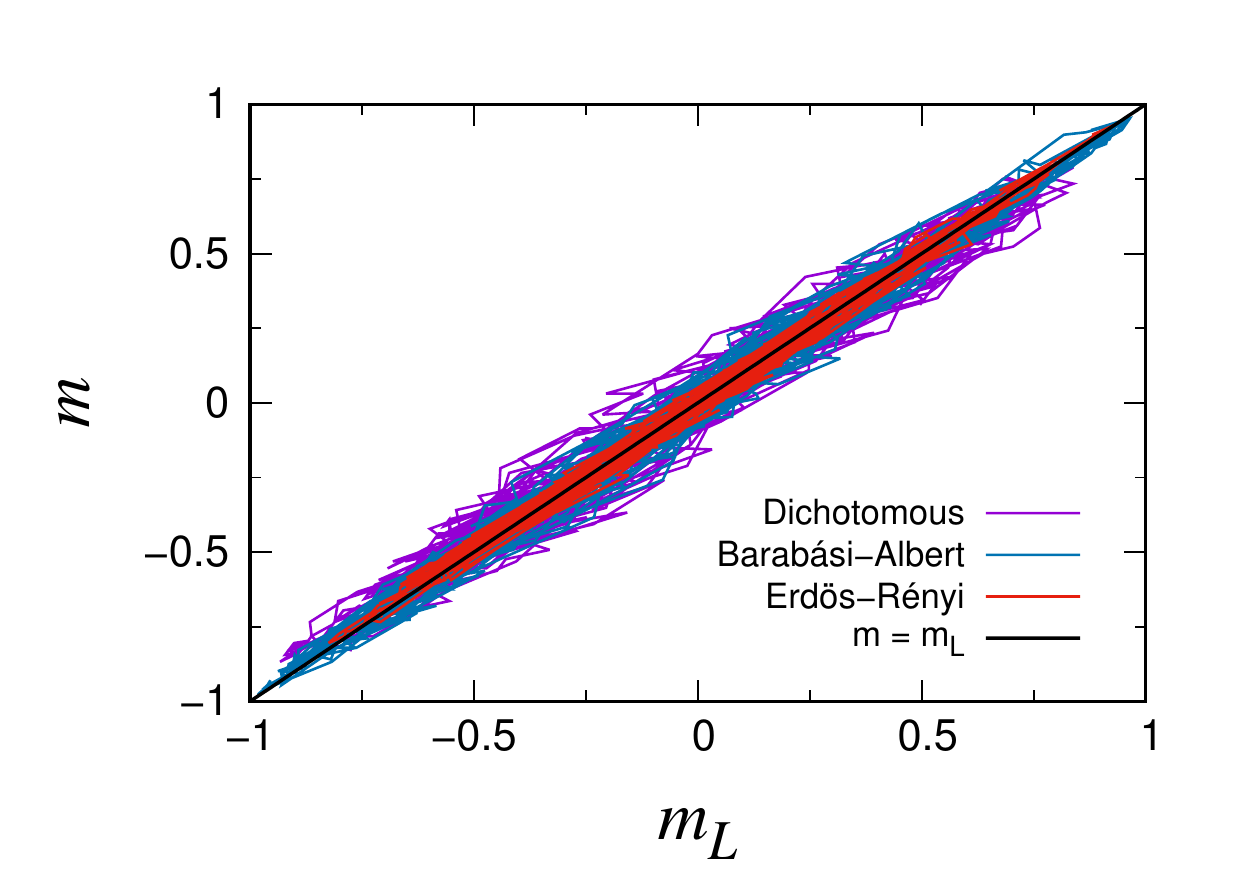}}
\subfloat[]{\label{fig:attractor:b}\includegraphics[width=0.45\textwidth]{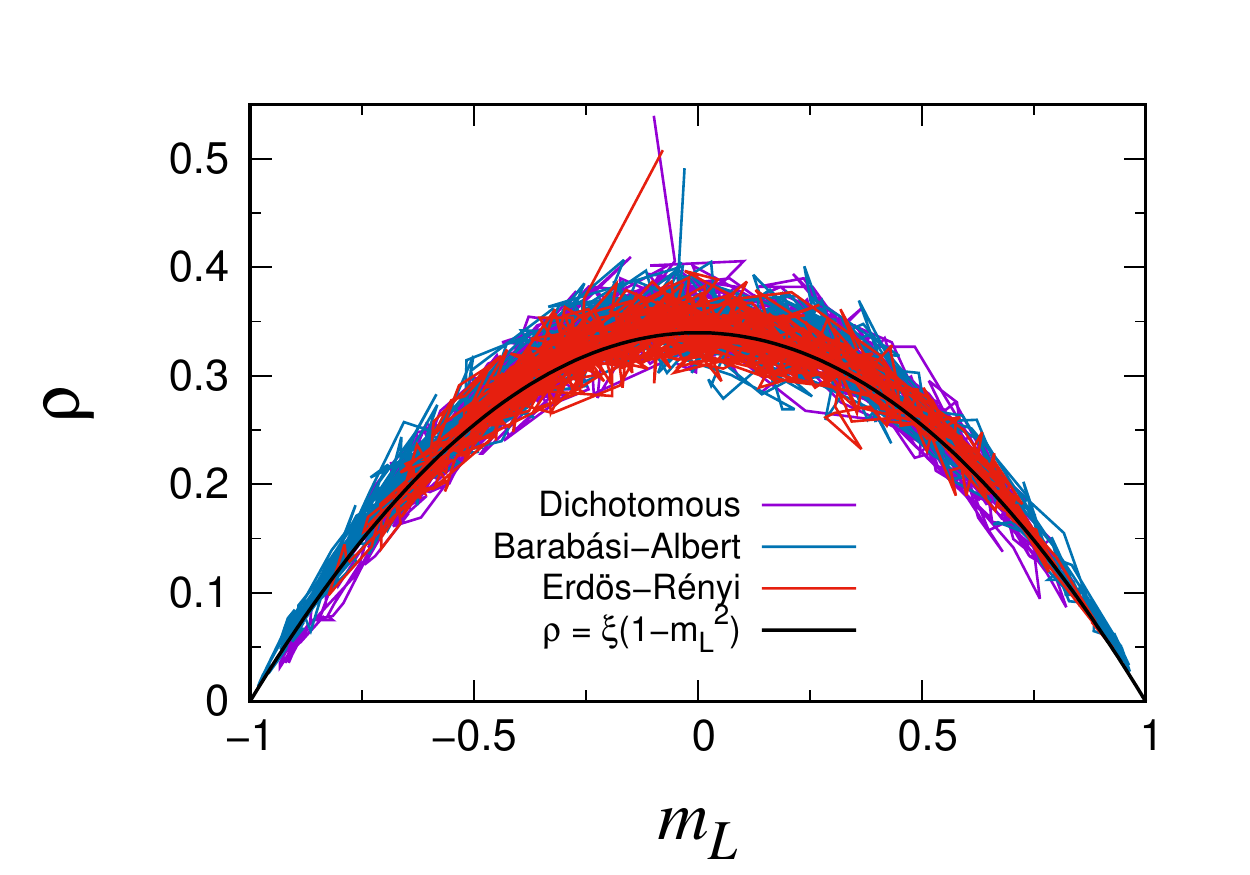}}
\caption{Single trajectories of the stochastic variables $m(t)$, $\rho(t)$ and $m_L(t)$ for three different networks with $a=0.01$, $h=1$, $\mu=4$ and $N=256$. Black solid lines correspond to the deterministic attractor part of Eqs. (\ref{expansion1}) and (\ref{expansion2}) while fluctuations around it correspond to $\varepsilon_{k}$ and $\varepsilon_{\rho}$.}
\label{fig:attractor}
\end{figure}

In Figs.~\ref{fig:attractor:a} and~\ref{fig:attractor:b} we plot, respectively, $m(t)$ and $\rho(t)$ versus $m_L(t)$ for different realizations of the stochastic process in different network structures. It is apparent from this figure that $m(t)$ and $\rho(t)$ fluctuate around the dynamical attractor $m(t)=m_L(t), \,\rho(t)= \phi^*_\rho(m_L(t))$. Based upon this picture of the phase space, our second approach to solve Eqs. (\ref{dtmk}--\ref{dtrho2}), which we name {\bfseries S2PA}, consists in splitting the $m_k(t)$ and $\rho(t)$ variables into the dynamical attractor contribution plus additional fluctuations as
\begin{eqnarray}
\label{expansion1}
m_k(t) &=& m_L(t) + \varepsilon_{k}(t) \, ,\\ \label{expansion2}
\rho(t) &=& \phi^*_\rho(m_L(t))+ \varepsilon_{\rho}(t) \, ,
\end{eqnarray}
where $m_L$, $\varepsilon_{k}$ and $\varepsilon_{\rho}$ are all stochastic variables. According to the discussion above, section \ref{sec:expansion2:a}, $\varepsilon_{k}$ and $\varepsilon_{\rho}$ are fast variables and we assume that they can be approximated by a van~Kampen type expansion
\begin{eqnarray}
\label{2expansion1}
\varepsilon_k(t) &=& \delta_{k}(t) + N^{-1/2} \lambda_{k}(t) + N^{-1} \nu_{k}(t)+\dots \, ,\\ \label{2expansion2}
\varepsilon_{\rho}(t) &=& \delta_{\rho}(t) + N^{-1/2} \lambda_{\rho}(t)+\dots \, ,
\end{eqnarray}
$\delta_{k}$ and $\delta_{\rho}$ being the deterministic equations (\ref{delta_k}) and (\ref{delta_rho}), while $\lambda_{k}$,$\lambda_{\rho}$, $\nu_{k}$ are stochastic. The order $O(N^{-1})$ of $\varepsilon_\rho$ will not be necessary for the expansion procedure in this case.

We now proceed to find the statistical properties of $\varepsilon_k$ and $\varepsilon_\rho$. Introducing Eqs. (\ref{expansion1}--\ref{2expansion2}) in the equations for the moments (\ref{dtmk}--\ref{dtrho2}) and equating powers of $N$, we find
\begin{eqnarray}
\label{eq_lambda_k}
\frac{d \langle \lambda_{k} \rangle}{d t} &=& - 2(a + h \xi) \langle \lambda_{k} \rangle \, , \\
\label{eq_lambda_rho}
\frac{d \langle \lambda_{\rho} \rangle}{d t} &=& - J_{\rho,\rho} \langle \lambda_{\rho} \rangle + \frac{4 h}{\mu} \xi (1-2\xi) \sum_{k} P_{k} \langle \lambda_{k} m_L \rangle \, , \\
\label{eq_nu_k}
\frac{d \langle \nu_{k} \rangle}{d t} &=& - 2(a + h \xi) \langle \nu_{k} \rangle - 2 h \left\langle \frac{\lambda_{k} \lambda_{\rho}}{1-m_L^2} \right\rangle \, ,
\end{eqnarray}
for the average values and
\begin{eqnarray}
\label{1corr_kk}
 \frac{d \langle \lambda_{k} m_{L} \rangle}{dt} = - 2(2 a + h \xi) \langle \lambda_{k} m_{L} \rangle, \\
\label{2corr_kk}
 \frac{d \langle \lambda_{k} \lambda_{k'} \rangle}{d t} + \frac{d \langle \nu_{k} m_{L} \rangle}{dt}+\frac{d \langle \nu_{k'} m_{L} \rangle}{dt}+N\frac{d \langle m_L^2 \rangle}{d t}= - 2(2 a + h \xi) \langle \nu_{k} m_{L} \rangle - 2(2 a + h \xi) \langle \nu_{k'} m_{L} \rangle \nonumber\\
 -4 aN \langle m_L^2 \rangle - 4 ( a + h \xi) \langle \lambda_{k} \lambda_{k'} \rangle + \frac{4}{P_{k}} \left( a + h \xi (1-\langle m_L^2 \rangle) \right) \delta_{k,k'} \nonumber\\
 - 2 h \left\langle \frac{(\lambda_{k}+\lambda_{k'}) \lambda_{\rho} m_L}{1-m_L^2} \right\rangle \, ,
\end{eqnarray}
for the correlations, where, for simplicity and brevity of the expressions, we have assumed that the deterministic contributions $\delta_{k}(t),\delta_{\rho}(t)$ are well captured by the linear system of equations (\ref{linearizing_a}) and (\ref{linearizing_b}) and that $\delta_{k}(0)=\delta_{\rho}(0)=0$. An important difference with respect to the expansion around the deterministic solution, Eqs. (\ref{expansion_detk}--\ref{expansion_dettho}), is that the equations for the moments are not closed, as indicated by the presence of the last term in Eq. (\ref{2corr_kk}). Nevertheless, we will argue later on in this same section that this term can be neglected at this level of approximation.

From these equations, it follows directly that $\langle \lambda_{k} \rangle_{{\rm st}} = \langle \lambda_{\rho} \rangle_{{\rm st}} = \langle \lambda_{k} m_L \rangle_{{\rm st}} = 0$. Furthermore, due to the up-down symmetry of the model, we know that $\langle \nu_{k} \rangle_{{\rm st}} = 0$, and thus it follows that $\left\langle \frac{\lambda_{k} \lambda_{\rho}}{1-m_L^2} \right\rangle_{{\rm st}} = 0$. Performing in Eq. (\ref{2corr_kk}) the double sum $\sum_{k,k'} k k'P_{k} P_{k'}$ and taking into account that, by definition, $\sum_{k} k P_{k} \varepsilon_{k}(t)=0$ (the same identity holds for $\delta_{k}(t)$, $\lambda_{k}(t)$ and $\nu_{k}(t)$ separately), we find
\begin{equation}
\label{mL_eq}
\frac{d \langle m_L^2 \rangle}{d t} = -4 \left(a+ \frac{\mu_2}{\mu^2} \frac{h \xi}{N} \right) \langle m_L^2 \rangle + \frac{\mu_2}{\mu^2} \frac{4 (a + h \xi)}{N} \, ,
\end{equation}
which, together with Eq. (\ref{dmL}), proves that $m_L(t)$ behaves as an autonomous stochastic variable. Furthermore, its first and second moments fulfill equivalent equations to those of $m(t)$ in the all-to-all scenario [see Eqs.~(\ref{first_FC_temp}) and~(\ref{second_FC_temp})] if we change $h \rightarrow 2 h \xi$ and $N \rightarrow N_{{\rm eff}}$. The stationary solution reads
\begin{equation}
\label{mL_st}
\langle m_L^2 \rangle_{{\rm st}} = \frac{a + h \xi}{a N_{{\rm eff}} + h \xi} \, ,
\end{equation}
and the transient regime
\begin{equation}
\label{mL_temp}
\langle m_L(t)^2 \rangle = \langle m_L^2 \rangle_{{\rm st}} + \left[ \langle m_L(0)^2 \rangle - \langle m_L^2 \rangle_{{\rm st}} \right] e^{-4 \left(a + \frac{h \xi}{N_{{\rm eff}}}\right) t } \, .
\end{equation}
Note that $a N \langle m_L^2 \rangle_{{\rm st}} = O(1)$, implying that the order of Eq. (\ref{2corr_kk}) is consistently $O(1)$ and also supporting the reasoning of \ref{app:riccati}. As shown in \ref{app:link}, it is also possible to obtain these statistical properties starting from a closed master equation for the link magnetization $m_L(t)$.

\begin{figure}[ht]
\centering
\subfloat[]{\label{fig:assumption:a}\includegraphics[width=0.45\textwidth]{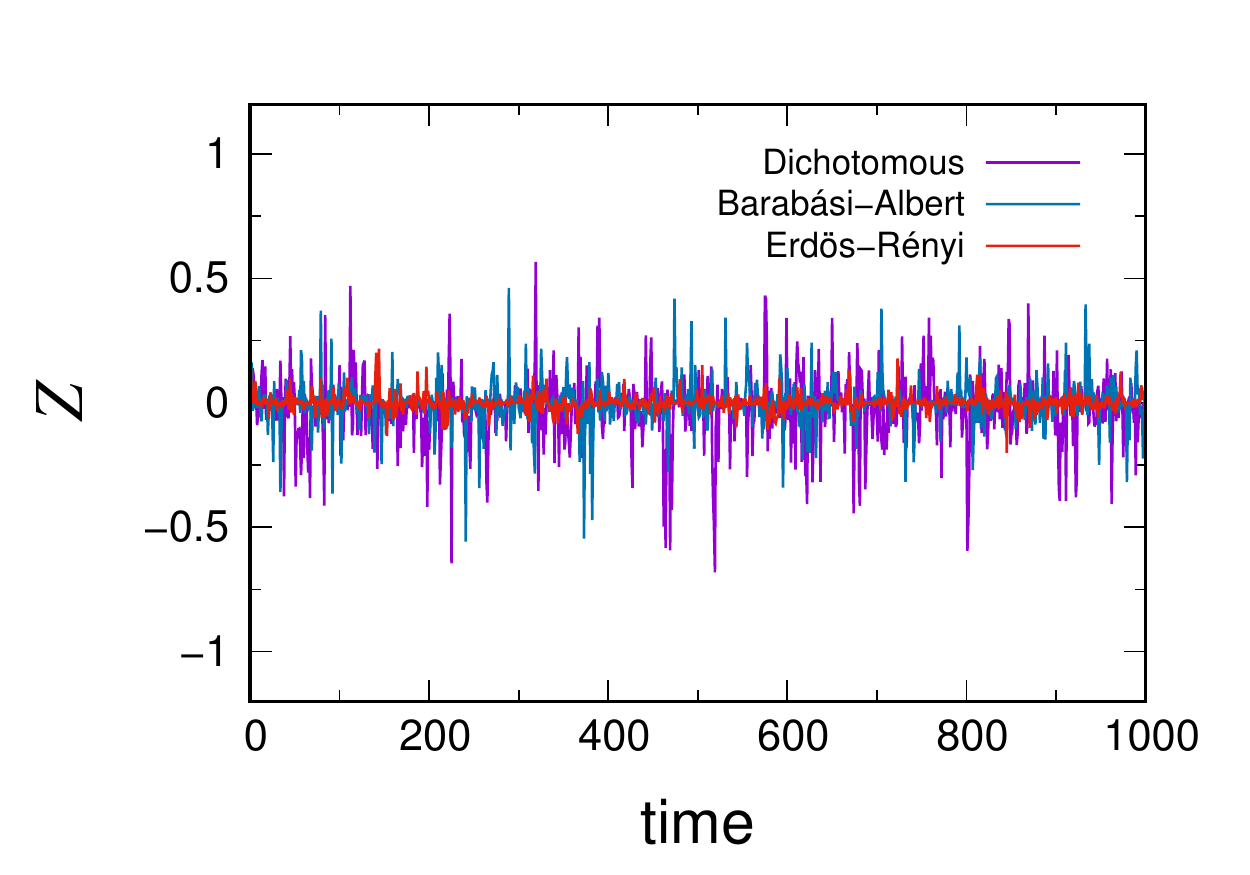}}
\subfloat[]{\label{fig:assumption:b}\includegraphics[width=0.45\textwidth]{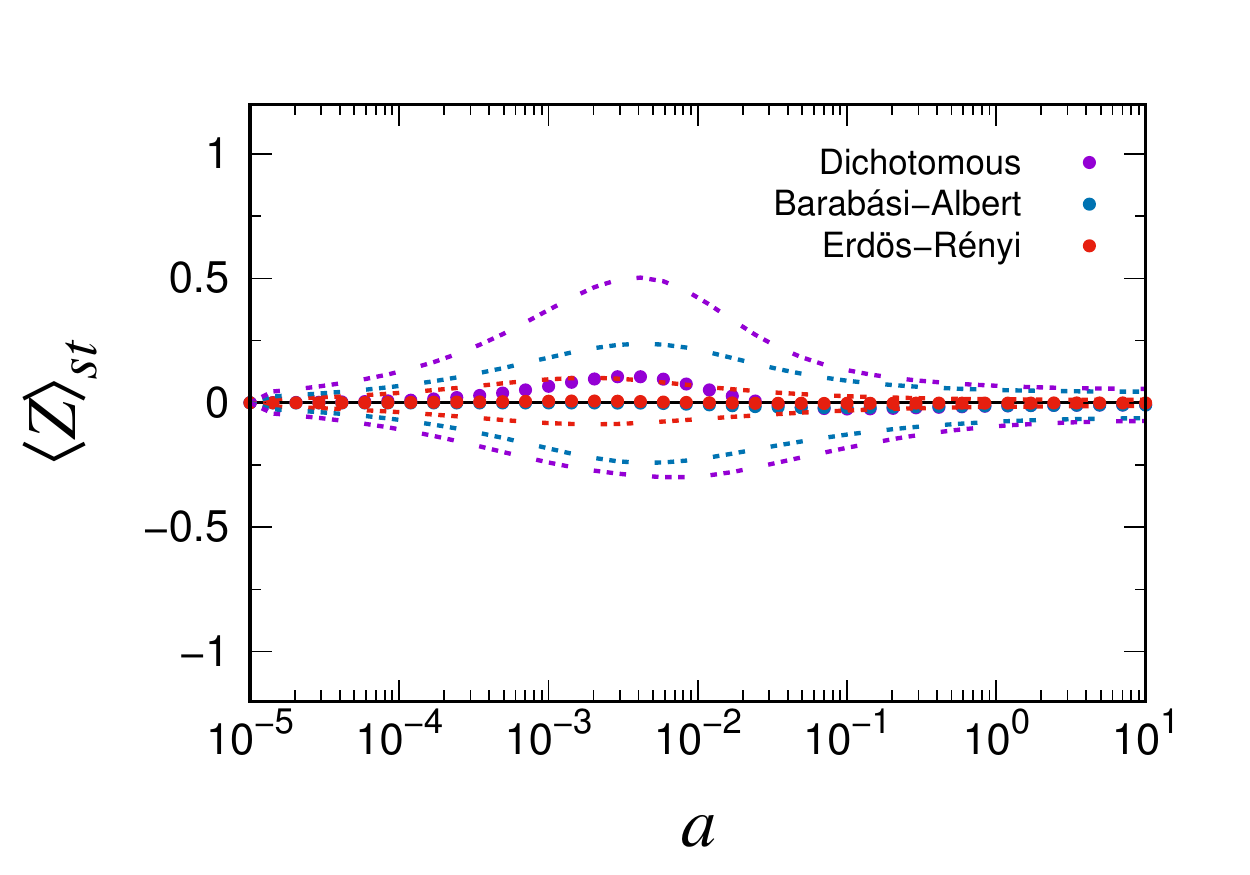}}
\caption{Statistical properties of the variable $Z\equiv\lambda_{k} \lambda_{\rho} m_L/(1-m_L^2)$. The left panel plots one representative time-evolution of $Z$ corresponding to the trajectories of Fig. \ref{fig:attractor} with $a=0.01$, $h=1$, $\mu=4$ and $N=256$. In practice, $Z$ has been computed using $\lambda_k=N^{1/2}(m_k-m_L)$ and $\lambda_\rho=N^{1/2}(\rho-\phi^*_\rho(m_L))$  coming from Eqs.(\ref{expansion1}-\ref{2expansion2}) in the steady state $\delta_k(t)=\delta_\rho(t)=0$. The average value $\langle Z\rangle_{\rm st}$ and its standard deviation $\sigma_Z$ are then plotted in the right panel as a function of $a$ for different network types (see section \ref{sec:numerical} for a precise definition). In this panel (b) the dots indicate $\left\langle Z \right\rangle_{\rm st}$ and the dashed lines are $\langle Z\rangle_{\rm st} \pm \sigma_Z$.}
\label{fig:assumption}
\end{figure}

Performing in Eq. (\ref{2corr_kk}) the partial sum $\sum_{k'} k' P_{k'}$ we obtain an expression for $\langle \nu_k m_L \rangle$ that depends on $\langle m_{L}^2 \rangle$,
\begin{eqnarray}
\label{mLnuk_eq}
 \frac{d \langle \nu_k m_L \rangle}{d t} +N\frac{d \langle m_L^2 \rangle}{d t}=& -4aN \langle m_L^2 \rangle - 2(2 a + h \xi) \langle \nu_{k} m_{L} \rangle \nonumber\\
 & + \frac{4k}{\mu}\left( a + h \xi (1-\langle m_L^2 \rangle) \right)- 2 h \left\langle \frac{\lambda_{k} \lambda_{\rho} m_L}{1-m_L^2} \right\rangle \, .
\end{eqnarray}
We now set
\begin{equation}
\label{simple}
\left\langle \frac{\lambda_{k} \lambda_{\rho} m_L}{1-m_L^2} \right\rangle_{\rm st} \approx 0 \, .
\end{equation}
A justification of this assumption relies on the use of the van Kampen expansion Eqs.(\ref{expansion_detk}, \ref{expansion_dettho}), that determines that at order $O(N^{-1/2})$ the average of Eq.(\ref{simple}) is proportional to $\langle \gamma_{k} \gamma_{k'} \gamma_{\rho} \rangle$, which is zero as the stochastic variables $\gamma_{a}$ are Gaussian \cite{moments}. This analysis is strictly valid only in the regime of large $a\gg a_c$, but in Fig. \ref{fig:assumption} we provide numerical evidence of its validity in the whole range of parameters of $a$. We observe that the assumption works very well for the Erd\"os--R\'enyi and Barab\'asi--Albert networks, while for the dichotomous network there is a narrow parameter region around the critical point where the average Eq.(\ref{simple}) is non-zero but still negligible.

Once this term has been neglected, the solution of Eq. (\ref{mLnuk_eq}) in the stationary state is
\begin{equation}
\label{mLnuk_st}
\langle \nu_k m_L \rangle_{{\rm st}} = - \frac{2 a N + 2 h \xi \frac{k}{\mu}}{2 a + h \xi}\langle m_L^2 \rangle_{{\rm st}} + \frac{2 k}{\mu} \frac{a + h \xi}{2 a + h \xi} \, .
\end{equation}
Replacing this result in Eq. (\ref{2corr_kk}), we obtain the following correlations in the stationary state
\begin{equation}
\label{corr_VK_st}
 \langle \lambda_{k} \lambda_{k'} \rangle_{{\rm st}} = -\frac{k+k'}{\mu}+ \frac{ aN\mu + h \xi(k+k')}{\mu(a + h \xi)}\langle m_L^2 \rangle_{{\rm st}} + \frac{\delta_{k,k'}}{P_{k}} \left( 1 - \frac{h \xi}{a + h \xi} \langle m_L^2 \rangle_{{\rm st}} \right).
\end{equation}

We have now all necessary ingredients to calculate the first and second moments of the global magnetization $m = \sum_{k} P_{k} m_{k}$, both in the stationary and transient regimes. For example, the stationary variance is
\begin{eqnarray}
\label{m2_st}
\sigma^2_{\rm st}[m] &=\langle m_L^2 \rangle_{{\rm st}} + N^{-1} \bigg( 2 \sum_{k} P_{k} \langle \nu_{k} m_L \rangle_{{\rm st}} + \sum_{k, k'} P_{k} P_{k'} \langle \lambda_{k} \lambda_{k'} \rangle_{{\rm st}} \bigg) \, ,
\end{eqnarray}
which leads to
\begin{eqnarray}
\label{av_mag2}
N\sigma^2_{\rm st}[m]&= \frac{h^2 \xi^2}{2 a + h \xi}\frac{N}{a N_{{\rm eff}} + h \xi} + \frac{2 h \xi}{2 a + h \xi} + \frac{a N_{{\rm eff}}}{a N_{{\rm eff}} + h \xi} \, .
\end{eqnarray}
The critical point is found by solving $\sigma^2_{\rm st}[m] =\frac{N+2}{3N}$, taking into account that $\xi$ depends on $a$,
\begin{equation}
\label{critical_point}
a_{c} = \frac{h}{N_{{\rm eff}}} \frac{\mu-2}{\mu-1}+O(N^{-2}) \, .
\end{equation}
The average value of $\rho$ can be determined in the stationary regime as
\begin{equation}
\label{rho_st}
\langle \rho \rangle_{{\rm st}} = \xi \left( 1-\langle m_L^2 \rangle_{{\rm st}} \right) + O(N^{-1}) \, .
\end{equation}
At variance with the results of the expansion around the deterministic solution [see Eqs. (\ref{rhoN1}) and (\ref{limit1})], the above expression is not divergent for $a \rightarrow 0$.

It is important to stress the similarities and differences of our stochastic pair approximation method with the one developed in Ref.~\cite{Vazquez2008} for the noiseless voter model. In that paper the authors use a master equation approach for the link-magnetization $m_L$ which can readily be extended to the noisy case (see \ref{app:link}) yielding for $m_L$ the same statistical properties as the ones obtained by our more sophisticated approach. However, the authors of Ref.~\cite{Vazquez2008} make, implicitly, the extra assumption that the link magnetization $m_L$ and the node magnetization $m$ share the same statistical properties. To check the validity of this assumption, and using $\langle m m_L \rangle_{{\rm st}} = \langle m_L^2 \rangle_{{\rm st}} + N^{-1} \sum_{k} P_{k} \langle \nu_{k} m_L \rangle_{{\rm st}}$, we obtain
\begin{equation}
\label{difm}
\langle (m-m_L)^2 \rangle_{{\rm st}} = \left( 1-\frac{N_{{\rm eff}}}{N} \right) \frac{a }{a +h \xi} \langle m_{L}^2 \rangle_{{\rm st}} \, ,
\end{equation}
a function of the degree heterogeneity and the parameters of the model (the ratio of time scales). From this expression we conclude that it is true that in the noiseless voter model, with $a=0$, $\langle (m-m_L)^2 \rangle_{{\rm st}}$ becomes zero (rather, of order $N^{-1}$, which is the accuracy of our expansion), validating the approach of Ref.~\cite{Vazquez2008} in that case. Remarkably, in the limit of a homogeneous network, with $N_{{\rm eff}}=N$, the statistical properties of $m$ and $m_L$ also coincide, consistently with the fact that $m$ and $m_L$ are the same quantity for those networks. Furthermore, in the critical zone $a=O(N^{-1})$ and for a heterogeneous network we obtain that $\langle (m-m_L)^2 \rangle_{{\rm st}}= O(N^{-1})$, meaning that when there is time scale separation $m$ does not differ significantly from $m_L$. Our analysis shows that in the noisy voter model the assumption that the statistical properties of the magnetization are those of the link magnetization can not be held for heterogeneous networks far away from the critical region, while for the voter model without noise ($a=0$) the statistical differences between both quantities turn out to be negligible. The relevance of this difference for other stochastic processes defined in heterogeneous networks is under study.

\section{Comparison with numerical simulations}\label{sec:numerical}

We now compare the different theoretical approaches based on the stochastic pair approximation and others with the results of numerical simulations. We focus on the steady state values $\sigma^2_{\rm st}[m]$ and $\langle \rho\rangle_{\rm st}$ as functions of the noise parameter $a$ and the degree heterogeneity $\kappa\equiv\mu_2/\mu^2-1$, for six different network structures keeping a constant value for the average connectivity $\mu=4$ and the total number of nodes $N=2500$, namely:
\begin{enumerate}[label=\arabic*)]
 \item an Erd\"os-R\'enyi random network with a fixed number of links $\mu N /2$ that are connected randomly between pairs of nodes, thus leading to a degree heterogeneity $\kappa = 0.19$;
 \item a Barab\'asi-Albert preferential-attachment network, with a power-law degree distribution $P_k\sim k^{-3}$ and a degree heterogeneity $\kappa=3.05$;
 \item a dichotomous network where a fraction $23/24$ of nodes has $k=2$ neighbors and the remaining fraction $1/24$ has $k=50$ neighbors, leading to a degree heterogeneity $\kappa=5.75$;
 \item a $z$-regular network where each node has exactly $k=4$ neighbors but the connections amongst them are chosen at random;
 \item a one-dimensional (1D) linear lattice where a node is connected to its closest 4 neighbors (2 to the right and 2 to the left);
 \item and a two-dimensional square lattice (2D) where each node is connected to its 4 nearest neighbors.
\end{enumerate}
The last three of these networks are called ``homogeneous", as $\kappa=0$. For each one of these networks we plot the theoretical predictions for the variance of the magnetization $\sigma^2_{\rm st}[m]$ and the average density of active links $\langle \rho\rangle_{\rm st}$ in the steady state as given by:
\begin{enumerate}[label=\alph*)]
 \item the global-state-approach as presented in Eqs. (\ref{second_FC}) and (\ref{rho_FC}), which is independent of all network details (the comparison has already been presented in Fig.~\ref{fig_var});
 \item the annealed-approximation as developed in Ref.~\cite{Carro2016};
 \item the expansion around the deterministic solution (S1PA) as given by Eqs. (\ref{variance_mag}) and (\ref{rhoN1});
 \item the expansion around the dynamical attractor (S2PA) as given by Eqs. (\ref{av_mag2}) and (\ref{rho_st});
 \item and, for comparison, we have also included the variance $\sigma^2_{\rm st}[m_L]$, which would have been the result of extending the theory developed at Ref.~\cite{Vazquez2008} to the noisy voter model, neglecting the difference between the magnetization $m(t)$ and the link-magnetization $m_L(t)$.
\end{enumerate}

\begin{figure}[ht]
\centering
\subfloat[]{\label{fig:stat_var_a:a}\includegraphics[width=0.49\textwidth]{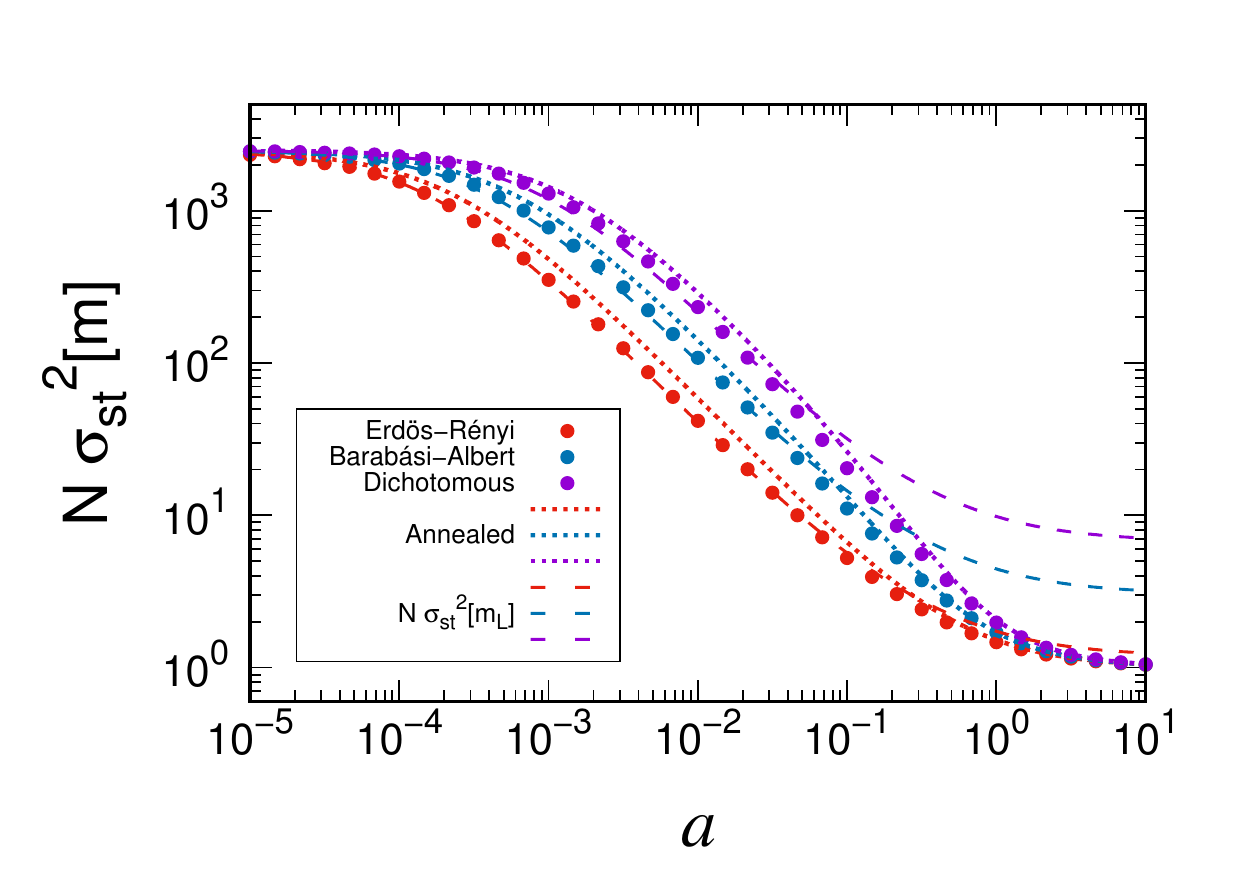}}\subfloat[]{\label{fig:stat_var_a:b}\includegraphics[width=0.49\textwidth]{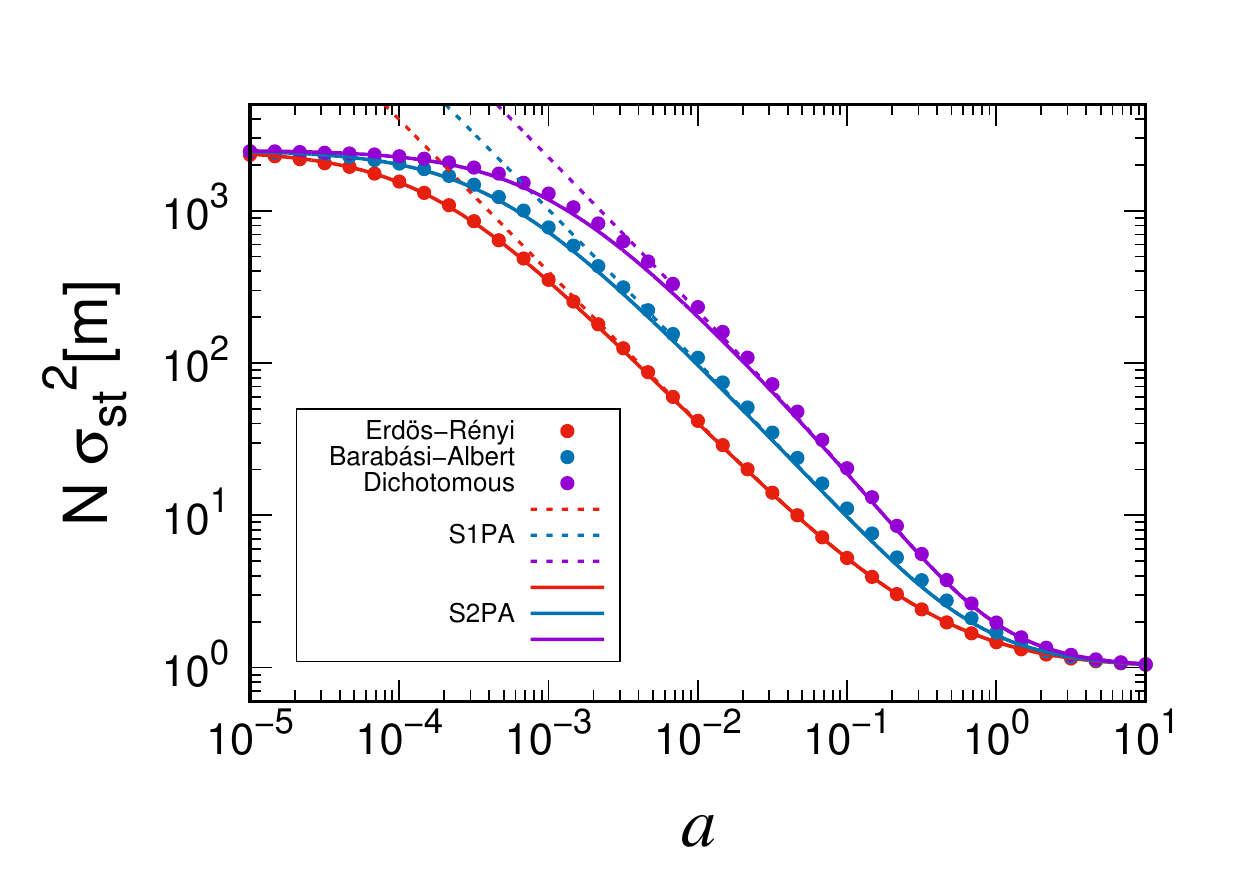}}\\
\subfloat[]{\label{fig:stat_var_a:c}\includegraphics[width=0.49\textwidth]{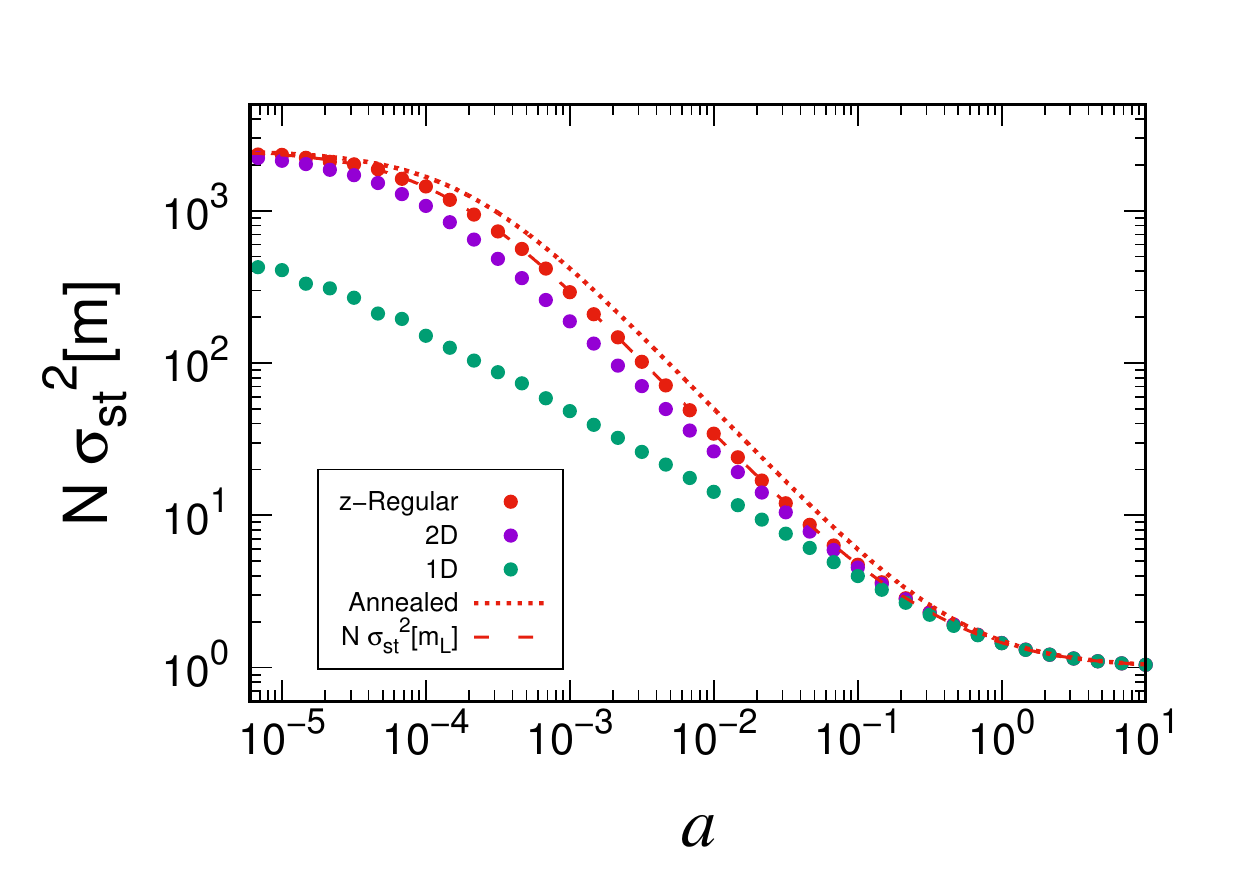}}\subfloat[]{\label{fig:stat_var_a:d}\includegraphics[width=0.49\textwidth]{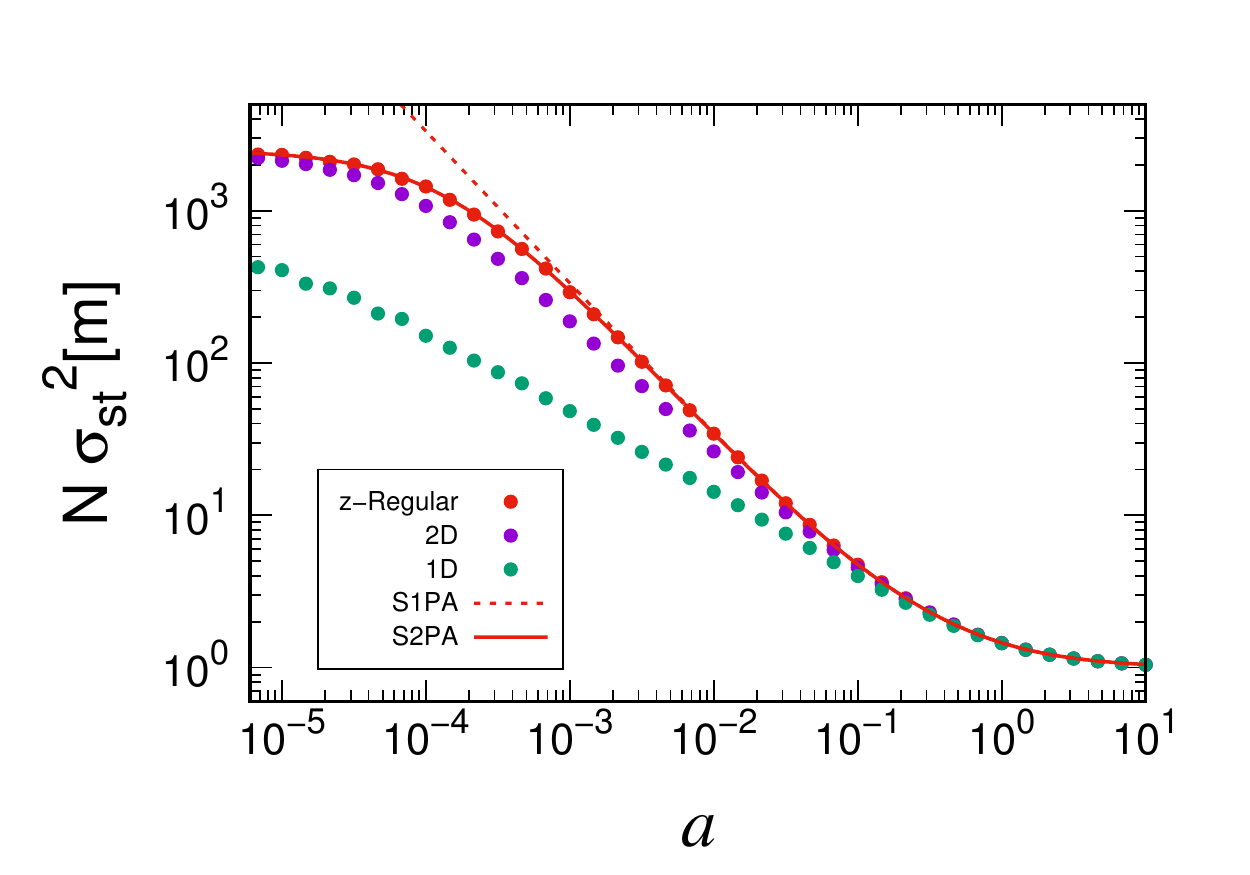}}
\caption{Stationary variance of the magnetization as a function of $a$ for different networks with fixed $\mu=4$ and $N=2500$. Figs.~\ref{fig:stat_var_a:b} and~\ref{fig:stat_var_a:d} show the comparison with the methods developed in this paper, S1PA and S2PA, for heterogeneous (Fig.~\ref{fig:stat_var_a:b}) and homogeneous (Fig.~\ref{fig:stat_var_a:d}) networks, while Figs.~\ref{fig:stat_var_a:a} and~\ref{fig:stat_var_a:c} show the corresponding comparison for the annealed approximation of Ref.~\cite{Carro2016}, and the results obtained neglecting the difference between the magnetization $m$ and the link magnetization $m_L$. Dots correspond to numerical simulations while lines are analytical results according to the legend.}
\label{fig:stat_var_a}
\end{figure}

\begin{figure}[ht]
\centering
\subfloat[]{\label{fig:stat_var_kappa:a}\includegraphics[width=0.49\textwidth]{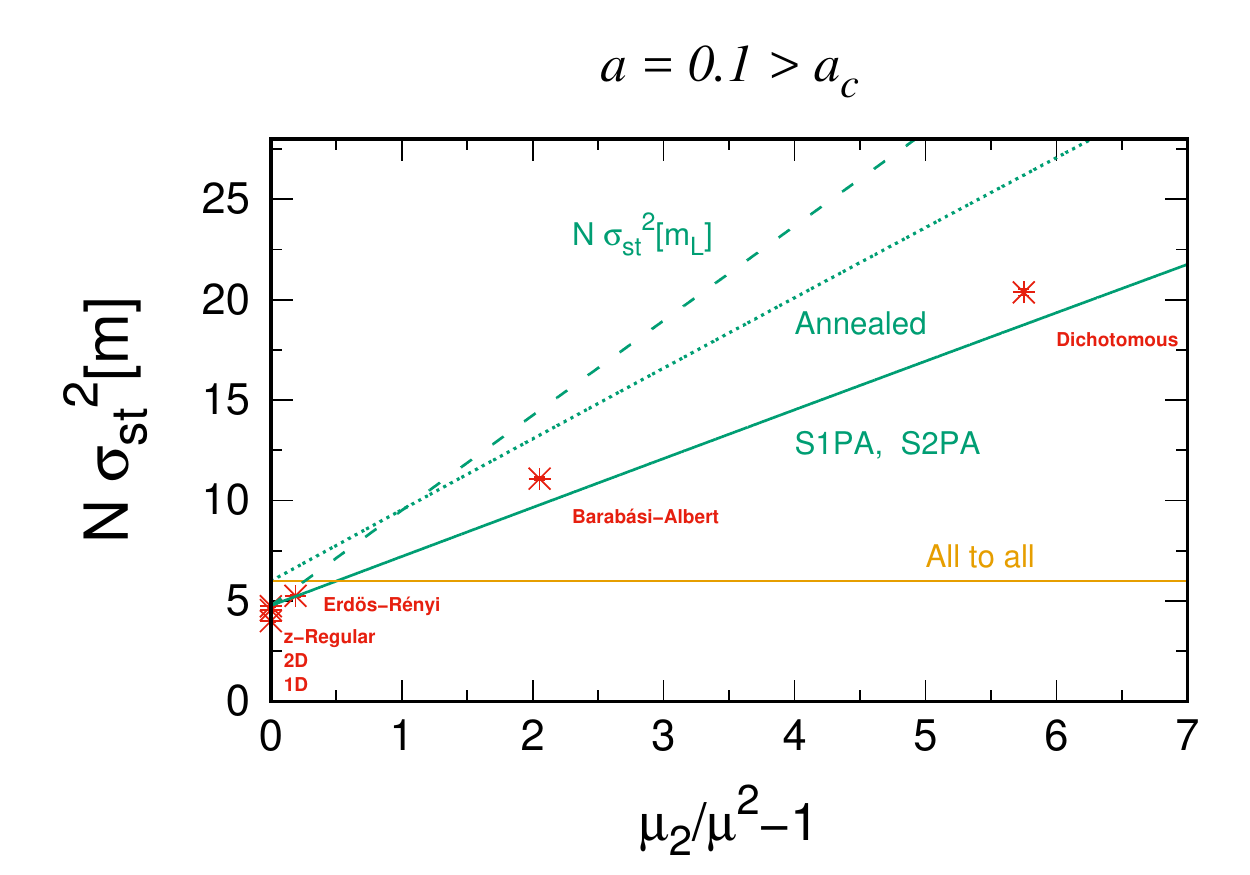}}
\subfloat[]{\label{fig:stat_var_kappa:b}\includegraphics[width=0.49\textwidth]{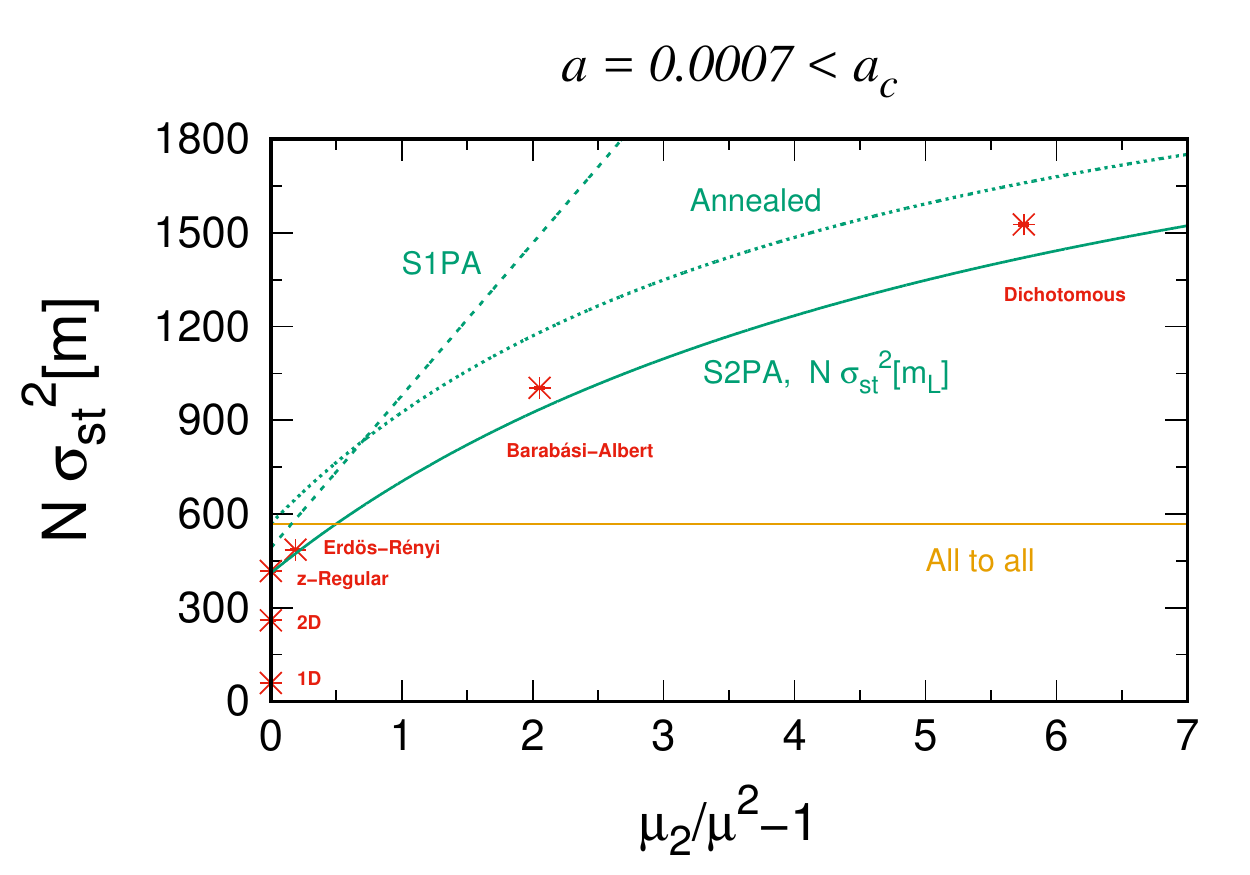}}
\caption{Stationary variance of the magnetization as a function of degree heterogeneity $\kappa = \mu_2/\mu^2-1$ for different networks with fixed $\mu=4$ and $N=2500$. Fig.~\ref{fig:stat_var_kappa:a} focuses on $a>a_c$, where the S1PA and the S2PA approaches coincide, while Fig.~\ref{fig:stat_var_kappa:b} focuses on $a<a_c$, where the S2PA and the assimilation of the magnetization $m$ and the link magnetization $m_L$ lead to similar results.}
\label{fig:stat_var_kappa}
\end{figure}

\begin{figure}[ht]
\centering
\includegraphics[width=0.7\textwidth]{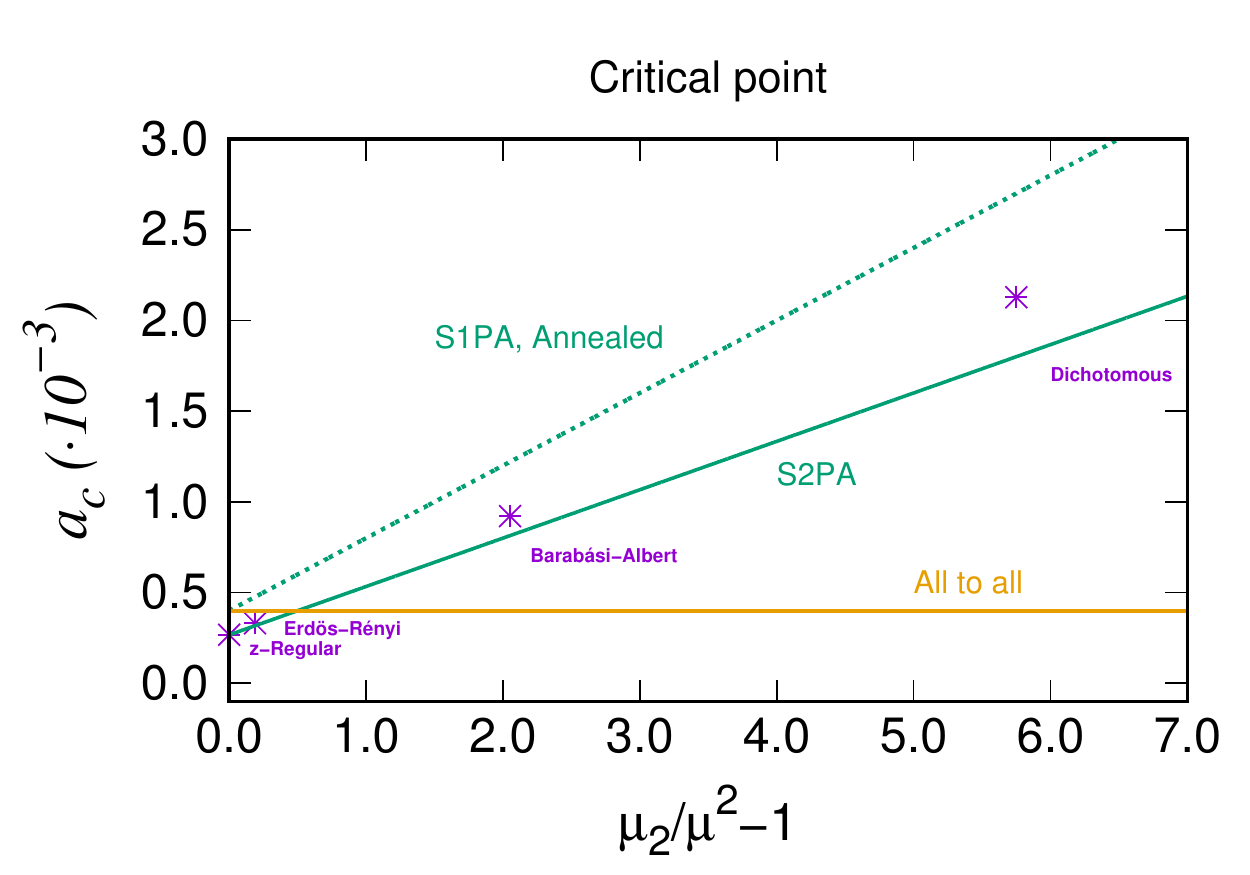}
\caption{The critical point as a function of degree heterogeneity $\mu_2/\mu_1^2-1$ for networks with fixed $\mu=4$ and $N=2500$. Dots correspond to numerical simulations while lines are analytical results: $a_c=h/N$ for the all-to-all approach and Eqs. (\ref{ac_S1PA}) and (\ref{critical_point}) for S1PA and S2PA, respectively. The annealed network approximation coincides with S1PA.}
\label{fig:critical}
\end{figure}

In Fig.~\ref{fig:stat_var_a} we plot $N\sigma^2_{\rm st}[m]$ as a function of the noise parameter $a$. For clarity, we have split the comparison in several panels. We can appreciate in Figs.~\ref{fig:stat_var_a:b} and~\ref{fig:stat_var_a:d} how S2PA reproduces very accurately the numerical results for the heterogeneous and the z-regular networks in the whole range of values of $a$, both above and below the critical point $a_c$. The results of S1PA are indistinguishable from those of S2PA above the critical point $a>a_{c}$, while below it $a<a_{c}$ the S1PA prediction for $N\sigma^2_{\rm st}[m]$ diverges as $a \rightarrow 0$, at odds with S2PA and the numerical simulations. We expect a full agreement between these two approaches only in the thermodynamic limit, in which $N\rightarrow \infty$ with $a$ fixed, as Eq. (\ref{av_mag2}) and Eq. (\ref{variance_mag}) would then coincide. The data is also analyzed in Fig.~\ref{fig:stat_var_kappa} as a function of the degree heterogeneity $\kappa$. While for $a > a_{c}$ (Fig.~\ref{fig:stat_var_kappa:a}) $\sigma^2_{\rm st}[m]$ is proportional to the degree heterogeneity in accordance with Eq.~(\ref{variance_mag}), for $a < a_{c}$ (Fig.~\ref{fig:stat_var_kappa:b}) the dependence is not linear but well captured by Eq.~(\ref{av_mag2}). On the other hand, $N\sigma^2_{\rm st}[m_L]$ coincides with $N\sigma^2_{\rm st}[m]$ only below the critical point $a<a_{c}$ or for homogeneous networks $\kappa \rightarrow 0$, where the difference between $m$ and $m_L$ is irrelevant, as discussed in the previous section. The annealed approximation developed in Ref.~\cite{Carro2016} reproduces correctly both the dependence on $a$ and $\kappa$, but it is less accurate in general than the pair approximation (S2PA). The global state approach Eq.~(\ref{second_FC}) has the correct scaling, but it is accurate only for networks with low levels of heterogeneity and does not predict any dependence with $\kappa$. In the limit $\mu_2=\mu^2\to\infty$, and for all values of $a$ below and above the critical point, the relative difference between Eqs.~(\ref{av_mag2}) and~(\ref{second_FC}) is of order $N^{-1}$. We also check the predicted values for the critical point $a_{c}$ in Fig.~\ref{fig:critical}. Again, the S2PA approach offers the best fit to the numerical data, while the S1PA and the annealed approximations coincide for all values of the degree heterogeneity.

\begin{figure}[ht]
\centering
\subfloat[]{\label{fig:rho_stat:a}\includegraphics[width=0.49\textwidth]{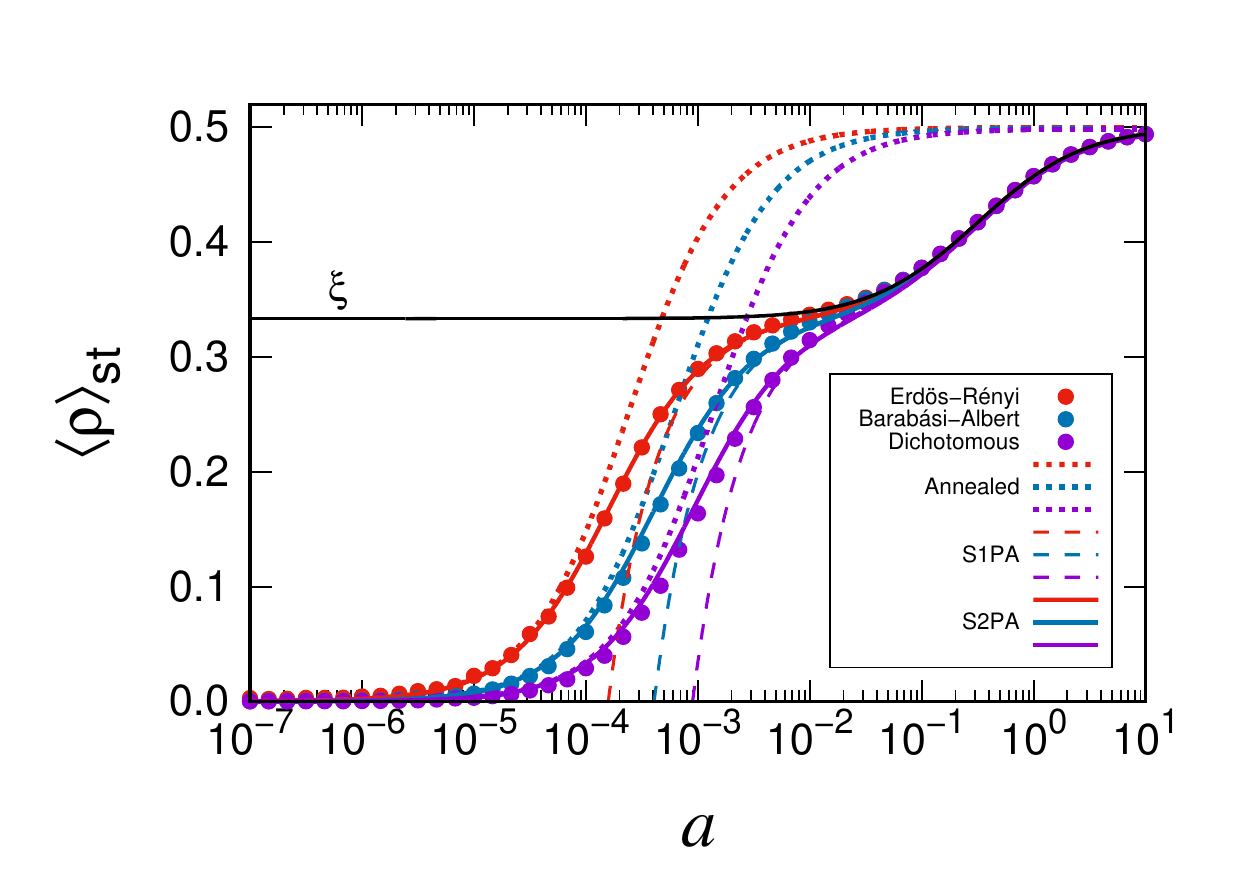}}
\subfloat[]{\label{fig:rho_stat:b}\includegraphics[width=0.49\textwidth]{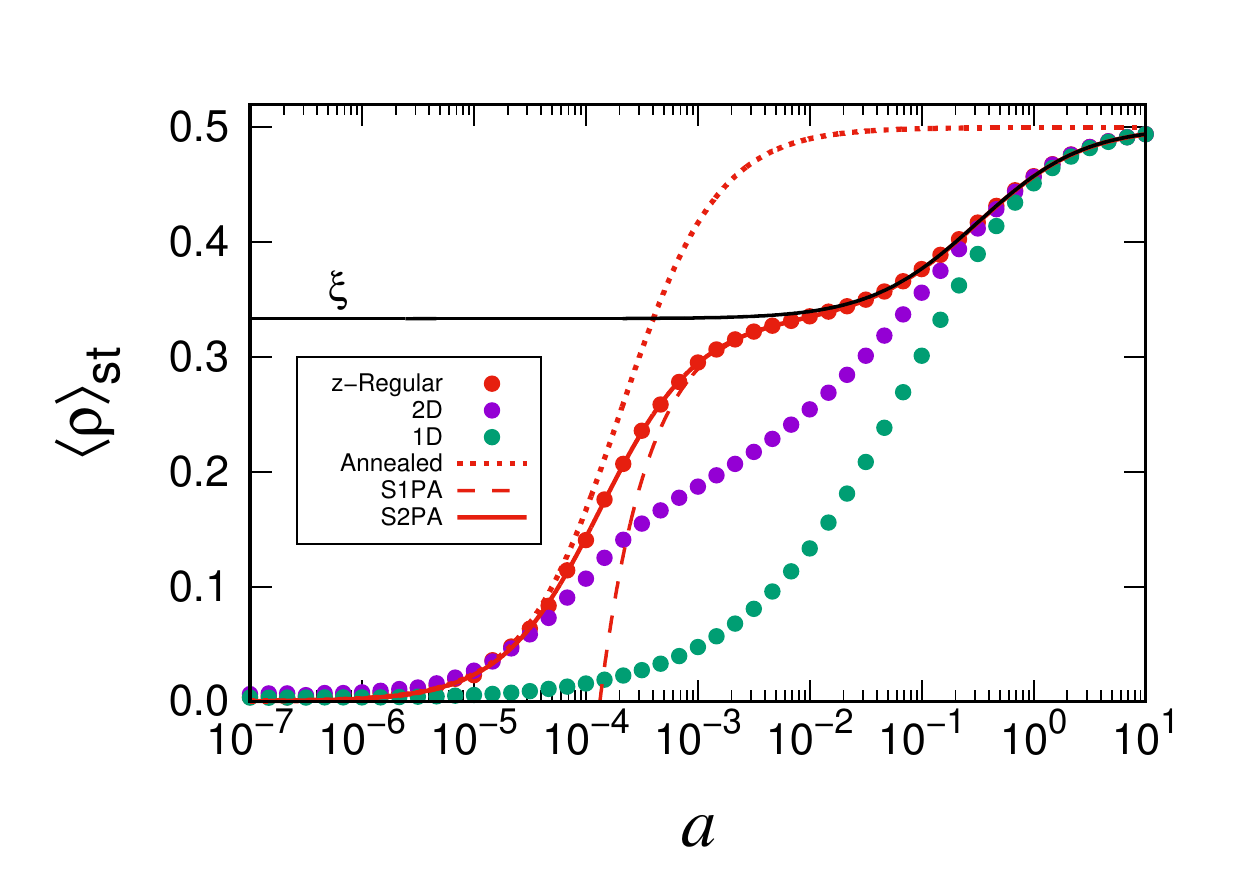}}
\caption{Stationary average density of active links as a function of $a$ for different networks with $\mu=4$ and $N=2500$. Fig.~\ref{fig:rho_stat:a} focuses on heterogeneous random networks and Fig.~\ref{fig:rho_stat:b} on homogeneous networks. Dots correspond to numerical simulations while lines are analytical results according to the legend. While for random networks (both heterogeneous or homogeneous) the S2PA offers a very good agreement for all values of $a$, the annealed approximation fails qualitatively for intermediate values and S1PA gives non-physical results $\langle \rho\rangle_{\rm st}<0$ for small $a$.}
\label{fig:rho_stat}
\end{figure}

In Fig.~\ref{fig:rho_stat} we plot the results of $\langle \rho \rangle_{{\rm st}}$ as a function of $a$. We have split the comparison for heterogeneous random (Fig.~\ref{fig:rho_stat:a}) and homogeneous (Fig.~\ref{fig:rho_stat:b}) networks. We can observe again a very high degree of accuracy of S2PA for random networks in the whole range of values of $a$. The results using the expansion around the deterministic solution S1PA are accurate above the critical point, but below it they show a divergence leading to non-physical results $\rho<0$. Comparing with other treatments, such as the annealed and all-to-all approximations, it is clear that the stochastic pair approximation S2PA developed here perfectly captures the shape of the curve $\langle \rho \rangle_{{\rm st}}$ vs $a$, while the annealed approximation discussed in Ref.~\cite{Carro2016} fails even qualitatively. This failure of the annealed and all-to-all approximations is directly related to the fact that the average degree $\mu$ is finite, as one can prove that in the limit $\kappa \rightarrow 0$ Eq.~(\ref{rho_st}) and Eq.~(\ref{rho_FC}) coincide only when $\mu \rightarrow \infty$ and thus $\xi \rightarrow 1/2$. Note that the expression of S2PA for $\langle \rho \rangle_{{\rm st}}$ given at Eq.~(\ref{rho_st}) is independent on whether the spin $m$ and link $m_L$ magnetizations are assumed to be equal or not, and thus this approximation leads to the same result as the extension of the theory developed at Ref.~\cite{Vazquez2008} for $\langle \rho \rangle_{{\rm st}}$.

Both the variance of the magnetization and the average density of active links show significant discrepancies between the results of numerical simulations and the proposed analytical methods in the case of regular networks of low dimensionality $d=1,\,2$. This is specially notorious for the 1D case, where even the scaling $\langle m^2 \rangle_{{\rm st}} \sim O(N^{0})$ if $a \sim O(N^{-1})$ does not apply and, instead, it is $\langle m^2 \rangle_{{\rm st}} \sim O(N^{-1/2})$. This is the reason why there is no finite size critical point in Fig.~\ref{fig:critical}, as defined previously, for the 1D and 2D cases. As the critical dimension of the noisy voter model is $d=2$, for any dimension above this value the same scaling properties are expected, in accordance to the all-to-all result. A detailed analysis of the finite-size scaling for regular lattices will be published elsewhere~\cite{regularD}.

\section{Time dependence}\label{sec:dynamics}

\begin{figure}[ht]
\centering
\includegraphics[width=0.7\textwidth]{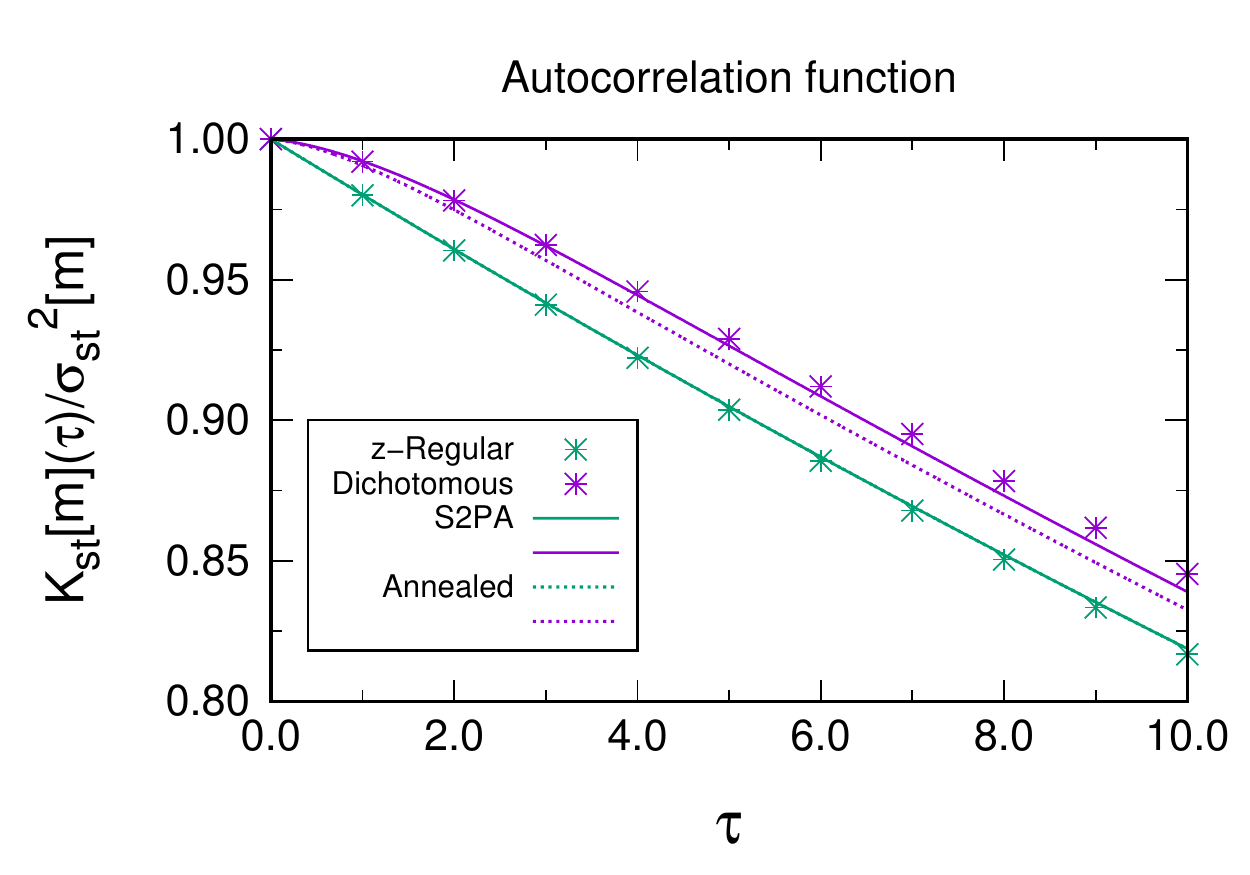}
\caption{Stationary autocorrelation function of the magnetization as a function of the time-lag $\tau$ for the dichotomous and z-regular networks with $a=0.01$, $\mu=4$ and $N=2500$. Dots correspond to numerical simulations while lines are analytical results: dashed lines are the annealed approximation~\cite{Carro2016}, while solid lines are the S2PA results, Eq.~(\ref{corr2}). The green-dotted analytic curve for the annealed approximation is superposed to the solid green line because in the absence of degree heterogeneity, i.e. for the z-regular network, $K_{{\rm st}}[m](t) = e^{-2 a t}$ in both approaches.}
\label{fig:corr}
\end{figure}

The method S2PA, developed in the previous sections, allows us to determine the time evolution of $\langle m(t) \rangle$ and $\langle \rho(t) \rangle$. From the construction of the variables we have $\langle m_k(t) \rangle = \delta_k(t) + \langle m_L(t) \rangle$, which leads to
\begin{equation}
\label{corr1}
\langle m(t) \rangle = m_L(0) e^{-2 a t} + \left[ m(0)-m_{L}(0) \right] e^{-2(a+h \xi) t} \, ,
\end{equation}
assuming that we start with a fixed initial condition $m(0)$ and $m_L(0)$. A preferable quantity to measure this time evolution, that does not depend on the initial conditions, is the stationary autocorrelation function
\begin{equation}
\label{corr2}
 K_{{\rm st}}[m](t) \equiv\langle m(t)m(0) \rangle_{{\rm st}} = \langle m m_L \rangle_{{\rm st}} e^{-2 a t} + \left[ \langle m^2 \rangle_{{\rm st}} - \langle m m_L \rangle_{{\rm st}} \right] e^{-2(a + h \xi) t} \, .
\end{equation}
The existence of two exponentials in this expression is directly related to the degree heterogeneity, as already pointed out with the annealed approximation in Ref.~\cite{Carro2016}. In Fig.~\ref{fig:corr} we plot the time evolution of $K_{{\rm st}}[m](t)$ for the two extreme cases of a network with no degree heterogeneity (z-regular random network) and a highly heterogeneous degree distribution (dichotomous network). While for the z-regular network there is only one slow exponential, for the dichotomous network the early stages are dominated by the fast exponential and later the slow part dominates, in accordance to Eq.~(\ref{corr2}). Hence, by fitting a double exponential to the autocorrelation data we are then capable of obtaining information about the network heterogeneity.

While it has not been studied before in the context of the noisy version of the model, the time evolution of the interface density of active links $\rho$ is commonly used to describe the dynamics of the (noiseless, $a=0$) voter model. In that case, the voter model is characterized by the existence of two absorbing states ($m=-1$ and $m=1$), both of them corresponding to full order or consensus ($\rho=0$). Therefore, the focus is on how the system approaches consensus and if it does so in the thermodynamic limit. In particular, the ordering process is found to be determined by the dimensionality $d$ of the underlying topology. For $d \leq 2$, and in the infinite size limit, the system orders by a coarsening process, i.e., by a continuous and unbounded growth of spatial domains of one of the opinions. The ensemble average interface density decays as $\langle \rho \rangle \sim t^{-1/2}$ for $d = 1$ and as $\langle \rho \rangle \sim (\log t)^{-1}$ for $d = 2$. After an initial transient, infinite systems with $d > 2$ are observed to fall into a metastable, highly disordered state ($0 < \rho < 1/2$) where coarsening processes have stopped. In these cases, the average interface density behaves as $\langle \rho \rangle \sim b - c \cdot t^{-d/2}$, and thus complete order is never reached. Finite-size systems, on the contrary, are always led to complete order by finite-size fluctuations. As a consequence, finite systems follow the above described behaviors only for a finite time, after which an exponential decay to complete order is observed. Notably, it has been shown that this characteristic decay time is proportional to the effective system size $N_{\rm eff}$ (for $d>2$).

According to the theoretical description presented in the previous Section~\ref{sec:expansion2}, the time dependence of $\langle \rho(t) \rangle$ can be determined as
\begin{equation}
\label{rho_temp}
\langle \rho(t) \rangle = \delta_{\rho}(0) e^{- J_{\rho,\rho} t} + \xi \left( 1-\langle m_L(t)^2 \rangle \right) + O(N^{-1}) \, ,
\end{equation}
where $\delta_{\rho}(0) = \langle \rho(0) \rangle - \xi$ and, for simplicity, we assume $\langle m_k(0) \rangle = \langle m_L(0) \rangle = 0$ so that the term $\delta_{\rho}(0) e^{- J_{\rho,\rho} t}$ is the only one that remains from Eq.~(\ref{delta_rho}). The form of this expression is in line with the qualitative description given in Ref.~\cite{Diakonova2015}, where the average interface density is said to be characterized by a short initial transient after which a plateau is reached at a value $\rho^* = \xi > 0$. In a more quantitative description, Eq.~(\ref{rho_temp}) contains the initial transient $\delta_{\rho}(0) e^{- J_{\rho,\rho} t}$, the exponential decay due to finite size effects after the plateau $\xi \langle m_L^2 \rangle_{{\rm st}} e^{-t/\tau}$ with $\tau=\frac{N_{{\rm eff}}}{4(a N_{{\rm eff}} +h \xi)}$ and the final steady state described by Eq.~(\ref{rho_st}). Note how the decay time of the initial transient is independent of the degree heterogeneity, unlike the one due to finite size effects. For $a<a_{c}$, $\tau \sim N_{{\rm eff}} \gg 1/J_{\rho,\rho}$, whereas $\tau$ and $1/J_{\rho,\rho}$ are both of the order $N^0$ for $a>a_c$. The difference between the plateau and the steady state $\xi \langle m_L(t)^2 \rangle$ is of order $1/N$ for $a>a_c$ and of order $aN$ for $a<a_c$.

\begin{figure}[ht]
\centering
\subfloat[]{\label{fig:rho_t:a}\includegraphics[width=0.49\textwidth]{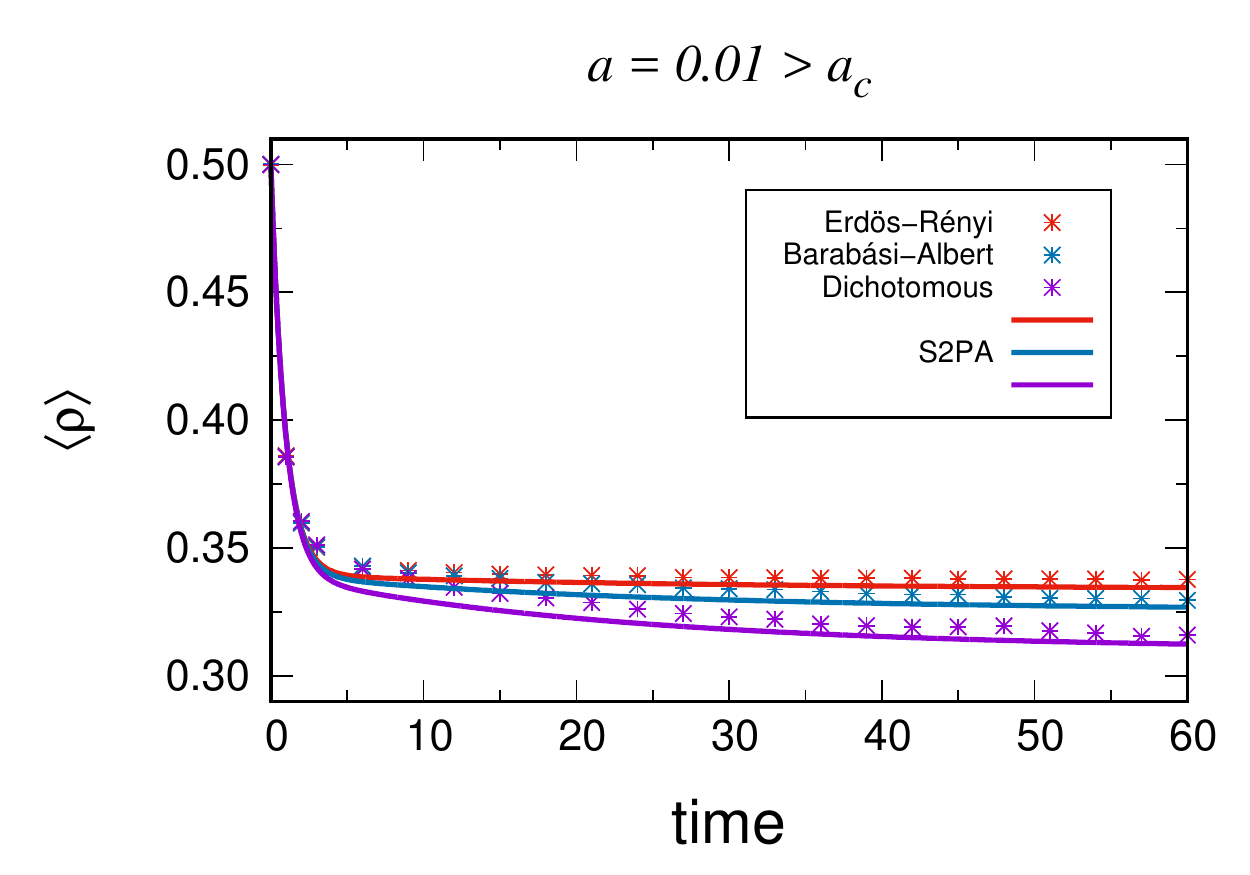}}
\subfloat[]{\label{fig:rho_t:b}\includegraphics[width=0.49\textwidth]{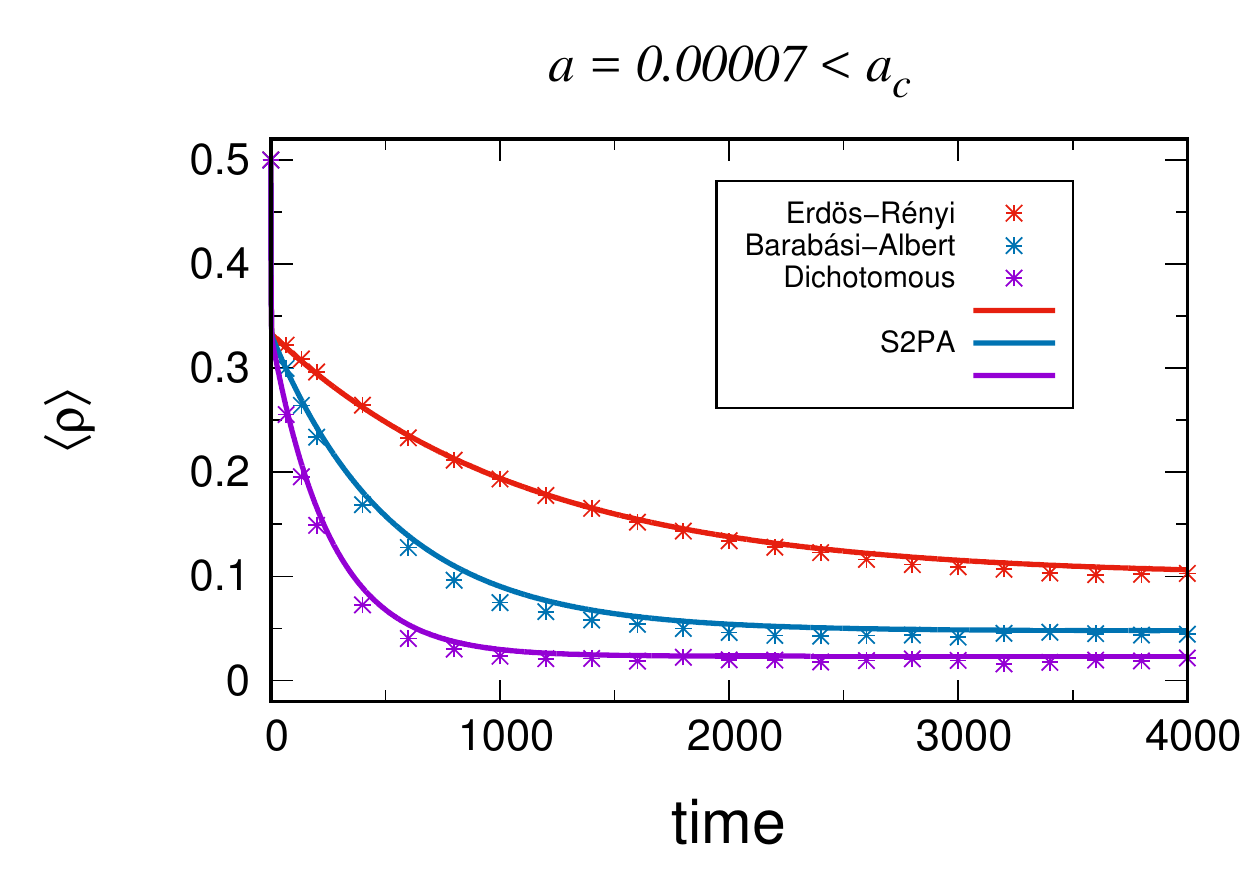}}\\
\subfloat[]{\label{fig:rho_t:c}\includegraphics[width=0.49\textwidth]{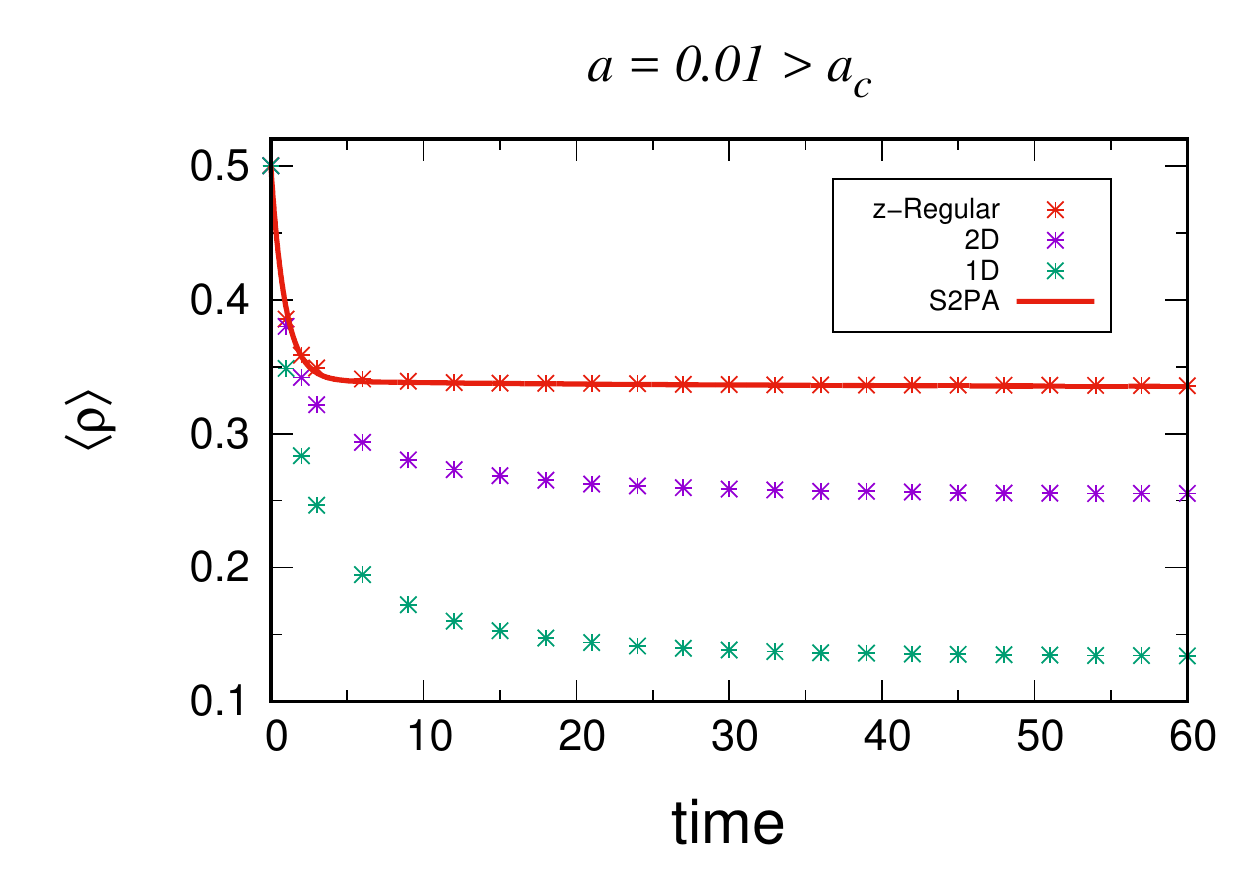}}
\subfloat[]{\label{fig:rho_t:d}\includegraphics[width=0.49\textwidth]{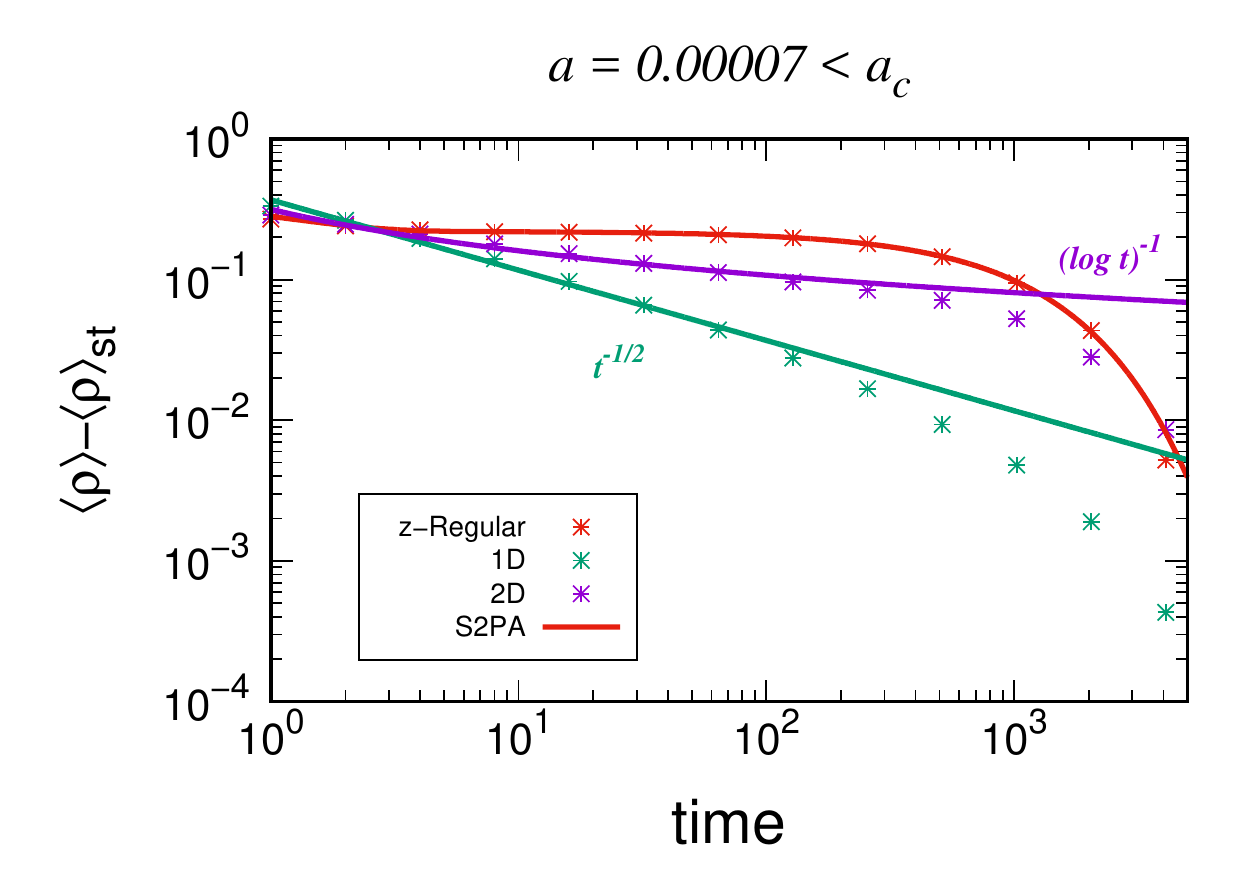}}
\caption{Temporal evolution of the average density of active links for different networks with $\mu=4$ and $N=2500$. Two fixed values of $a$ are taken, one below the critical point, Fig.~\ref{fig:rho_t:a} and~\ref{fig:rho_t:c}, and one above, Fig.~\ref{fig:rho_t:b} and~\ref{fig:rho_t:d}. Dots correspond to numerical simulations while lines are the analytical result Eq.~(\ref{rho_temp}).}
\label{fig:rho_t}
\end{figure}

Numerical results for the time evolution of the average order parameter are presented in Fig.~\ref{fig:rho_t} for the six different networks studied. Note that, due to their particular ordering dynamics and for comparison with the voter model, we have included both one- and two-dimensional regular networks. In this figure, we can clearly distinguish the two stages in the dynamics of the average interface density identified in Ref.~\cite{Diakonova2015} and described by Eq.~(\ref{rho_temp}): an initial ordering transient, with a significant decrease of the order parameter from its initial, close to complete disorder value $\rho = 1/2$, corresponding to a random initial distribution of states; and a final steady state, where the order parameter fluctuates around a characteristic level of disorder. As shown in the figure, for $a<a_c$ the initial ordering behavior of the noisy voter model is remarkably similar to the ordering dynamics of the voter model described above: a power-law decay as $\langle \rho \rangle \sim t^{-1/2}$ for $d = 1$ (regular 1D), a logarithmic decay as $\langle \rho \rangle \sim (\ln t)^{-1}$ for $d = 2$ (regular 2D), and a fast decay to the plateau followed by a slow exponential decay $\langle \rho \rangle \sim \rho^* e^{-t/\tau}$ for $d > 2$ (z-regular, Erd\"os-R\'enyi, Barab\'asi-Albert, dichotomous). Therefore, as in the voter model, the initial ordering of systems with dimension $d \leq 2$ is characterized by a coarsening process with growth of spacial domains, while the ordering of systems with dimension $d > 2$ occurs mainly by finite-size fluctuations after a period in a metastable, highly disordered state. We can also note that, apart from coarsening, finite-size fluctuations are present too for $d \leq 2$, being responsible for the exponential decay interrupting the coarsening process. It is important to note, however, that, while in the voter model any ordering process in a finite system is a transient towards complete order, in the noisy voter model it is a transient towards a non-zero steady state of the order parameter. Thus, the voter model-like initial ordering of the noisy voter model is interrupted whenever the average order parameter reaches its steady state value. In general, the behavior for $a>a_{c}$ is quite different, as one can notice from the figure. This is because the influence of finite size effects on $\langle \rho \rangle$ becomes weaker as $a$ is increased. Thus, the value at the plateau $\rho=\rho^*$ is closer to the steady state solution $\langle \rho \rangle_{{\rm st}}$ and the time scale of the fast initial transient is not that different from the one of finite size effects $\tau$, which mixes the two behaviors. Furthermore, for $d \leq 2$ the previous voter-like time dependence is no longer true.

\section{Conclusions}\label{sec:conclusions}

We have developed a full stochastic approach to the pair approximation scheme to study binary state dynamics on heterogeneous networks. Our starting point is the master equation for the set of variables $\{L,N_{1,k}\}$, being $N_{1,k}$ the number of nodes with degree $k$ in the state up and $L$ the number of active links. The effective rates of the master equation are obtained using the pair approximation as discussed in Ref.~\cite{Vazquez2008} in the context of the noiseless voter model. For the noisy voter model, however, we show that the spin magnetization and the link magnetization cannot be identified. Once the equations for the time evolution of the moments and correlations are formulated, we proceed to close the hierarchy by using two different closure schemes. The first one, S1PA, is based on a van~Kampen type of expansion around the deterministic solution of the dynamics corresponding to the thermodynamic limit, while the second one, S2PA, is based on an expansion around a stochastic dynamical attractor.

In the first approach, we expand the solution around the deterministic fixed point. In this way we are able to derive a linear set of equations for the correlation matrix and first corrections to the average values. Although our methodology is very general for any set of rates, we have focused on the noisy voter model, for which we have been able to carry out a full analytical treatment. This first analysis, based on the ansatz that the fluctuations of the magnetization scale with system size as $\sigma_{{\rm st}}^2[m] \sim N^{-1}$, predicts that the fluctuations depend linearly on the degree heterogeneity of the network. We also obtained an expression for the average density of active links $\langle \rho \rangle_{{\rm st}}$. Both expressions have a mathematical divergence as the noise intensity approaches $a \rightarrow 0$.
Compared to numerical simulations, these results turn out to be very accurate well above the critical point of the noisy 
voter model, $a/h\gg N^{-1}$, but fail below or close to it. The failure of this approach is ultimately linked to the ansatz of the scaling of the correlations with system size.
As we approach the critical point $a \rightarrow 0$, it is not true that fluctuations diverge but there is a change in the scaling with $N$, in the noisy voter case becoming $\sigma_{{\rm st}}^2[m] \sim N^{0}$.

Our second approach is based on a system size expansion around a stochastic dynamical attractor. Although this method turns out to be more precise than the expansion around the deterministic solution and fits much better the numerical data for the noisy voter model, it is much more involved and its applicability must be analyzed for the particular model under study. Exploring the phase space of deterministic solutions, we were able to identify a special trajectory $m(t) = m_L(t)$ and $\rho(t) \approx \xi (1-m_L(t)^2)$ ($m_L$ being the link magnetization) around which the dynamics is fast and the fluctuations do fulfill a van~Kampen type of scaling, namely $\sigma_{{\rm st}}^2[m-m_L] \sim N^{-1}$, for all values of $a$. Using this expansion we are able to obtain expressions for all the desired quantities, not limited to the stationary values but including the dynamics of $\langle \rho(t) \rangle$ and $\langle m(t) \rangle$, which match perfectly compared to numerical simulations and work better than previous approaches to the same model, such as all-to-all or annealed network approximations. We highlight that the results of $\sigma_{{\rm st}}^2[m_L]$ are the same as the all-to-all network if we rescale the herding as $h \rightarrow 2 h \xi$ and the system-size as $N \rightarrow N_{{\rm eff}} = N \frac{\mu^2}{\mu_2}$ in terms of the first $\mu$ and second $\mu_2$ moments of the degree distribution, a result that was previously reported for the noiseless voter model. 
However, due to the effect of noise $a \neq 0$, it is not true that $\sigma_{{\rm st}}^2[m] \approx \sigma_{{\rm st}}^2[m_L]$, 
which implies that finite size effects in a complex network can not be reduced to replacing the system size $N$ by an effective 
one $N_{{\rm eff}}$.

\section*{Acknowledgments}
We acknowledge financial support from the Agencia Estatal de Investigacion (AEI, Spain) and Fondo Europeo de Desarrollo Regional under project ESOTECOS FIS2015-63628-C2-2-R (MINECO/AEI/FEDER, UE) and the Spanish State Research Agency, through the Mar{\'\i}a de Maeztu Program for Units of Excellence in R\&D (MDM-2017-0711). A. F. P. acknowledges support by the FPU program of MECD (Spain).

\appendix
\section{Solution of Eqs.~(\ref{correlationskk}--\ref{correlationsrhorho})}\label{app:ckk}

Replacing the Jacobian Eqs. (\ref{Jacobian_vkk}--\ref{Jacobian_rho2}) and the steady state value Eqs. (\ref{G_{kk}}--\ref{G_{krho}}) into Eqs. (\ref{correlationskk}--\ref{correlationsrhorho}), these latter can be reduced to
\begin{eqnarray}
\label{correlationskk_s}
4(a+h\xi)C_{k,k'} - 2h\xi \sum_{k''} \frac{P_{k''}k''}{\mu} \left( C_{k,k''} + C_{k',k''} \right) &=& \frac{4(a+h\xi)}{P_{k}} \delta_{k,k'} \, ,\\
\label{correlationskrho_s}
(2a+2h\xi+J_{\rho,\rho})C_{k,\rho} - 2h\xi \sum_{k''} \frac{P_{k''}k''}{\mu} C_{k'',\rho} &=& 0 \, ,\\
\label{correlationsrhorho_s}
2 C_{\rho,\rho} J_{\rho,\rho} &=& G_{\rho} \, .
\end{eqnarray}
From Eq. (\ref{correlationsrhorho_s}) we get $C_{\rho,\rho}=G_{\rho}/2J_{\rho,\rho}$, while from Eq. (\ref{correlationskrho_s}), by inspection, we note that $C_{k,\rho}$ is a constant that does not depend on $k$. Reinserting this independence of $k$ back into the equation, we notice that the only possibility is the trivial solution $C_{k,\rho}=0$. Eq. (\ref{correlationskk_s}) is more involved. Let us define $D_{k}=\sum_{k'} \frac{P_{k'} k'}{\mu} C_{k,k'}$ and $D=\sum_{k} \frac{P_{k} k}{\mu} D_k$, so the equation becomes
\begin{equation}
\label{correlationskk_s2}
4(a+h\xi)C_{k,k'} - 2h\xi (D_k+D_{k'})= \frac{4(a+h\xi)}{P_{k}} \delta_{k,k'} \, .
\end{equation}
We now perform the double sum $\sum_{k} \frac{P_{k} k}{\mu} \sum_{k'} \frac{P_{k'} k'}{\mu}$, obtaining in this way a closed relation for $D$
\begin{equation}
\label{D}
D = \frac{a+h \xi}{a} \frac{\mu_2}{\mu^2} \, .
\end{equation}
If we perform instead the single sum $\sum_{k'} \frac{P_{k'} k'}{\mu}$, we obtain an expression for $D_{k}$ as a function of $D$
\begin{equation}
\label{Dk}
D_{k} = \frac{h\xi}{2a+h\xi} D + \frac{k}{\mu} \frac{2(a+h\xi)}{2a+h\xi} \, ,
\end{equation}
and we then solve $C_{k,k'}$
\begin{equation}
C_{k,k'} = \frac{h\xi}{2(a+h\xi)} \left( D_{k} + D_{k'} \right) + \frac{1}{P_{k}}\delta_{k,k'} \, .
\label{Ckk_B}
\end{equation}
Finally, replacing Eq. (\ref{D}) into Eq. (\ref{Dk}) and this latter into Eq. (\ref{Ckk_B}), we obtain Eq. (\ref{Ckk}).

\section{A linear noise example}\label{app:linear}

Consider the simple example of two coupled linear stochastic equations for variables $x(t)$ and $y(t)$
\begin{eqnarray}
\label{lin_ex_slow}
\dot x &= -\frac{1}{\tau_{x}} x + N^{-1/2} \xi_{x}(t) \, , \\
\label{lin_ex_fast}
\dot y &= -\frac{1}{\tau_{x}} y + \frac{1}{\tau_{y}} (x-y) + N^{-1/2} \xi_{y}(t) \, ,
\end{eqnarray}
where $\xi_{x,y}(t)$ are two uncorrelated Gaussian white noise variables $\langle \xi_{a}(t) \xi_{b}(t') \rangle = \delta_{a,b} \delta(t-t')$. The $x$ variable is independent of $y$ and has a stationary variance $\langle x^2 \rangle_{{\rm st}} = \tau_{x}/(2N)$. This variance fulfills the van~Kampen ansatz as long as $\tau_{x} = O(1)$, while when the time scale diverges $\tau_{x} = O(N^{\alpha})$ with $0 \leq \alpha \leq 1$, the variance has an anomalous scaling $\langle x^2 \rangle_{{\rm st}} = O(N^{\alpha-1})$. The deterministic part of the equation has the solution $y(t) = x(t)$ which is an attractor of trajectories. We then propose to split the $y$ variable as $y(t) = x(t) + \varepsilon(t)$, with two stochastic parts: a slow one $x(t)$, based on the solution of the deterministic attractor, and a fast one $\varepsilon(t)$. Introducing this change of variables in Eq. (\ref{lin_ex_fast}), we find
\begin{equation}
\label{lin_ex_change}
\dot \varepsilon = - \left( \frac{1}{\tau_{x}} + \frac{1}{\tau_{y}} \right) \varepsilon + N^{-1/2} \xi_{y}(t) - N^{-1/2} \xi_{x}(t) \, ,
\end{equation}
whose variance is $\langle \varepsilon^2 \rangle_{{\rm st}} = N^{-1}\frac{\tau_{x} \tau_{y}}{\tau_{x}+\tau_{y}}$. Hence, as long as $\tau_{y}=O(1)$, and regardless of the scaling of $\tau_{x}$, the variance of $\varepsilon$ always scales as $\langle \varepsilon^2 \rangle_{{\rm st}}=O(N^{-1})$, which justifies the use of a van~Kampen type of expansion for this variable in the whole range of values of $\tau_{x}$.

\section{Solution of Eq. (\ref{determ_rho2})}\label{app:riccati}

The equation is
\begin{equation}
-\phi_L \frac{d\phi^*_\rho}{d\phi_L}=1+\alpha \phi^*_\rho-\beta\frac{{\phi^*_\rho}^2}{1-\phi_L^2} \, ,
\label{eq:riccati}
\end{equation}
with $\alpha=\frac{h}{a}(1-2/\mu)-2$, $\beta=\frac{2h}{a}(1-1/\mu)$ and boundary conditions $\phi^*_\rho(\phi_L=\pm 1)=0$. Note that, the value $\phi^*_\rho(\phi_L=0)\equiv y_0$ is determined by the (positive) solution of $1+\alpha y_0-\beta y_0^2=0$ or $\phi^*_\rho(0)=\xi$, where $\xi$ is given by Eq. (\ref{xi_eq}).

\begin{figure}[ht]
\centering
\subfloat[]{\label{fig_riccati:a}\includegraphics[width=0.45\textwidth]{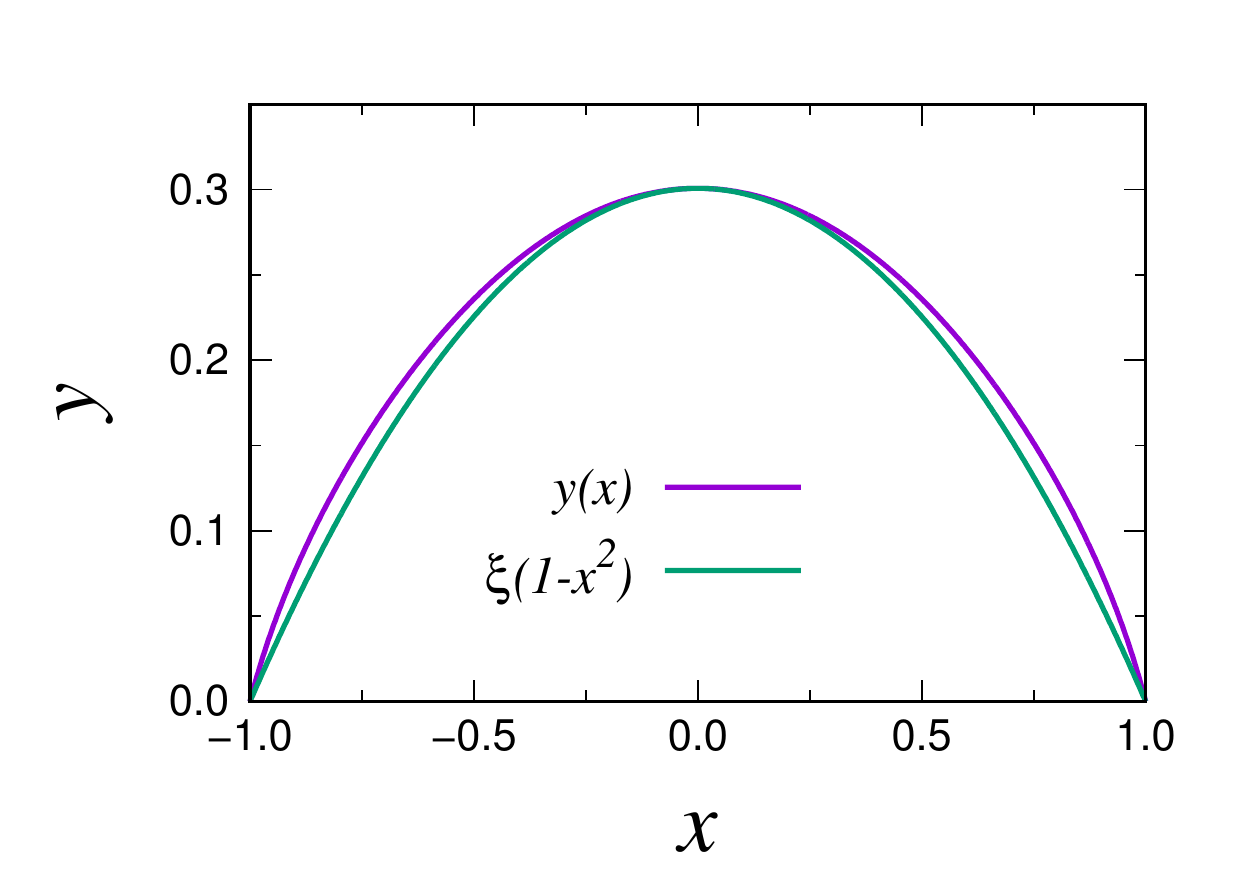}}
\subfloat[]{\label{fig_riccati:b}\includegraphics[width=0.45\textwidth]{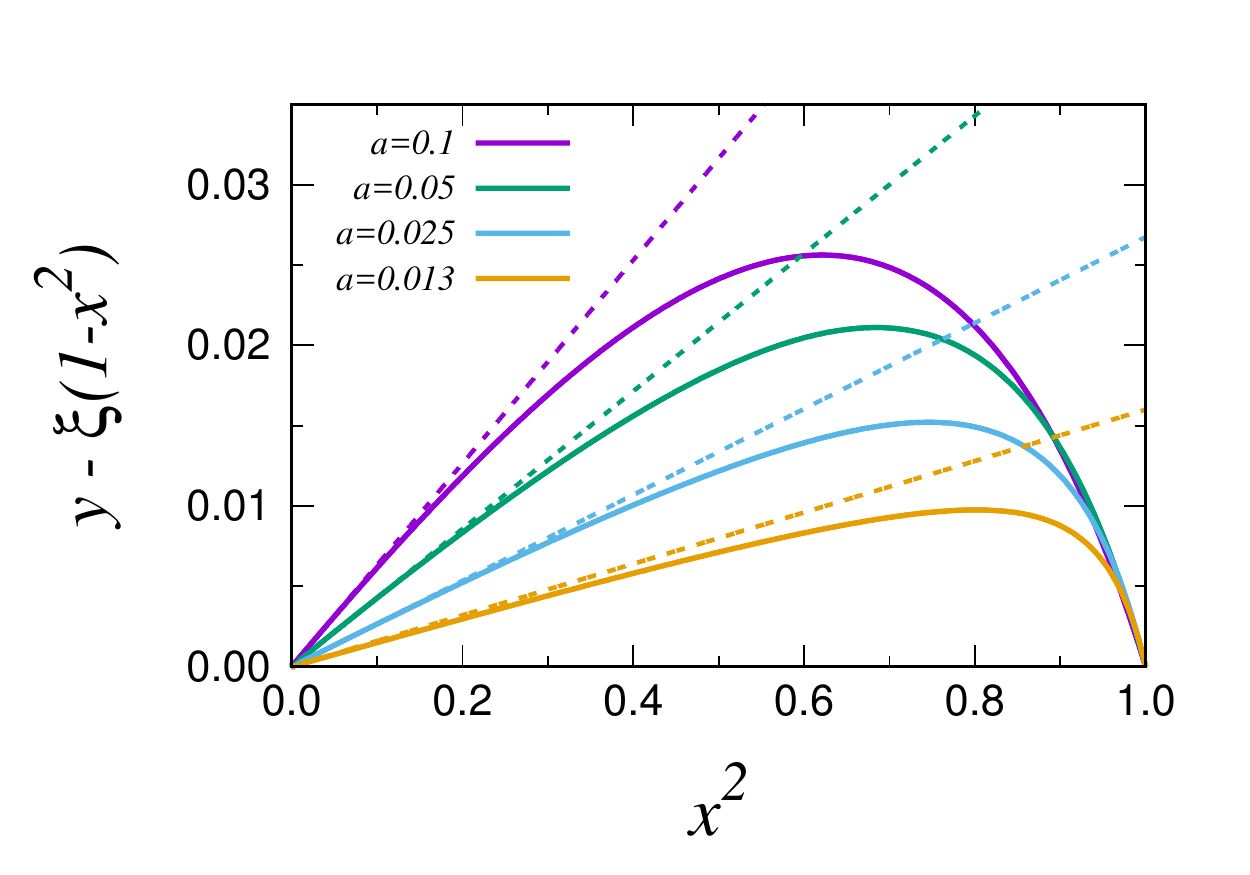}}
\caption{Exact solution Eq.~(\ref{Sol_Rica}) compared to the approximation Eq.~(\ref{rho_proposed}). The two expressions are shown in Fig.~\ref{fig_riccati:a} for $a=0.1$, $h=1$ and $\mu=3$ and their difference in Fig.~\ref{fig_riccati:b} for different values of $a$ with the same other parameters. The dashed lines correspond to the upper bound proportional to $\Lambda a\phi_L^2$.}
\label{fig_riccati}
\end{figure}

To find the general solution of the Ricatti-type equation (\ref{eq:riccati}) we follow a standard procedure. We first introduce the change of variables $s=\phi_L^2$, which reduces the equation to
\begin{equation}
-2s \frac{d\phi^*_\rho}{ds}=1+\alpha \phi^*_\rho-\beta\frac{{\phi^*_\rho}^2}{1-s} \, .
\label{eq:riccati2}
\end{equation}
Then the change of variables
\begin{equation}
\phi^*_\rho(s)=-\frac{2s(1-s)}{\beta z(s)}\frac{dz(s)}{ds} \, ,
\label{eq:change}
\end{equation}
leads to the second-order linear differential equation
\begin{equation}
\frac{d^2z}{ds^2}+\frac{2+\alpha-(4+\alpha)s}{2s(1-s)}\frac{dz}{ds}-\frac{\beta}{4(1-s)s^2}z=0 \, ,
\label{eq:riccatil}
\end{equation}
whose solution is
\begin{equation}
 z(s)=Cs^{\lambda_1}F(\lambda_1,1+\lambda_2;1+\lambda_1+\lambda_2;s)+s^{-\lambda_2}F(-\lambda_2,1-\lambda_1;1-\lambda_1-\lambda_2;s) \, ,
\label{eq:solz}
\end{equation}
where $C$ is an integration constant (another irrelevant global multiplicative constant has been set equal to one). Note that $F(a,b;c;s)$ is the Gauss hypergeometric function and that we have introduced the notation
\begin{eqnarray}
\lambda_1&=&\frac14(-\alpha+\sqrt{\alpha^2+4\beta}) \, ,\\
\lambda_2&=&\frac14(\alpha+\sqrt{\alpha^2+4\beta})=\frac{\beta}{2}\xi \, .
\end{eqnarray}
We note that, independently of the value of the constant $C$, the solution verifies $\phi^*_\rho(0)=\xi$. The constant $C$ is obtained by demanding that $\phi^*_\rho(1)=0$. Using the known expansion~\cite{wolfram} of the hypergeometric function
\begin{eqnarray}
F(a,b;a+b;s)&=&-\frac{\Gamma(a+b)}{\Gamma(a)\Gamma(b)}\log(1-s)+O((1-s)^1) \, ,
\end{eqnarray}
which is valid for $s\to 1$, we arrive at
\begin{equation}
C=-\frac{\Gamma(1-\lambda_1-\lambda_2)\Gamma(\lambda_1)\Gamma(1+\lambda_2)}{\Gamma(1+\lambda_1+\lambda_2)\Gamma(1-\lambda_1)\Gamma(-\lambda_2)} \, .
\end{equation}
The solution is finally written as
\begin{equation}
\label{Sol_Rica}
\phi^*_\rho(\phi_L)=\xi(1-\phi_L^2)G(\phi_L^2) \, ,
\end{equation}
where
\begin{equation}
G(s)=-\frac{sz'(s)}{\lambda_2z(s)} \, .
\end{equation}
For a more explicit expression of $\phi^*_\rho(\phi_L)$ in terms of hypergeometric functions we could perform the derivative $z'(s)=\frac{dz(s)}{ds}$ using
\begin{equation}
\frac{dF(a,b;c;s)}{ds}=\frac{a b}{c}F(a+1,b+1;c+1;s) \, .
\end{equation}
By expanding $G(s)$ around $s=0$ it is possible to find the following bound
\begin{equation}
\left|\phi^*_\rho(\phi_L)-\xi(1-\phi_L^2)\right|=\xi(1-\phi_L^2)(G(\phi_L^2)-1)<\Lambda a\phi_L^2 \, ,
\end{equation}
with
\begin{equation}
 \Lambda=\frac{2\mu}{h(\mu-2)^2+2a\mu^2+\mu\sqrt{4(1-\mu)h^2+\mu^2(h+2a)^2}}=\frac{1}{2a+h}\frac{1}{\mu}+O(\mu^{-2}) \, .
\end{equation}
Note that the difference $\left|\phi^*_\rho(\phi_L)-\xi(1-\phi_L^2)\right|$ is a measure of the error introduced by using the approximate form $\xi(1-\phi_L^2)$ instead of the exact function $\phi^*_\rho(\phi_L)$, as done in Eq.~(\ref{rho_proposed}), and thus the previous equation establishes an upper bound for this error. The validity of the approximation and its error bound are graphically proven in Fig.~\ref{fig_riccati}, where we plot the difference $\left|\phi^*_\rho(\phi_L)-\xi(1-\phi_L^2)\right|$ as a function of $m_L^2$ and the straight line $\Lambda a m_L^2$.

If we now assume that $m_L$ fulfills the same scaling properties as $m$ described earlier in Eqs.~(\ref{first_FC}--\ref{second_FC}), then $\langle m_L^2 \rangle_{{\rm st}} \sim O(N^{0})$ if $a\sim O(N^{-1})$ and $\langle m_L^2 \rangle_{{\rm st}} \sim O(N^{-1})$ if $a\sim O(N^{0})$, and it turns out that the difference between the exact function $\phi_\rho^*$ and the approximation Eq.~(\ref{rho_proposed}) is of order $N^{-1}$ and thus it can be neglected at the order of our approximation.

\section{Master equation for the link magnetization}\label{app:link}

We present in this appendix a simpler method, based on a master equation approach, to compute the statistical properties of the link magnetization, including fluctuations and finite-size effects \cite{Peralta2018}. This approach is very similar to the all-to-all approximation, and it can be derived following the same steps. We only consider one description variable $m_L$ to be relevant. The possible processes are $m_L \rightarrow m_L \pm \Delta_{k}$, with $\Delta_{k}= \frac{2 k}{\mu N}$, where the sign $-$ (resp. $+$) corresponds to the switching of a node of degree $k$ from the up (resp. down) to the down (resp. up) state. The master equation then reads
\begin{equation}
\label{master_link}
\frac{\partial P(m_L;t)}{\partial t} = \sum_{k} \left( E_{m_L}^{\Delta_{k}}-1 \right) \left[ W^{-}_{k} P \right] + \left( E_{m_L}^{-\Delta_{k}}-1 \right) \left[ W^{+}_{k} P \right] \, ,
\end{equation}
where $W^{\pm}_{k}$ are the effective rates of the proposed processes, that only depend on the description variable $m_L$. We assume that all nodes with degree $k$ have the same rate of switching to the other state, replacing the local density $\frac{\sum_{j} A_{ij} n_{j}}{k_i}$ by the global density of up nodes connected to nodes in state up or down, which coincides with the $c_{0/1}$ calculated in Eq.~(\ref{binom_prob}) as $\frac{\rho}{1 \pm m_L}$. Additionally, and this is the key point, we assume that all the other variables follow strictly the deterministic attractor and do not deviate at all, i.e., $m_k=m_L$, $\rho=\xi (1-m_L^2)$. We then have, for the noisy voter model model,
\begin{eqnarray}
\label{effective_link_rates}
W^{+}_{k} &=& \frac{N_{k}}{2} (1-m_L) \left( a + h \xi (1+m_L) \right) \, , \\
W^{-}_{k} &=& \frac{N_{k}}{2} (1+m_L) \left( a + h \xi (1-m_L) \right) \, .
\end{eqnarray}
Following~\cite{Vazquez2008}, we take the continuum limit by expanding the master equation~(\ref{master_link}) in powers of $\Delta_{k}$ to second order
\begin{equation}
\label{FP_link}
\frac{\partial P(m_L;t)}{\partial t} = -\frac{\partial}{\partial m_L} \big[ f_{L}(m_L) P\big] + \frac{1}{2} \frac{\partial^2}{\partial m_L^2}\big[ g_{L}(m_L) P\big] \, ,
\end{equation}
where now $P(m_L;t)$ must be understood as a probability density function and the functions $f_L(m_L)$ and $g_L(m_L)$ are defined as
\begin{eqnarray}
f_L(m_L) &=& \sum_{k} \Delta_{k} [ W^{+}_{k} - W^{-}_{k}]=-2am_L \, ,\\
g_L(m_L) &=& \sum_{k} \Delta_{k}^2 [ W^{+}_{k} + W^{-}_{k}]=4( a + h\xi (1-m_L^2))/N_{\rm eff} \, .
\end{eqnarray}
The time evolution of the moments $\langle m_L(t) \rangle$ and $\langle m_L(t)^2 \rangle$ can be obtained by integrating both sides of the Fokker-Planck equation~(\ref{FP_link}) respectively as $\int m_L dm_L$ and $\int m_L^2 dm_L$, such that one reproduces the results obtained in Eqs.~(\ref{dmL}) and~(\ref{mL_eq}).

If we take the same continuum limit in the master equation~(\ref{me2}) of the all-to-all approach, we find, for the probability $P(m;t)$, an equation similar to Eq.~(\ref{FP_link}), but dependent now on functions $f(m)$ and $g(m)$, defined as
\begin{eqnarray}
f(m) &=& \frac{2}{N} [ W^{+} - W^{-} ]=-2am \, ,\\
g(m) &=& \frac{4}{N^2} [ W^{+} + W^{-} ]=2(2 a + h (1-m^2))/N \, .
\end{eqnarray}
These functions are equivalent to $f_L$ and $g_L$ if we replace $h \rightarrow 2 h \xi$, $N \rightarrow N_{{\rm eff}}$ and $m \rightarrow m_L$, in agreement with the equivalence of Eq.~(\ref{mL_st}) and Eq.~(\ref{second_FC}) under the same replacement.

\bibliographystyle{unsrt}

\begin{thebibliography}{10}

\bibitem{Gunton1983}
J. D. Gunton, M. San Miguel, and P. S. Sahni.
\newblock {The Dynamics of First Order Phase Transitions}.
\newblock In {\em Phase Transitions and Critical Phenomena} {\bf 8}, 269 (1983).

\bibitem{Marro1999}
J. Marro and R. Dickman.
\newblock {\em {Nonequilibrium phase transitions in lattice models}}.
\newblock Cambridge University Press, 1999.

\bibitem{Hopfield1982}
J. J. Hopfield.
\newblock {Neural networks and physical systems with emergent collective
 computational abilities}.
\newblock {\em Proceedings of the National Academy of Sciences} {\bf 79}, 2554 (1982).

\bibitem{Clifford1973}
P. Clifford and A. Sudbury.
\newblock {A model for spatial conflict}.
\newblock {\em Biometrika} {\bf 60}, 581 (1973).

\bibitem{Crawley1987}
M.~J. Crawley and R.~M. May.
\newblock {Population dynamics and plant community structure: Competition
 between annuals and perennials}.
\newblock {\em Journal of Theoretical Biology} {\bf 125}, 475 (1987).

\bibitem{Anderson1991}
R.~M. Anderson, R.~M. May, and B. Anderson.
\newblock {\em {Infectious Diseases of Humans: Dynamics and Control}}.
\newblock Oxford University Press, Oxford, 1991.

\bibitem{PastorSatorras2001}
R. Pastor-Satorras and A. Vespignani.
\newblock {Epidemic Spreading in Scale-Free Networks}.
\newblock {\em Physical Review Letters} {\bf 86}, 3200 (2001).

\bibitem{Watts2002}
D.~J. Watts.
\newblock {A simple model of global cascades on random networks}.
\newblock {\em Proceedings of the National Academy of Sciences} {\bf 99}, 5766 (2002).

\bibitem{Castellano2009}
C. Castellano, S. Fortunato, and V. Loreto.
\newblock {Statistical physics of social dynamics}.
\newblock {\em Reviews of Modern Physics} {\bf 81}, 591 (2009).

\bibitem{Castellano2012}
C. Castellano and R. Pastor-Satorras.
\newblock {Competing activation mechanisms in epidemics on networks}.
\newblock {\em Scientific Reports} {\bf 2}, 371 (2012).

\bibitem{Albert2002}
R. Albert and A.-L. Barab{\'{a}}si.
\newblock {Statistical mechanics of complex networks}.
\newblock {\em Reviews of Modern Physics} {\bf 74}, 47 (2002).

\bibitem{Newman2003}
M. E. J. Newman and J. Park.
\newblock {Why social networks are different from other types of networks}.
\newblock {\em Physical Review E} {\bf 68}, 36122 (2003).

\bibitem{Barrat2008}
A. Barrat, M. Barthelemy, and A. Vespignani.
\newblock {\em {Dynamical processes on complex networks}}.
\newblock Cambridge University Press, 2008.

\bibitem{Newman2010}
M. E. J. Newman.
\newblock {\em {Networks: an introduction}}.
\newblock Oxford University Press, Oxford, 2010.

\bibitem{Lambiotte2007}
R. Lambiotte.
\newblock {How does degree heterogeneity affect an order-disorder transition?}
\newblock {\em Europhysics Letters} {\bf 78}, 68002 (2007).

\bibitem{Gleeson2011}
J. P. Gleeson.
\newblock {High-Accuracy Approximation of Binary-State Dynamics on Networks}.
\newblock {\em Physical Review Letters} {\bf 107}, 68701 (2011).

\bibitem{Gleeson2013}
J. P. Gleeson.
\newblock {Binary-State Dynamics on Complex Networks: Pair Approximation and Beyond}.
\newblock {\em Physical Review X} {\bf 3}, 21004 (2013).

\bibitem{Dorogovtsev2002}
S. N. Dorogovtsev, A. V. Goltsev, and J. F. F Mendes.
\newblock {Ising model on networks with an arbitrary distribution of
 connections}.
\newblock {\em Physical Review E} {\bf 66}, 16104 (2002).

\bibitem{Leone2002}
M. Leone, A. V{\'{a}}zquez, A. Vespignani, and R. Zecchina.
\newblock {Ferromagnetic ordering in graphs with arbitrary degree
 distribution}.
\newblock {\em European Physical Journal B} {\bf 28}, 191 (2002).

\bibitem{VianaLopes2004}
J. Viana Lopes, Y. G. Pogorelov, J. M. B. dos Santos, and R. Toral.
\newblock {Exact solution of Ising model on a small-world network}.
\newblock {\em Physical Review E} {\bf 70}, 26112 (2004).

\bibitem{Boguna2003}
M. Bogu{\~{n}}{\'{a}}, R. Pastor-Satorras, and A. Vespignani.
\newblock {Absence of Epidemic Threshold in Scale-Free Networks with Degree
 Correlations}.
\newblock {\em Physical Review Letters} {\bf 90}, 28701 (2003).

\bibitem{Durrett2010}
R. Durrett.
\newblock {Some features of the spread of epidemics and information on a random
 graph}.
\newblock {\em Proceedings of the National Academy of Sciences} {\bf 107}, 4491 (2010).

\bibitem{Castellano2010}
C. Castellano and R. Pastor-Satorras.
\newblock {Thresholds for Epidemic Spreading in Networks}.
\newblock {\em Physical Review Letters} {\bf 105}, 218701 (2010).

\bibitem{Parshani2010}
R. Parshani, S. Carmi, and S. Havlin.
\newblock {Epidemic Threshold for the Susceptible-Infectious-Susceptible Model on Random Networks}.
\newblock {\em Physical Review Letters} {\bf 104}, 258701 (2010).

\bibitem{Satorras2015}
R. Pastor-Satorras, C. Castellano, P. Van~Mieghem, and A. Vespignani.
\newblock {Epidemic processes in complex networks.}
\newblock {\em Reviews of Modern Physics} {\bf 87}, 925 (2015).

\bibitem{Alfarano2009}
S. Alfarano and M. Milakovi{\'{c}}.
\newblock {Network structure and N-dependence in agent-based herding models}.
\newblock {\em Journal of Economic Dynamics and Control} {\bf 33}, 78 (2009).

\bibitem{Chakrabarti2008}
D. Chakrabarti, Y. Wang, C. Wang, J. Leskovec, and C. Faloutsos.
\newblock {Epidemic thresholds in real networks.}
\newblock {\em ACM Trans. Inf. Syst. Secur.} {\bf 10}, 1 (2008).

\bibitem{Gomez2010}
S. G\'omez, A. Arenas, J. Borge-Holthoefer, S. Meloni, and Y.~Moreno.
\newblock Discrete-time markov chain approach to contact-based disease
 spreading in complex networks.
\newblock {\em Europhysics Letters} {\bf 89}, 38009 (2010).

\bibitem{vanKampen1981}
N. G. Van Kampen.
\newblock {\em {Stochastic Processes in Physics and Chemistry}}.
\newblock North-Holland, Amsterdam, 1981.

\bibitem{Lafuerza2013}
L. F. Lafuerza and R. Toral.
\newblock {On the effect of heterogeneity in stochastic interacting-particle systems}.
\newblock {\em Scientific Reports} {\bf 3}, 1189 (2013).

\bibitem{Keeling}
M. J. Keeling.
\newblock {Multiplicative moments and measures of persistence in ecology}.
\newblock {\em J. Theor. Biol.} {\bf 205}, 269 (2000).

\bibitem{Rand}
D. A. Rand.
\newblock {Correlation equations and pair approximations for spatial ecologies}.
\newblock {\em CWI Quarterly} {\bf 12}, 329 (1999).

\bibitem{Demirel}
G. Demirel, F. Vazquez, G. A. B\"ohme, and T. Gross
\newblock {Moment-closure approximations for discrete adaptive networks}.
\newblock {\em Physica D} {\bf 267}, 68 (2014).

\bibitem{Bauch}
C. T. Bauch.
\newblock {The spread of infectious diseases in spatially structured populations: an invasory pair approximation}.
\newblock {\em Math. Biosci.} {\bf 198}, 217 (2005).

\bibitem{Kuehn}
C. Kuehn.
\newblock {Moment Closure - A Brief Review}.
\newblock {\emph{Control of Self-Organizing Nonlinear Systems. Understanding Complex Systems.} Springer, 2016.}

\bibitem{Sood2005B}
V. Sood and S. Redner.
\newblock {Voter Model on Heterogeneous Graphs}.
\newblock {\em Physical Review Letters} {\bf 94}, 178701 (2005).

\bibitem{Sood2008}
V. Sood, T. Antal, and S. Redner.
\newblock {Voter models on heterogeneous networks}.
\newblock {\em Physical Review E} {\bf 77}, 41121 (2008).

\bibitem{Vilone2004}
D. Vilone and C. Castellano.
\newblock {Solution of voter model dynamics on annealed small-world networks}.
\newblock {\em Physical Review E} {\bf 69}, 16109 (2004).

\bibitem{Dorogovtsev2008}
S. N. Dorogovtsev, A. V. Goltsev, and J. F. F. Mendes.
\newblock {Critical phenomena in complex networks}.
\newblock {\em Reviews of Modern Physics} {\bf 80}, 1275 (2008).

\bibitem{Guerra2010}
B. Guerra and J. G{\'{o}}mez-Garde{\~{n}}es.
\newblock {Annealed and mean-field formulations of disease dynamics on static
 and adaptive networks}.
\newblock {\em Physical Review E} {\bf 82}, 35101 (2010).

\bibitem{Sonnenschein2012}
B. Sonnenschein and L. Schimansky-Geier.
\newblock {Onset of synchronization in complex networks of noisy oscillators}.
\newblock {\em Physical Review E} {\bf 85}, 51116 (2012).

\bibitem{Carro2016}
A. Carro, R. Toral, and M. San Miguel.
\newblock The noisy voter model on complex networks.
\newblock {\em Scientific Reports} {\bf 6}, 24775 (2016).

\bibitem{Vazquez2008}
F. Vazquez and V. M. Egu{\'{\i}}luz.
\newblock {Analytical Solution of the Voter Model on Uncorrelated Networks}.
\newblock {\em New Journal of Physics} {\bf 10}, 63011 (2008).

\bibitem{Vazquez2008b}
F. Vazquez, V. M. Egu{\'{\i}}luz, and M. San Miguel.
\newblock {Generic Absorbing Transition in Coevolution Dynamics}.
\newblock {\em Physical Review Letters} {\bf 100}, 108702 (2008).

\bibitem{Vazquez2010}
F. Vazquez, X. Castell{\'{o}}, and M. San Miguel.
\newblock {Agent based models of language competition: macroscopic descriptions
 and order–disorder transitions}.
\newblock {\em Journal of Statistical Mechanics: Theory and Experiment} {\bf 2010}, P04007 (2010).

\bibitem{Oliveira1993}
M. J. de~Oliveira, J. F. F. Mendes, and M. A. Santos.
\newblock Nonequilibrium spin models with ising universal behaviour.
\newblock {\em J. Phys. A} {\bf 26}, 2317 (1993).

\bibitem{Dickman1986}
R. Dickman.
\newblock Kinetic phase transitions in a surface-reaction model: Mean-field theory.
\newblock {\em Phys. Rev. A} {\bf 34}, 4246 (1986).

\bibitem{Pugliese2009}
E. Pugliese and C. Castellano.
\newblock Heterogeneous pair approximation for voter models on networks.
\newblock {\em Europhysics Letters} {\bf 88}, 58004 (2009).

\bibitem{Ferreira2014}
A. S. Mata, R. S. Ferreira, and S. C. Ferreira.
\newblock Heterogeneous pair-approximation for the contact process on complex networks.
\newblock {\em New Journal of Physics} {\bf 16}, 053006 (2014).

\bibitem{Vespignani2011}
A. Vespignani.
\newblock Modelling dynamical processes in complex socio-technical systems.
\newblock {\em Nature Physics} {\bf 8}, 32 (2011).

\bibitem{Ferreira2011}
S. C. Ferreira, R. S. Ferreira, and R. Pastor-Satorras.
\newblock Quasistationary analysis of the contact process on annealed scale-free networks.
\newblock {\em Phys. Rev. E} {\bf 83}, 066113 (2011).

\bibitem{Ferreira2011B}
S. C. Ferreira, R. S. Ferreira, C. Castellano, and R. Pastor-Satorras.
\newblock Quasistationary simulations of the contact process on quenched networks.
\newblock {\em Phys. Rev. E} {\bf 84}, 066102 (2011).

\bibitem{Peralta2018}
A. F. {Peralta}, A. {Carro}, M. {San Miguel}, and R. {Toral}.
\newblock {Analytical and numerical study of the non-linear noisy voter model
 on complex networks}.
\newblock {\em Chaos} {\bf 28}, 075516 (2018).

\bibitem{moments}
A. F. Peralta and R. Toral.
\newblock System-size expansion of the moments of a master equation.
\newblock To appear in \emph{Chaos} (October 2018).

\bibitem{Holley1975}
R. A. Holley and T. M. Liggett.
\newblock {Ergodic Theorems for Weakly Interacting Infinite Systems and the Voter Model}.
\newblock {\em The Annals of Probability} {\bf 3}, 643 (1975).

\bibitem{Moran1958}
P. A. P. Moran.
\newblock Random processes in genetics.
\newblock {\em Mathematical Proceedings of the Cambridge Philosophical Society} {\bf 54}, 60 (1958).

\bibitem{Lebowitz1986}
J. L. Lebowitz and H. Saleur.
\newblock {Percolation in strongly correlated systems}.
\newblock {\em Physica A: Statistical Mechanics and its Applications} {\bf 138}, 194 (1986).

\bibitem{Fichthorn1989}
K. Fichthorn, E. Gulari, and R. Ziff.
\newblock {Noise-induced bistability in a Monte Carlo surface-reaction model}.
\newblock {\em Physical Review Letters} {\bf 63}, 1527 (1989).

\bibitem{Considine1989}
D. Considine, S. Redner, and H. Takayasu.
\newblock {Comment on ``Noise-induced bistability in a Monte Carlo surface-reaction model''}.
\newblock {\em Physical Review Letters} {\bf 63}, 2857 (1989).

\bibitem{Kirman1993}
A. Kirman.
\newblock {Ants, rationality and recruitment}.
\newblock {\em Quarterly Journal of Economics} {\bf 108}, 137 (1993).

\bibitem{Granovsky1995}
B. L. Granovsky and N. Madras.
\newblock {The noisy voter model}.
\newblock {\em Stochastic Processes and their Applications} {\bf 55}, 23 (1995).

\bibitem{Alfarano2008}
S. Alfarano, T. Lux, and F. Wagner.
\newblock {Time variation of higher moments in a financial market with heterogeneous agents: An analytical approach}.
\newblock {\em Journal of Economic Dynamics and Control} {\bf 32}, 101 (2008).

\bibitem{Alfarano2013}
S. Alfarano, M. Milakovi{\'{c}}, and M. Raddant.
\newblock {A note on institutional hierarchy and volatility in financial markets}.
\newblock {\em The European Journal of Finance} {\bf 19}, 449 (2013).

\bibitem{Diakonova2015}
M. Diakonova, V. M. Egu{\'{\i}}luz, and M. San Miguel.
\newblock {Noise in coevolving networks}.
\newblock {\em Physical Review E} {\bf 92}, 32803 (2015).

\bibitem{Nagi}
N. Khalil, R. Toral, and M. San~Miguel.
\newblock Zealots in the mean-field noisy voter model.
\newblock {\em Phys. Rev. E} {\bf 97}, 012310 (2018).

\bibitem{Oriol}
O. Artime, A. F. Peralta, R. Toral, J. Ramasco, and M. San~Miguel.
\newblock {Aging-induced continuous phase transition}.
\newblock {\em Phys. Rev. E} {\bf 98}, 032104 (2018).

\bibitem{Toral2014}
R. Toral and P. Colet.
\newblock {\em Stochastic numerical methods: an introduction for students and scientists}.
\newblock John Wiley \& Sons, 2014.

\bibitem{Newman2003B}
M. E. J. Newman.
\newblock {The Structure and Function of Complex Networks}.
\newblock {\em SIAM Review} {\bf 45}, 167 (2003).

\bibitem{Boguna2004}
M. Bogu{\~{n}}{\'{a}}, R. Pastor-Satorras, and A. Vespignani.
\newblock {Cut-offs and finite size effects in scale-free networks}.
\newblock {\em European Physical Journal B} {\bf 38}, 205 (2004).

\bibitem{Bianconi2009}
G. Bianconi.
\newblock {Entropy of network ensembles}.
\newblock {\em Physical Review E} {\bf 79}, 36114 (2009).

\bibitem{Toral2007}
R. Toral and C. J. Tessone.
\newblock {Finite size effects in the dynamics of opinion formation}.
\newblock {\em Communications in Computational Physics} {\bf 2}, 177 (2007).

\bibitem{regularD}
A. F. Peralta and R. Toral.
\newblock Work in progress, 2018.

\bibitem{wolfram}
{Wolfram Research, Inc.}
\newblock Expansion of the {Gauss} hypergeometric function as a generalized power series around \mbox{$z=1$}.
\newblock \mbox{\tt{http://functions.wolfram.com/HypergeometricFunctions/}}
\mbox{\tt{Hypergeometric2F1/06/01/04/01/03/}}.
\newblock Accessed on: March 1st, 2018.

\end{thebibliography}

\end{document}